\documentclass[prc,aps,final,twocolumn,nofootinbib]{revtex4-1}
\usepackage{graphicx}
\usepackage{mathrsfs}
\usepackage{bm}
\def\r{{\bf r}}
\def\p{{\bf p}}
\def\d{\hbox{d}}
\def\be{\begin{equation}}
\def\ee{\end{equation}}
\def\bea{\begin{eqnarray}}
\def\eea{\end{eqnarray}}
\def\l{\label}
\def\Im{\mathop{\rm Im}}
\def\Re{\mathop{\rm Re}}
\def\Tr{\mathop{\rm Tr}}

\def\vareps{E}
\def\epsi{\mathcal{E}}
\def\simg{\gtrsim}
\def\siml{\lesssim}
\def\cK{\mathcal{K}}

\begin{document}



%
\title{SHELLS, ORBIT BIFURCATIONS AND SYMMETRY RESTORATIONS \\ IN FERMI SYSTEMS}

%
\author{A.G. Magner}
\email{Email: magner@kinr.kiev.ua}
\author{M.V. Koliesnik}
\affiliation{\it Institute for Nuclear Research, 03680 Kyiv, Ukraine} 
%
%
\author{K.\ Arita}
\affiliation{\it Department of Physics, Nagoya Institute of 
Technology, Nagoya 466-8555, Japan} 

\date{April 20, 2016}


\bigskip

%





%


\begin{abstract}
The periodic-orbit
theory based on the improved stationary-phase method
within the phase-space path integral approach 
is presented for the 
semiclassical description of the nuclear
shell structure, concerning the main topics of the fruitful activity
of V.\ G.\ Solovjov.
We apply this theory to 
study bifurcations and symmetry breaking phenomena in a
radial power-law potential
which is 
close to the realistic Woods-Saxon one up to about the
Fermi energy.
Using the realistic parametrization of nuclear shapes
we explain  the origin 
of the double-humped fission barrier 
and the
asymmetry in the fission isomer shapes
by the bifurcations of periodic orbits.  
The semiclassical origin of 
the oblate-prolate shape asymmetry
and tetrahedral shapes is also suggested
within the improved periodic-orbit approach.
The enhancement 
of shell structures at some surface diffuseness and deformation 
parameters of such shapes 
are explained by existence of the 
simple local bifurcations and 
new non-local bridge-orbit bifurcations 
in integrable and partially 
integrable
Fermi-systems.
We obtained good agreement between the 
semiclassical and quantum shell-structure
components of the level density and energy for several
surface diffuseness and deformation parameters of the potentials, 
including their symmetry breaking and 
 bifurcation values.
\end{abstract}
\maketitle



\section{INTRODUCTION}

Semiclassical periodic-orbit theory (POT) is a convenient tool for analytical 
studies of the shell structure in the single-particle level density of finite
fermionic systems near the Fermi surface 
\cite{gutz,bablo,strumag,bt76,smod,creagh,sclbook,migdalrev}. This theory
relates the oscillating level density and shell-correction energy 
to the sum of 
periodic orbits and their stability characteristics, and thus, gives 
the analytical quantum-classical correspondence. 
According to the shell-correction method (SCM) \cite{strut,fuhi}, 
the oscillating part of the total energy of a finite 
fermion system, the so-called shell-correction energy, is associated 
with an inhomogeneity of the single-particle (s.p.) energy levels near the 
Fermi surface. The SCM is based on Strutinsky's smoothing procedure 
to extract the shell components of the level density and energy, which has to
be added to the macroscopic parts, in particular, within the 
Liquid Drop Model (LDM) \cite{myswann69} or Extended Thomas-Fermi (ETF)
approach \cite{brguehak}. Deep foundations of the 
relation of a quasiparticle spectrum
near the Fermi surface to the finite many-body fermionic systems
with a strong particles' interaction, 
such as atomic nuclei, can be found in 
Refs.\ \cite{migdal,khodsap} which are based on the Landau quasiparticles' 
theory of Fermi liquids \cite{landau,abrikha}. 
Depending on the level density at the Fermi energy -- and with 
it the shell-correction energy -- being a maximum or a minimum, 
the nucleus is particularly unstable or stable, respectively. This
situation varies with particle numbers and deformation parameters
of the nucleus, and other parameters of its mean-field potential. 
In consequence, the shapes of stable nuclei depend strongly 
on particle numbers and deformations. 
The SCM was successfully used to describe nuclear masses and deformation
energies and, in particular, fission barriers of heavy nuclei, see  
the early review by Strutinsky's group, in which also the miscroscopic 
foundations of the SCM are discussed in \cite{fuhi}. 
Numerous other phenomena for which the shell effects
in deformed nuclei play a crucial role were studied in many other publications 
\cite{sclbook,migdalrev,bormot,ringschuck,hofbook}. One of them 
is related, e.g., to the description 
of the rotational bands at high nuclear spins \cite{fraupash,mix,sfraurev},
in particular by using the semiclassical POT  
\cite{magkolstr,kolmagstr,magkolstrizv1979,richter,fkmsprb1998,dfcpuprc2004,%
mskbg,mskbPRC2010,belyaev,GMBBPS2015,GMBBPS2016,belyaevzel}. 
The shell effects are always in a center of attention 
in the description of the collective nuclear dynamics, 
within the semi-microscopic approaches \cite{bormot,ringschuck,hofbook},
also within the physically transparent Quasipaticle-Phonon Model (QPM) 
\cite{SOLQPMI-EPNP1978,SOLQPMII-EPNP1980,SOLbook1981,SOLQPMIII-EPNP1983,%
SOLQPMIV-EPNP1983,SOLQPMV-EPNP1985,SOLbook1992}. 
The collective modes were intensively studied within this model, especially
in the complex deformed nuclei 
\cite{SOLbook1981,solovjov1,SOLbook1976}, see also the pioneer works
\cite{SOLsuperfluidity-NPA1958,belyaev61,SOLbook1963,belzel} concerning the
the pairing-correlation effects in nuclear physics, as well as
 within the more microscopical Hartree-Fock (HF) \cite{brink} and 
Hartree-Fock-Bogoliubov (HFB) approaches  (see e.g., 
\cite{ringschuck,brackquenNPA1981,abrpairing} and references therein),
and within the theory of finite Fermi systems \cite{migdal}.  
In addition, concerning the main content of this review article, 
we should mention calculations of the transport coefficients,
such as the inertia, friction and moments of inertia within the
response function theory 
\cite{hofbook,hofivyam,ivhofpasyam}, 
by applying the POT  
\cite{sfraurev,dfcpuprc2004,mskbg,mskbPRC2010,belyaev,GMBBPS2015,GMBBPS2016,magvvhof,GMFprc2007,MGFyaf2007,BMYijmpe2012,BMps2013}.

The idea was to use 
the POT for a deeper understanding, based on classical pictures, of the 
origin of the nuclear shell structure and its relation to a possible chaotic 
nature of the nucleons' dynamics
\cite{strumag,smod,nishioka,annphysik97,sclbook,migdalrev}.
This provides us with a transparent description in terms of the 
classical periodic orbits (POs) for answering, 
sometimes even analytically, some fundamental questions
concerning the physical origins of 
the double-humped fission barrier and, 
in particular, of the existence of the isomer minimum 
\cite{smod,sclbook,migdalrev,brreisie,spheroidpre,spheroidptp,magNPAE2010}.  
Gutzwiller was the first \cite{gutz} who suggested the Feynman path
integral representation for the Green's function to apply POT  in the case
of absence of any symmetry of the Hamiltonian, in addition to its time
independence.  In this case, the energy $E$ of the particle moving
in a mean-field potential is the only one integral of motion.  
For a given $E$, the POs 
are isolated, i.e.\ ,
any variation of the classical trajectory (CT) leads to a change of the
classical action along the CT. The Gutzwiller POT
was  extended to the continuous symmetries, as rotational and high 
(e.g., harmonic oscillator) ones, in \cite{strumag,smod,magosc} on the basis
of the early investigation of billiard-like systems within another 
Green's multiple-reflection method
by Balian and Bloch \cite{bablo}. Independently, Berry and Tabor \cite{bt76}
developed their direct Poisson-summation method using the
phase-space variables 
for POs having a high degeneracy\footnote{
The classical degeneracy is defined by the number of
independent parameters $\mathcal{K}$ for a continuous 
family of the classical periodic
orbits at a given energy of the particle.}
in integrable systems. 

Some applications of the 
POT to nuclear deformation energies with pronounced shell effects 
were presented 
and discussed by using the phase space variables 
\cite{bt76,smod,creagh,sclbook} and Maslov-Fedoriuk catastrophe
(turning- and caustic-point) theories
\cite{fedoryuk_pr,maslov,fedoryuk_book1,chester,MAFptp2006}.  
Within the improved stationary-phase method 
\cite{migdalrev,spheroidpre,spheroidptp,MAFptp2006,ellipseptp,kkmabPS2015}
(improved SPM, or ISPM), one can  solve the symmetry-breaking and 
 bifurcation problems\footnote{The simplest PO bifurcation is a change
of the number of solutions of the classical dynamic equations from one PO
(or, PO family) to two POs (or PO families) with a variation of potential
parameters.}. 
See also 
\cite{sclbook,ozoriobook,ullmo,creagh97,sie97,zphD,ssun,schomerus,hhun,%
brack2001,bmtjpa2001,kaidelbrack,bridge,arita2012,fmbpre2008}
concerning the bifurcation and normal-form theories and 
semi-analytical uniform approximations. The divergences and discontinuities
of the standard SPM (SSPM)
\cite{bablo,strumag,bt76,creagh,sclbook} near the symmetry-breaking 
and bifurcation points were removed, in particular within the 
analytical ISPM.

In the way to a more realistic semiclassical calculation, it is important
also to account for a diffuseness of the nuclear edge. As found in 
\cite{arita2012,aritapap}, 
the shell structure in the radial power-law 
potential (RPLP) is a good approximation 
to that of the familiar Woods-Saxon (WS)
potential for nuclei 
in the spatial domain where the particles are bound. We shall 
generalize the ISPM trace formulas \cite{migdalrev,MVApre2013}
for this potential from two to three dimensions, 
and discussed various limits to 
other known potentials as to the harmonic oscillator (HO) and the 
spherical billiard and to the SSPM results far from the symmetry-breaking and
bifurcation points.

In Section II, we outlook the main ingredients of the POT 
within the extended Gutzwiller
approach (EGA)
accounting for the bifurcation phenomenon by the ISPM.
Some general points of the phase space trace formulas are  
studied for families of the
maximal degeneracy in arbitrary spherical potentials.
The POT shell
components of the s.p. level density and energy for any Hamiltonians,
in particular, for a non-integrable potential 
like the H\'enon-Heiles (HH) Hamiltonian is presented too.
The isomer shapes within the fission cavity model with the
realistic deformation parameters are discussed  in Section III.
In Section IV we extend the 
semiclassical  ISPM
for the RPLP from the two- to the 
three-dimensional case.
The trace formulas for the level densities and energy shell
corrections will be derived for all PO families 
found by scaling of the RPLP classical dynamics.
Several asymptotics of the ISPM to the well known 
SSPM, billard 
and HO limits will be obtained.
The semiclassical calculations of the level density and energy shell
corrections
are
compared with the quantum results 
for the RPLP for different radial powers. 
The extensions of the RPLP POT to more realistic 
 deformed potentials with the surface diffuseness is given in Section V.
The POT 
shell structures  in these potentials with
the oblate-prolate and tetrahedral deformations are analyzed 
and comparison of the semiclassical and quantum results 
is displaced in this Section.
This review article is summarized in Section VI.
Some technical details of our POT calculations
are given in Appendix A.

\section{GENERAL POT INGREDIENTS}

In this section,
we shall outlook the POT within the extended
Gutzwiller approach by using the phase-space variables. The trace formulae
for the semiclassical level density Section IIA), 
the ISPM
and bifurcations (Section IIB), level density
averaging (Section IIC),
 and the shell-correction
energy (Section IID) 
will be presented in terms 
of the PO sums. 
Sections IIE and IIF 
will be devoted to the specific trace formulae 
and classical dynamics for
the integrable and non-integrable Hamiltonians, respectively.

\subsection{Phase-space trace formula} 

The level density, $g(\vareps)=\sum_i \delta(\vareps-\vareps_i)$, 
determined by spectrum levels $\vareps_i$ for the Hamiltonian $\hat{H}$
can be obtained approximately semiclassically by 
using the phase-space trace formula in $\mathcal{D}$ dimensions 
\cite{spheroidpre,spheroidptp,MAFptp2006,ellipseptp}:
\bea\l{pstrace}
&g_{\rm scl}(\vareps)=\frac{1}{(2\pi\hbar)^{\mathcal{D}}}\Re\sum_{\rm CT}
\int \d {\bf r}^{\prime\prime} \int \d {\bf p}^{\prime} \times\\ 
&\quad \times
\delta \left(\vareps - H({\bf r}^{\prime},{\bf p}^{\prime})\right)
\left|\mathcal{J}_{\rm CT}({\bf p}^\prime_\perp,
{\bf p}^{\prime\prime}_\perp)\right|^{1/2}\times\nonumber\\
&\quad \times\exp\left\{\frac{i}{\hbar}\;
\Phi_{\rm CT}- i \frac{\pi}{2} \mu^{}_{\rm CT}\right\}\;,\nonumber
\eea
where $H({\bf r},{\bf p})$ is the classical Hamiltonian in the phase space 
variables ${\bf r},{\bf p}$, $\Phi_{\rm CT}$ the phase integral,
\bea\l{legendtrans}
\!\Phi_{\rm CT} \!&\equiv&\! 
S_{\rm CT}({\bf p}^\prime,{\bf p}^{\prime\prime},t^{}_{\rm CT}) +
\left({\bf p}^{\prime\prime}- {\bf p}^{\prime}\right) 
 \cdot {\bf r}^{\prime\prime} =\\
&=& S_{\rm CT}({\bf r}^\prime,{\bf r}^{\prime\prime},\vareps) +
{\bf p}^{\prime} \cdot \left({\bf r}^{\prime} - {\bf r}^{\prime\prime}\right)\,,
\nonumber  
\eea
see the derivations in Appendix A2. 
In (\ref{pstrace}), the sum is taken over all discrete CT
manifolds for a particle motion from the initial
point $({\bf r}^\prime,{\bf p}^\prime)$ to the final 
point $({\bf r}^{\prime\prime},{\bf p}^{\prime\prime})$ with a given energy
$\vareps$ \cite{MAFptp2006}.
A CT can  uniquely be specified by fixing, for instance, 
the initial condition ${\bf r}^{\prime\prime}$ and the
final momentum ${\bf p}^{\prime}$ for a given time
$t^{}_{\rm CT}$ of the motion along the CT. 
$S_{\rm CT}({\bf p}^\prime,{\bf p}^{\prime\prime},t^{}_{\rm CT})$ 
is the action in the
momentum representation, 
\be\l{actionp}
S_{\rm CT}({\bf p}^\prime,{\bf p}^{\prime\prime},t^{}_{\rm CT}) = 
-\int_{{\bf p}^\prime}^{{\bf p}^{\prime\prime}}
\d {\bf p} \cdot {\bf r}({\bf p})\;. 
\ee
The integration by
parts relates (\ref{actionp}) to the action
\be\l{actionr}
S_{\rm CT}(\r^\prime,\r^{\prime\prime},\vareps) = 
\int_{\r^\prime}^{\r^{\prime\prime}}
\d \r \cdot \p(\r)\; 
\ee
[or other generating functions, see (\ref{phishat})]
in the spatial coordinate space by the Legendre transformation.
The Maslov phase  $\mu^{}_{\rm CT}$  is determined by the number 
of  conjugate ( turning and caustics) points
along the CT \cite{fedoryuk_pr,maslov,fedoryuk_book1}).
We introduced here a local phase-space 3D coordinate system,
$\r=\{x,y,z\}$, $\,\p=\{p_x,p_y,p_z\}$, related to a
PO which gives the main contribution into the trace integral 
among the CTs. The variables $x, p_x$ are locally the parallel and 
$\{\r_\perp,\p_\perp\}$ the 
perpendicular ( with respect to a CT) phase-space coordinates 
specified more below ($\r_\perp=\{y,z\} $, 
$\p_\perp=\{p_y,p_z\}$) \cite{gutz,smod,sclbook}. In (\ref{pstrace}),
$\mathcal{J}_{\rm CT}(\p^\prime_\perp,\p^{\prime\prime}_\perp)$ 
is the Jacobian
for the transformation from an initial perpendicular-to-CT momentum component
$\p_\perp^\prime$ to a final one $\p_\perp^{\prime\prime}$.
We can take first the integral over $p_{\parallel}^\prime$ 
of the momentum integration by using the energy conservation
$\delta$-function, 
\bea\l{pstraceppar}
&\delta g_{\rm scl}(\vareps)=
\frac{m}{(2\pi\hbar)^3} \Re\sum^{}_{\rm CT}
\int \d {\bf r}^{\prime\prime}
\int \frac{\d {\bf p}_\perp^\prime}{p_\parallel^{\prime}}\times
\\
&\quad \times\left|\mathcal{J}_{\rm CT}({\bf p}^\prime_\perp,
{\bf p}^{\prime\prime}_\perp)\right|^{1/2}
\exp\left[\frac{i}{\hbar}
\Phi_{\rm CT} - \frac{i\pi}{2}
\mu^{}_{\rm CT}\right]\;.\nonumber
\eea
The CT is determined by the Hamilton equations with 
the energy conservation condition, 
$\vareps=H({\bf r}'',{\bf p}'')=H({\bf r}',{\bf p}')$. 

For calculations of the trace integral by the SPM, one may write
the stationary phase conditions in both ${\bf p}^{\prime}$ 
and ${\bf r}^{\prime\prime}$ 
variables. According to the definitions (\ref{legendtrans}) 
and (\ref{actionp}), the stationary phase condition for the
${\bf p}^{\prime}$ variable
is a closing condition in the spacial coordinates:
\be\l{statcondp}
\left(\frac{\partial \Phi_{\rm CT}}{
\partial {\bf p}^{\prime}}\right)^* \equiv 
\left({\bf r}^\prime - {\bf r}^{\prime\prime}\right)^*=0\;.
\ee
The star means that any quantity in the circle brackets is taken at
the stationary point, e.g.,  
${\bf p}^{\prime}={\bf p}^{\prime\;*}$. 
In the next integration over ${\bf r}^{\prime\prime}$ in (\ref{pstraceppar})  
by the SPM 
we use the Legendre transformation (\ref{legendtrans}). Thus, 
according to (\ref{statcondp}), the
closing condition leads to the expression 
$\Phi^\ast_{\rm PO}=S_{\rm PO}({\bf r}^\prime,{\bf r}^{\prime\prime},\vareps)$.
The stationary-phase equation
for this integration over  spacial coordinates, 
${\bf r}={\bf r}^\prime={\bf r}^{\prime\prime}$, writes \cite{gutz}
\bea\l{statcondr}
\!\!\left(\frac{\partial \Phi_{\rm CT}}{
\partial {\bf r}^\prime}+\frac{\partial \Phi_{\rm CT}}{
\partial {\bf r}^{\prime\prime}}\right)^* 
\!&\equiv &\! \left(\frac{\partial S_{\rm CT}}{
\partial {\bf r}^\prime}+\frac{\partial S_{\rm CT}}{
\partial {\bf r}^{\prime\prime}}\right)^* \!\equiv \\
&\equiv &
-\left({\bf p}^\prime - {\bf p}^{\prime\prime}\right)^*=0\;,\nonumber 
\eea
where the star means ${\bf r}^\prime={\bf r}^{\prime\prime}={\bf r}^*$ along with 
${\bf p}^{\prime\prime}={\bf p}^{\prime\prime\;*}={\bf p}^*$. Equations 
(\ref{statcondp}) and (\ref{statcondr}) are the closing conditions for a CT 
in the phase space, too. 
Therefore, the {\it stationary phase}
conditions are equivalent to 
these {\it periodic-orbit} equations.
One of the SPM integrations in 
(\ref{pstrace}), for instance
over the parallel momentum $p_{\parallel}'$ in the local Cartesian
coordinate system introduced above, is identity due the energy conservation,
and therefore, can be taken exactly. 
The PO conditions (\ref{statcondp}) and (\ref{statcondr})
can be sometimes conveniently written 
in a more symmetric equivalent form,
\bea\l{statcondpr}
&&\left(\frac{\partial \Phi_{\rm CT}}{
\partial {\bf p}^{\prime\prime}}\right)^* \equiv 
\left({\bf r}^\prime - {\bf r}^{\prime\prime}\right)^*=0\;,\\
&&\left(\frac{\partial \Phi_{\rm CT}}{
\partial {\bf r}^{\prime\prime}}\right)^* \equiv 
\left({\bf p}^{\prime\prime} - {\bf p}^{\prime}\right)^*=0\;.\nonumber
\eea
After applying these PO equations [(\ref{statcondp}) and 
(\ref{statcondr}), or (\ref{statcondpr})], with accounting for the breaking
of symmetries
one may arrive at the trace formula in terms of the sum over POs 
\cite{migdalrev,sclbook}.

The total ISPM trace formula is the sum of the contribution of
all POs (families with the classical degeneracy
${\cal K}\ge 1$ and isolated orbits (${\cal K}=0$),
\begin{equation}\label{deltadenstot}
\delta g_{\rm scl}(\vareps)=\sum_{\rm PO}
 \delta g^{}_{\rm PO}(\vareps)\;,
\end{equation}
where
\begin{equation}\label{dgPO}
 \delta g^{}_{\rm PO}(\vareps)\!=\! \Re 
\!\left\{\!A_{\rm PO} 
\exp\left[\frac{i}{\hbar} S_{\rm PO}(\vareps) \!-\!
\frac{i\pi}{2}\mu^{}_{\rm PO} \right]\right\},
\end{equation}
with $A_{\rm PO}$ being the amplitude of density oscillations
depending on the PO classical degeneracy $\mathcal{K}$  and stability
factors.
 $S_{\rm PO}(\vareps)$ is the action and $\mu^{}_{\rm PO}$ the Maslov phase 
along the PO
\cite{gutz,strumag,sclbook,migdalrev,belyaev}.

\subsection{BIFURCATIONS AND AMPLITUDE ENHANCEMENT}

For solving bifurcation problems in integrable and 
non-integrable systems, more exact integrations are required.
In the SPM, after performing the exact integrations over the ``cyclic'' 
(``parallel'') phase-space variables related to the integrals of motion 
(the energy, angular momentum and others corresponding to the symmetries
of the Hamiltonian), one uses an expansion of
the action phase $\Phi_{\rm CT}$ in the remaining ``perpendicular'' variables
$\xi=\{\r',\p''\}_\perp$ in the
integrand of (\ref{pstrace}) over $\xi$ near the stationary point
$\xi^\ast$, 
\bea\l{ispmexp}
\!\!&\Phi_{\rm CT}(\xi)=\Phi^{}_{\rm PO} + \frac12\Phi_{\rm PO}''(\xi^\ast) 
(\xi-\xi^{\ast})^2 +\\ 
&\!+\! \frac16\Phi_{\rm PO}'''(\xi^\ast) (\xi-\xi^{\ast})^3\!+\!
\dots,\,\,\,
\Phi_{\rm PO}\!=\!\Phi^\ast_{\rm CT}\!=\!\Phi_{\rm CT}(\xi^\ast).\nonumber
\eea
To demonstrate the key point of our derivations of the trace formula, 
we consider here only one (one-dimensional) variable, called $\xi$
again, from the  phase space integration variables in (\ref{pstrace}), 
on which we meet a bifurcation (catastrophe) point in applying the SPM.
(We shall give comments if this might lead to a misunderstanding.)
In the standard SPM, the above expansion is truncated at the 2nd order term
and the integration over the variable $\xi$ is extended to $\pm\infty$.
The integration can be performed analytically and yields a Fresnel integral,
see e.g.\ \cite{sclbook}.

However, one meets singularities using the standard SPM which are related to 
zeros or infinities of $\Phi_{\rm PO}''(\xi^\ast)$ 
(or of eigenvalues of the corresponding 
matrix in the case of several integration variables $\xi$) 
while $\Phi_{\rm PO}'''(\xi^\ast)$ remains finite in the simplest case. 
These singularities occur when a PO (isolated or family) 
undergoes a bifurcation at the stationary point $\xi^\ast$ under the 
variation of some parameter (e.g., energy or deformation). The 
Fresnel integrals 
of the standard SPM sketched above will then diverge. In order to avoid such 
singularities, we observe that the bifurcation problem is similar to the 
caustic singularity, as two closed stationary points, 
considered by Fedoriuk within the catastrophe theory 
\cite{fedoryuk_pr,fedoryuk_book1}, adopted to its specific position at the 
edge of the phase-space volume accessible for the classical motion 
(see also Appendix A in \cite{MAFptp2006}). 
Therefore, we employ what we call the ``improved 
stationary-phase method'' (ISPM) 
\cite{ellipseptp,spheroidpre,spheroidptp,MAFptp2006,migdalrev}. 
Hereby the integration over $\xi$ in (\ref{pstrace}) is restricted 
to the {\it finite limits} defined by the classically allowed phase space 
region through the energy-conserving delta function  
in the integrand of (\ref{pstrace}). The expansion 
(\ref{ispmexp}) of the action phases, and similarly, of the amplitudes in (3) 
is generally used up to the second- and zero-order 
terms, respectively, and if necessary, to higher order terms 
in $\xi -\xi^*$.

In the simplest version of the ISPM, the expansion of the phase is truncated
at 2nd order, keeping the finite integration limits $\xi_{-}$ and $\xi_{+}$ 
given by the accessible region of the classical motion in (\ref{pstrace}).  
It will lead to a factor like 
\bea\l{errfuns}
&e^{i\Phi_{\rm PO}/\hbar}
\int_{\xi_{-}}^{\xi_{+}} \d \xi\;\exp\left[\frac{i}{2\hbar}\Phi_{\rm PO}''\; 
(\xi-\xi^{\ast})^2\right]\propto \\
&\propto \frac{1}{
\sqrt{\Phi_{\rm PO}''}}\; 
e^{i\Phi_{\rm PO}/\hbar}\;
{\rm erf}\left[\mathcal{Z}_{-},\mathcal{Z}_{+}\right],\nonumber
\eea
where ${\rm erf}(z_1,z_2)$ is the generalized error function with complex arguments
\bea\l{ispmlimits}
&{\rm erf}(z_1,z_2)=\frac{2}{\sqrt{\pi}}\int_{z_1}^{z_2}e^{-z^2}{\mbox d}z\;, \\
&\mathcal{Z}_{\pm}=\left(\xi_{\pm}-\xi^*\right)
\sqrt{-\frac{i}{2\hbar}\Phi_{\rm PO}''}\;.\nonumber
\eea
Note that the above expression (\ref{errfuns}) has no divergence at the 
bifurcation point where $\Phi_{\rm PO}''(\xi^\ast)=0$, since the error function 
(\ref{ispmlimits}) also goes to zero linearly in $\sqrt{\Phi_{\rm PO}''}$
[cf.\ the second equation in 
(\ref{ispmlimits})], which keeps the result finite. 
[For the case of several variables $\xi$ for which we find zeros or 
infinities of eigenvalues of the matrix $\Phi_{\rm PO}''(\xi^\ast)$, we 
diagonalize this matrix and reduce the Fresnel-like integrals to products of 
error functions similar to (\ref{errfuns}).]

This procedure is proved to be valid in the semiclassical limit by the
Maslov-Fedoriuk theorem \cite{fedoryuk_pr,maslov,fedoryuk_book1}.
In this way, we can derive contributions from each periodic orbit free of
divergences at any bifurcation point, and the oscillating part of the
level density can be approximated by the following {\it semiclassical 
trace formula}: 
\be\l{dgsc}
\delta g(\vareps) \simeq \delta g_{scl}(\vareps)
=\sum_{\rm PO} \delta g^{}_{\rm PO}(\vareps)\,,
\ee
with (\ref{dgPO}) for the PO contribution  $\delta g^{}_{\rm PO}(\vareps)$.
The amplitude $A_{\rm PO}(\vareps)$ in (\ref{dgPO})  
(complex, in general) is 
of the order of the phase space volume occupied by CTs, 
and 
the factor given in (\ref{errfuns}) which depends on the degeneracies
and stabilities of the POs, respectively. Sometimes,
it is convenient to split the Maslov phase which is invariant along the PO,
$\mu^{}_{\rm PO}=\sigma^{}_{\rm PO} +\phi_d$, 
into two terms where $\sigma^{}_{\rm PO}$ is called 
the Maslov 
index. $\phi_d$ is an extra phase that depends on the dimensionality of the 
system and degeneracy $\mathcal{K}$ of the PO manifold. 
($\phi_d$ is zero when all orbits are isolated ($\cK$=0), 
as defined in \cite{gutz}). The sum in (\ref{dgsc}) is an 
asymptotic one, correct to leading order in $1/\hbar^{1/2}$, and in 
non-integrable systems it is hampered by convergence  problems \cite{gutz}. 
For systems in which all orbits are isolated in phase space, Gutzwiller 
\cite{gutz} expressed the amplitudes $A_{\rm PO}(\vareps)$ 
(which are real in 
this case) explicitly in terms of the periods and stability matrices of the 
POs, see some examples below. His trace formula has become famous,
in particular in connection with {\it ``quantum chaos''} \cite{gutz}.
Notice that according to (\ref{errfuns}), any more exact integration
in (\ref{pstrace}) over a bifurcation/catastrophe variable $\xi$ of the 
improved SPM leads to an enhancement of the amplitude $A_{\rm PO}$
in the transition interval
from the bifurcation point to the region of the asymptotic 
(SSPM) behaviour of $A_{\rm PO}$. 
The height of this maximum is of order $1/\hbar^{1/2}$ as compared to the 
result of the standard SPM integration 
(integrable or non-integrable system; see more 
specific examples in Sections
IIE, IIF and IIIE).  Thus, for the family with
the degeneracy $\mathcal{K}$, one has the enhancement of the level
density amplitude,
\be\l{ampenhance}
A^{(\mathcal{K})}_{\rm PO} \sim A^{\rm G}_{\rm PO} \;
\hbar^{-\mathcal{K}/2}\;,
\ee
where $A^{\rm G}_{\rm PO}$ is the Gutzwiller trace-formula amplitude
for the contributions of isolated POs. 
In addition, for non-integrable
systems, one finds a local enhancement of the PO amplitude 
$A_{\rm PO}$ as compared to the Gutzwiller amplitude, see examples
below (Sections
IIF and V).

The trace formula (\ref{dgsc}) thus relates the quantum oscillations 
in the level 
density to quantities that are determined purely by the classical system.
Strutinsky and his collaborators, in 
their search for 
simple physical explanations of shell effects,
realized that this kind of approach could help to understand the shell effects
in terms of classical pictures \cite{smod,migdalrev}. 
However, in the application to nuclear physics,
Gutzwiller's expression for the amplitudes 
$A_{\rm PO}(\vareps)$ could not 
be used,
because they diverge when the POs are not isolated in phase space. This happens
whenever a system has continuous (e.g., rotational) symmetries, 
and hence, for 
most typical shell-model potentials (except in non-axially deformed 
situations). Therefore,
Gutzwiller's theory was extended to systems with continuous symmetries 
\cite{strumag,sclbook,migdalrev}.

Trace formulae for systems with all kinds of mixed symmetries, 
including the integrable cases, were also developed later by various 
other authors. 
The treatment of bifurcations is still an on-going subject of current research.
Uniform approximations were constructed for orbit bifurcations and 
symmetry breaking under the variation of the energy or a potential parameter; 
references to most of these developments are given in 
\cite{sclbook,migdalrev}. 

\subsection{Averaged level density}

For comparison with quantum densities we need also to use a local 
averaging of the trace
formula  (\ref{deltadenstot}) over the spectrum. 
As this trace formula has the simple form
as a sum of separating PO terms everywhere, including
the bifurcations, one can take approximately analytically
the integral over energies 
with Gaussian weight factor (folding
integral) 
\cite{strumag,migdalrev,sclbook,MVApre2013}. 
As the result, for this
averaged density $\delta g^{\rm scl}_\Gamma(\vareps)$ with 
the averaging parameter $\Gamma$, which is much smaller than
the Fermi energy $\vareps^{}_F$, one obtains
\be\l{avdeltadentot}
\!\delta g^{\rm scl}_{\Gamma}(\vareps)\!=\!\sum_{\rm PO} \delta g^{}_{\rm PO}(\vareps) 
\exp\left[\!-\left(t_{\rm PO} \Gamma/\hbar\right)^2\right],
\ee
where $t_{\rm PO}$ is the period for a PO, $t_{\rm PO}=M T_{\rm PO}$,
$M$ the PO repetition number,
$T_{\rm PO}$ the period for a primitive ($M=1$) PO. 

The total ISPM level density  as function of the 
energy $\vareps$ is given by 
\be\l{scldenstyra}
g^{\rm scl}_{\Gamma}(\vareps)= g^{}_{\rm ETF}(\vareps) + 
\delta g^{\rm scl}_{\Gamma}(\vareps)\;,
\ee
where $g^{}_{\rm ETF}(\vareps)$ is the 
average part obtained within the extended Thomas-Fermi (ETF)
approximation \cite{sclbook}.

\subsection{Shell-correction energy}

The PO expansion  for the shell-correction energy
$\delta U_{\rm scl}$ can be expressed 
in terms of the oscillating PO component 
of level density $\delta g^{}_{\rm PO}(\vareps)$ (\ref{dgPO})
at the Fermi energy $\vareps^{}_F$
\cite{strumag,sclbook,ellipseptp,spheroidptp,MVApre2013}
\bea\l{escscl}
&&\delta U_{\rm scl} = 2\sum_{\rm PO} \frac{\hbar^2}{t_{\rm PO}^2}\,
\delta g^{}_{\rm PO}(\vareps^{}_F)=\\
&=& 2\Re \sum_{\rm PO} \left(\frac{\hbar}{M T_{\rm PO}}\right)^2 A_{\rm PO} 
\times \nonumber\\
&\times& \exp\left[\frac{i}{\hbar} S_{\rm PO}(\vareps^{}_F) -\frac{i \pi}{2} 
\mu^{}_{\rm PO}\right],\nonumber
\eea
where 
$t^{}_{\rm PO}=M T_{\rm PO}(\vareps^{}_F)$ is the time of particle motion
along the PO (taking into
account its repetition number $M$) at the Fermi energy $\vareps=\vareps^{}_F$
as $\delta g^{}_{\rm PO}(\vareps^{}_F)$, $T_{\rm PO}$ 
is the period for the 
primitive ($M=1$) PO.
The factor 2 takes into account the spin degeneracy for neutron or proton
Fermi systems.
 The Fermi energy $\vareps^{}_F$ is related to the conservation of the particle
number $N$ through the equation:
\be\l{partnum}
N=\int_0^{\vareps^{}_F} \d \vareps\,g(\vareps)\;. 
\ee

Note that the shell-correction energy $\delta U$ is
the observed physical quantity independent of any artificial
averaging parameter $\Gamma$, in contrast to the
level density $g^{\rm scl}_{\Gamma}(\vareps)$. 
The convergence of the PO sum (\ref{escscl}) is ensured by the
factor, $\hbar^2/t^{2}_{\rm PO}$, in addition to
the amplitude $A_{\rm PO}$ of the oscillating level density
$\delta g^{}_{\rm PO}(\vareps^{}_F)$ (\ref{dgPO}).
Therefore, the short-time POs (their families) yield the main contributions
into the PO sum (\ref{escscl})
if they occupy enough large
phase-space volume.

\subsection{Spherical potentials}

\subsubsection{Trace formula in action-angle variables}

 We now transform the phase space trace formula (\ref{pstrace})
from the Cartesian phase space variables $\{\r;\p\}$ to the
canonical angle-action coordinates $\{{\bf \Theta};{\bf I}\}$. 
They  
are specified in
the spherical angle-action variables as
$\{\Theta_r,\Theta_\theta(=\Theta),\Theta_\varphi(=\varphi);
I_r,I_\theta,I_\varphi\}$, and then, 
$\{\Theta_r,\theta,\varphi;I_r,L,L_z\}$. 
The last variables have immediately physical meaning, and therefore,
simpler to use
 for the integrable spherically symmetric Hamiltonian (\ref{Hsph}).
They parametrize the action variables 
$I_\theta=I_\theta(L,L_z)$ and $I_\varphi=L_z$ so that the
CT characteristics, such as the curvature, are simplified for these potentials.
In particular, for integrable systems the action-angle variables are
preferably useful because in this case the Hamiltonian $H$ does
not depend on the angle variables ${\bf \Theta}$, i.e., 
$H=H({\bf I})=H(I_r,I_\theta,I_\varphi)=H(I_r,L)$.  From (\ref{pstrace})
one simply has
\bea\l{pstraceactang}
&&g_{\rm scl}(\vareps)
\!=\!\frac{1}{(2\pi\hbar)^3}\! \Re\sum_{\rm PO}
\int {\rm d} \Theta_r' {\rm d} \theta'
 {\rm d} \varphi'\! \times\\
&&\times \int {\rm d} I_r {\rm d}L
 {\rm d}L_z  
\delta\left(\vareps-H(I_r,L)\right)\times \nonumber\\
&\times& 
\left|\mathcal{J}_{\rm CT}(\p''_\perp,\p'_\perp)\right|^{1/2}\;
\exp\left[\frac{i}{\hbar}
\Phi_{\rm CT} - \frac{i \pi}{2}
\mu^{}_{\rm CT}\right]\;. \nonumber
\eea
The phase $\Phi_{\rm CT}$ (\ref{legendtrans}) expressed 
in terms of the
action-angle variables through the actions 
(\ref{actionp}) or (\ref{actionr}) (standard generating functions)
are considered in the mixed representation. 
The Jacobian ${\cal J}(\p''_\perp,\p'_\perp)$ is 
also transformed to the new variables. 
We took also into account 
explicitly that the actions ${\bf I}$ (
$\{I_r,I_\theta,I_\varphi\}$ or $\{I_r,L,L_z\}$) are constants
of motion for the spherical integrable Hamiltonian omitting the upper 
subscripts
in ${\bf I}$ as related to their initial (prime) and final (double
prime) values. As usual, one also employs some Jacobian transformations,
taking into account that there is no variations in the parallel
$x$ direction along the PO. The Jacobian of canonical 
transformations equals one, and 
$\partial I_\theta/\partial L=1$ and 
$\d I_\theta\d I_\varphi = \d L \d L_z$ from the spherical symmetry. The
integration limits for $L_z$ are obviously $-L \leq L_z \leq L$, and for
$L$, one has $0 \leq L\leq L_{\rm max}$ where $L_{\rm max}$ depends
of the energy $E$ and will be specified below.
Note that, in spite of non-orthogonality of the angle-action
coordinate system, there are still
the definite relations 
between the parallel (or perpendicular) components of quantities
in actions $S_{\rm CT}(\r',\r'',\vareps)$
in the Cartesian and the angle-action coordinate system.
They serve the conservation of actions $I_i$ for integrable
Hamiltonians along the trajectory CT \cite{MAFptp2006}. Therefore,
it makes sense to relate $x$ components  
$I_x$, $\Theta_x$ 
and corresponding $y,z$ components  of 
actions and angles $\hat{q}$ to the ``parallel'' and ``perpendicular'' ones with
respect to the reference POs in the trace formula 
(\ref{pstraceactang}), respectively.
Similar relations between the corresponding spherical
components as $r,p_r$ and $\Theta_r,I_r$ can be found too.

The PO
 solutions to the
stationary phase equations (\ref{statcondp}) and (\ref{statcondr})
are also invariants
with respect to the considered canonical transformation as the Hamiltonian
which altogether always can be expressed through both the 
Cartesian,  and the angle-action 
coordinates, also in the canonical spherical coordinates, 
by using the suitable transformation equations.

The main strategy in the next derivations of the trace formulas is following.
First, for any spherical Hamiltonians one has no dependence
of the integrand in the phase space trace
formulas (\ref{pstraceactang}) on the angular momentum projection
$L_z$ and corresponding azimuthal angle $\varphi''$.
Then, the integral over $L_z$ is $2L$ and the integral over
$\varphi''$ equals $2\pi$ , except for the diametrical
contribution for which $L=0$, see below. 
Second, in
(\ref{pstraceactang}), as 
in the Cartesian phase-space variables considered above,
for any spherical Hamiltonians
we take exactly the integral over the parallel actions, 
$I_r$ or $L$, 
and get $\int {\rm d}\Theta_r'/\omega_r=T_r$ or 
$\int {\rm d}\Theta'/\omega_\theta=T_\theta$
for  $\mathcal{K}=3$ families  
and diameter $\mathcal{K}=2$ POs 
or circle ($\mathcal{K}=2$) orbits, 
respectively,  because of the $\delta$-function conserving
the particle energy $\vareps$, and the invariance of the action along
a PO \cite{MAFptp2006}. The diameter family contribution
into the trace formula is exclusive
case due to the zero angular momentum, $L=L_z=0$. This case will be considered
separately for the RPLP. 
Third, we are left with the perpendicular action and angle variables.
For the derivation of the leading family (${\cal K}=3$) terms we have no
dependence of the angle variables and obtain the 
semiclassical Poisson summation trace formula \cite{bt76,sclbook}. Then,
we shall apply the ISPM for the calculation of the integral
over the ``perpendicular'' action variable 
with the corresponding stationary phase condition.
For the contribution of the circular-orbit
families with ${\cal K}=2$, there is the isolated stationary point
($r''=r^{\prime\prime\;*}=r_c, 
p_r'=p_r^{\prime\;*}=0$) in one of center 
planes in the ``perpendicular'' 
$\Theta_r'$ and $I_r$
integration variables  in the spherical 
phase-space variables $r''$ and $p_r'$. 

\subsubsection{Classical dynamics}

For any spherical potentials $V(r)$ the Hamiltonian $H$
in the spherical canonical phase-space variables 
$\{r,\theta,\varphi,p_r,p_\theta,p_\varphi\}$  writes
\be\l{Hsph}
H= \frac{1}{2m}\,\left(p_r^2 + \frac{p_\theta^2}{r^2}+
\frac{p_\varphi^2}{r^2 \sin^2\theta}\right)+ V(r)
= \vareps\;. 
\ee
Here, $p_\varphi=L_z$ is the projection of the angular momentum on
arbitrary $z$ axis, 
$~p_\theta= \left(L^2-L_z^2/\sin^2\theta\right)\;$,
$L=|{\bf L}|$ is the angular momentum, $|L_z| \leq L$, and $p_r(r)$
is defined by
\bea\l{pr}
p_r(r)&\equiv& \sqrt{p^2(r) -
  \frac{L^2}{r^2}}\;, \\
p(r)&=&\sqrt{2m\left[\vareps-V(r)\right]}\;.\nonumber
\eea
As the angular moment $L$ is conserved for the motion of a
particle in a spherically symmetric mean field $V(r)$, all CTs are
lying in a plane crossing the center $r=0$.
Integrating the differential equations (\ref{prteq}), 
one obtains the radial $r=r(t)$ and the angle 
$\theta=\theta(t)$ CT, i.e., a CT $r=r(\theta)$ in the azimuthal plane.
For one period $t=T$ along the PO
at the boundary condition $r=r''$ for $\Theta=\Theta''$, one has 
(Appendix A5)
\be\l{tetar1}
\theta''-\theta'=-\pi \frac{\partial I_r}{\partial L}\;,
\ee
where $I_r$ is the radial action in 
the spherical action-angle variables :
\bea\l{actionvar}
I_r&=&\frac{1}{2\pi}\,\oint p_r \,{\rm d}r = \\
&=&
\frac{1}{\pi}\int_{r_{min}}^{r_{max}}{\rm d}r
\sqrt{2m\left[E-V(r)\right] - \frac{L^2}{r^2}}\;, \nonumber\\
I_\theta&=&\frac{1}{2\pi}\,\oint p_\theta \,{\rm d}\theta 
=\frac{1}{\pi}\,\int_{\theta_{min}}^{\theta_{max}}{\rm d}\theta
\sqrt{L^2-\frac{L_z^2}{\sin^2\theta}}\;,\nonumber\\
I_\varphi&=&\frac{1}{2\pi}\,\oint p_\varphi \,{\rm d}\varphi
=L_z\;.\nonumber
\eea
The turning points $r_{min}$, $r_{max}$ and $\theta_{min}$, $\theta_{max}$
are the solutions of equations: 
\bea\l{turneq}
p_r^2(r,L) &\equiv& 2m\left[E-V(r)\right] - \frac{L^2}{r^2}=0\;,\\
&&L^2-\frac{L_z^2}{\sin^2\theta}=0\;. \nonumber
\eea
The Hamiltonian $H({\bf r},{\bf p})$ (\ref{Hsph}), 
as expressed through the spherical 
action-angle variables, does not depend on the cyclic angle variables.
From (\ref{freq}) for the frequencies 
$\omega_{\theta}=\partial H/\partial I_\theta$ and 
$\omega_r=\partial H/\partial I_r$,  
the periodic-orbit equations (\ref{statcondp}) and (\ref{statcondr}) 
take the form of the following resonance 
conditions:
\bea\l{percond}
f(L) &\equiv& \omega_\theta/\omega_r \equiv 
-\partial I_r(E,L)/\partial L \equiv \\
&\equiv&
\frac{L}{\pi} \,\int_{r_{min}}^{r_{max}}\frac{{\rm d}r}
{r^2\sqrt{2m\left[E-V(r)\right] - L^2/r^2}}= \nonumber\\ 
&=&
n_\theta/n_r\;,\quad \omega_\theta/\omega_\varphi \equiv 1\;, \nonumber
\eea
where $n_\theta$ and $n_r$ are co-primitive integers. The energy surface
$I_r=I_r(E,L)$ (\ref{actionvar}) is simplified to a function 
of one variable $L$ for a 
given energy $E$ of the particle  because the second
equation for the ratio of frequencies,  
$\omega_\theta/\omega_\varphi$, 
is the identity [$I_\theta=L$
for $L_z=0$, according to (\ref{actionvar})]. The solutions
to the first PO equation in (\ref{percond}) 
for $L=L^*(n_r,n_\theta)$ 
define the three-parametric families ($\mathcal{K}=3$) 
of orbits $M(n_\theta,n_r)$. 
The angular momentum projection $L_z$, 
and the two single-valued
integrals of motion related to the two fixed rationals for
$\omega_\theta/\omega_r$ and $\omega_\theta/\omega_\varphi$ 
from the PO conditions can be taken as parameters \cite{strumag,smod}.

Except for obvious two-parametric ($\mathcal{K}=2$) diameter families
at $L=0$,
there is also the specific $\mathcal{K}=2$ 
family of circle orbits at another edge 
$L=L_C$ of the energy surface 
$I_r=I_r(E,L)$.  In a fixed plane crossing the center with radius
$r=r_{min}=r_{max}=r_c$, one has the isolated circle 
PO. The first equation in (\ref{turneq}) determines 
the turning points $r_{min}(L)$ and $r_{max}(L)$ 
as functions of $L$ at the fixed energy $\vareps$ and their cross 
gives the specific $L=L_{\rm C}=p(r^{}_{\rm C}) r^{}_{\rm C}$.

Note that according
to the expression for the radial momentum $p_r$ from (\ref{pr})
which must be real, 
the maximal angular momentum $L$  is namely this $L_{\rm C}$, which is
related to the zero $p_r$.  
With the zero minimal value of the angular
momentum $L$, one finds
$0 \leq   L \leq r^{}_{{\rm C}}\,p(r^{}_{\rm C}) 
= r^{}_{\rm C} \sqrt{2m\left(\vareps-V(r^{}_{\rm C})\right)}
=L_{\rm C}\;$.
The maximal value of the angular momentum 
$L=L_{\rm C}$ is related to
a circular orbit and the minimal one $L=0$ corresponds to a diameter.
The critical values $r=r^{}_{\rm C}$ and  $L=L_{\rm C}$ are determined
as the solution of the system of the two equations with respect to $r$
and $L$:
\be\l{rc}
p_r^2(r,L)=0,\qquad \frac{{\rm d}}{{\rm d}r} p_r^2(r,L)=0\;,
\ee 
with ${\rm d}^2 p_r^2(r,L) /{\rm d}r^2\neq 0\;$,
as assumed to be the case for the spherical potentials as considered
below.
The first equation claims that there is no radial velocity, $\dot{r}=0$,
and the next equation is that the radial force is equilibrated by the
centrifugal force.

Another general key quantity in the POT
is the curvature $K$  
of the energy surface $I_r=I_r(\vareps,L)$,
\be\l{curv}
K=\frac{\partial^2 I_r}{\partial I_\theta^2}= 
\frac{\partial^2 I_r(E,L)}{\partial L^2}= 
-\frac{\partial f(L)}{\partial L}\;,
\ee
where $f(L)$ is the ratio of frequencies defined in (\ref{percond}).

\subsection{Symmetry breaking and bifurcations in a non-integrable potential} 
 
Recent studies of the POT are focused on overcoming catastrophe problems in the 
derivation of the semiclassical trace formulae arising in connection with
symmetry breaking and bifurcation phenomena, where the standard 
stationary-phase method  fails (see \cite{gutz}). Semi-analytical 
uniform approximations 
solving these problems for the case of well separated pitchfork bifurcations
in the non-integrable H\'enon-Heiles (HH) potential were suggested 
in \cite{kaidelbrack,bracktanaka}, using the normal-form 
theory of non-linear dynamics \cite{sie97,ssun,schomerus}.  

In this subsection, we derive an analytical trace formula 
for the semiclassical level density of the HH potential, employing the improved 
stationary-phase method (Section IIB)
\cite{MAFptp2006} valid for arbitrarily dense 
sequences of pitchfork bifurcations near the saddle-point energy and for 
harmonic-oscillator symmetry breaking in the limit of small energies. 
In this respect, the regular-to-chaotic transition in Fermi systems
becomes important for the understanding of its influence on shell
correction amplitudes.  Figure\ \ref{fig1} shows transparently
such a transition through Poincar\'e Surfaces of Section (PSS) 
of the nonlinear classical dynamics for the 
HH potential as a simple nontrivial example \cite{sclbook,hhprl}, 
see also
\cite{BMYnpae2010,BMYijmp2011,BMprc2012} for the PSS and Lyapunov 
exponents in the three-dimensional 
axially symmetric Legendre-polynomial and spheroidal billiards. 
The PSS is a
successive crossing points of a classical trajectory with a given
plane (surface of section) 
in the phase space.  The regular
trajectory is confined in a torus, and the crossing points of such
 a trajectory with the PSS 
will accumulate on a certain
closed curve.  On the other hand, the chaotic trajectory will make a
scattered plot where a certain area is randomly filled by the crossing
points.
As shown in this figure, the obvious transition from chaos to order occurs
with dimensional 
decreasing energy $e$ of the particle (in units of the saddle energy) 
from the saddle ($e=1$) to 
a small-energy (harmonic-oscillator) limit $e \rightarrow 0$.
We show below the relation of this behavior of the PSS to the amplitudes of 
oscillations in the level density (density of states) and total energy of
fermion systems. 

\subsubsection{Trace formulae, symmetry breaking and bifurcations 
}

The level density $g(E)$ (\ref{pstrace}) is obtained from the semiclassical
Green's function \cite{gutz} by taking its trace
 in the phase-space Poincar\'e variables $Q,p$  
\cite{MAFptp2006,sie97,ssun,schomerus}:
\bea\l{pstraceqP}
&\hspace{-0.4cm}
g_{\rm scl}(\vareps)= \frac{1}{(2\pi\hbar)^2}\Re\sum^{}_{\rm CT}
\int {\rm d}Q\int{\rm d}p\; t^{}_{y\,{\rm CT}}
\times\\
&\times\left|\mathcal{J}_{\rm CT}(p,P)\right|^{1/2}\times\nonumber\\ 
&\times \exp\left\{\frac{i}{\hbar}
\left[\widehat{S}_{\rm CT}(Q,p,\vareps) - Q p\right] - 
\frac{i \pi}{2} \mu^{}_{\rm CT}\right\}. \nonumber
\eea
Here $Q$ and $p$ are the final $x''$ and initial $p_x'$ coordinates 
in the phase-space variables 
$x,y,p_x,p_y$ perpendicular to a reference classical
trajectory  in two dimensions,
  $t_{y\,{\rm CT}}=m\oint d y/p_y$ is the primitive partial period of 
the $y$ motion along the CT, $\widehat{S}_{\rm ct}(Q,p,\vareps)$ the generating
function, $\mu^{}_{\rm CT}$ the Maslov phase, and 
$\mathcal{J}_{\rm CT}(p,P)$ is the Jacobian for the transformation between the 
variables shown as its arguments. The ISPM generating function  
$\widehat{S}_{\rm CT}(Q,p,\vareps)$ is defined by
\be\l{genfunact}
{\widehat S}_{\rm CT}(Q,p,\vareps) = S_{\rm CT}(Q,p,\vareps)  + qp ,
\ee
where  $~S_{\rm CT}(Q,p,\vareps)~$ is the action 
$~~~S_{\rm CT}(\r',\r'',\vareps)=$
$=\int_{\r'}^{\r''} \p\cdot\d \r$ 
expressed in terms of
the Poincar\'e variables $Q$ and $p$ through the mapping transformation
equations $Q=Q(q,p)$ and $P=P(q,p)$ along a CT ($\r'$ and $\r''$ are the
initial and final spatial coordinates of the CT).
It can be replaced by a (truncated) fourth-order expansion 
around the stationary points $Q^*, p^*$  which correspond to the POs, 
$Q^*=q~$, $p^*=P$  
\cite{MAFptp2006}. For 
pitchfork bifurcations, the expansion of the generating function
${\widehat S}_{\rm CT}(Q,p,\vareps)$ (\ref{genfunact}) is  
similar to the normal forms  
\cite{sie97,ssun,schomerus} with 
the following power series in $Q-Q^*$ and $p-p^*$:
\bea\l{genfunexp}
\!\!S_{\rm CT}(Q,p,\vareps) \!&=&\!
S_{\rm PO}(\vareps) \!+\! 
\epsilon_{{}_{{\rm PO}}}^{(Q)} (Q-Q^*)^2 \!+\! \\
 +a_{{}_{{\rm PO}}}^{(Q)}(Q\!-\!Q^*)^4 &+& 
 \epsilon_{{}_{{\rm PO}}}^{(p)} (p-p^*)^2
 + a_{{}_{{\rm PO}}}^{(p)} (p-p^*)^4,\; \nonumber
\eea
where $S_{\rm PO}(\vareps)$ is the action along the PO.
 Performing also more exact integrations 
over $Q$ and $p$ in (\ref{pstraceqP}), one obtains
for the case of pitchfork bifurcations 
\bea\l{dgscl}
&&\hspace{-0.4cm}\delta g_{\rm scl}(\vareps)\!=\!\frac{1}{(2 \pi \hbar)^2}\; 
\!\Re\sum_{\rm PO} \frac{T_{\rm PO}}{[\hbar^2 a_{\rm PO}^{(Q)}a_{\rm PO}^{(p)}]^{1/4}}\; 
\!\times\! \\
&\times& \!\mathcal{A}\left(\xi_{\rm PO}^{(Q)}\right)\!
\mathcal{A}\left(\xi_{\rm PO}^{(p)}\right)
\exp\!\left[\frac{i}{\hbar}S_{\rm PO}(\vareps) \!-\!\frac{i\pi}{2} 
\sigma_{{}_{{\rm PO}}} \!-\!i \phi
\right]\!, \nonumber
\eea
where $T_{\rm PO}$ is the period for a primitive PO, $S_{\rm PO}(\vareps)$ is 
its full action
(including repetitions) at energy $\vareps$, and 
\bea\l{amphh}
\mathcal{A}(\xi)&=&\int_{z_{-}}^{z_{+}} \d z\;
\exp\left[i \left(\xi z^2 + z^4\right)\right], \\
\quad \xi&=&
\epsilon/(\hbar a)^{1/2},\nonumber
\eea
is the amplitude factor; $\epsilon$ and $a$ (for $a>0$) are 
the coefficients in the power expansion  
of the generating function $\widehat{S}_{\rm CT}(Q,p,\vareps)$ 
(\ref{genfunact}) with (\ref{genfunexp}) 
in $Q-Q^*$ and $p-p^*$, which are proportional to the 2nd and 4th  
derivatives of 
$\widehat{S}_{\rm CT}(Q,p,\vareps)$ 
at the stationary points $Q^*$ and $p^*$;
$\sigma^{}_{\rm PO}$ is 
the Maslov index related to the turning and caustic points 
along the POs, $\phi$ a constant phase independent of the PO. The integration 
(\ref{amphh}) is performed over the finite classically accessible region of the
Poincar\'e variables $Q$ and $p$, denoted here as 
\be\l{zQpeq}
z=\frac{Q-Q^*}{(a^{(Q)}/\hbar)^{1/4}},\quad \mbox{or}\quad
z=\frac{p-p^*}{(a^{(p)}/\hbar)^{1/4}},
\ee
i.e.\  from $z_{-}^{(Q)}$ to $z_{+}^{(Q)}$ and from $z_{-}^{(p)}$ to $z_{+}^{(p)}$, 
respectively, with
\be\l{zpmQp}
\hspace{-0.2cm}z_{\pm}^{(Q)}=\frac{Q_\pm-Q^*}{(a^{(Q)}/\hbar)^{1/4}},\,\,\,
z_{\pm}^{(p)}=\frac{p_\pm-p^*}{(a^{(Q)}/\hbar)^{1/4}}.
\ee
In (\ref{dgscl}), the sum runs over the straight-line orbits $A_\sigma$, the 
rotational orbits $R_\sigma$, and the librational orbits $L_\sigma$ 
of the standard 
HH Hamiltonian \cite{sclbook,bmtjpa2001,kaidelbrack} 
(here in units with $m=\omega=\hbar=1$):
\be\l{hhpoten}
H=\frac12\left(\dot{x}^2+\dot{y}^2\right) + 
\frac12\left(x^2+y^2\right) + \alpha\left(x^2 y -
\frac{y^3}{3}\right).
\ee

Using the barrier energy $\vareps_{\rm barr}=1/(6\alpha^2)$ 
as dimensionless energy unit,
$e=\vareps/\vareps_{\rm barr}=6 \alpha^2 \vareps$, and the 
following scaled variables
\bea\l{scalinghh}
p_u&=&\alpha p_x\,,\,\,\, p_v=\alpha p_y\,,\,\,\, u=\alpha x\,, 
\,\,\, v=\alpha y\,, \\
{}\hspace{-2.0ex}h&=&3(\dot{u}^2+\dot{v}^2)+ 3(u^2+v^2)+6vu^2-2v^3, \nonumber
\eea
one obtains classical dynamic equations independent of the parameter $\alpha$
\be\l{eqmot}
\ddot{u}=-u-2 u v\,,\qquad \ddot{v}=-v + v^2 - u^2.
\ee

The scaled HH potential is shown in Fig.\ \ref{fig2} 
as equipotential lines, and the orbits A, B and C (at $e=1$) are presented, too.
The HH potential is invariant under rotations about 120 degrees, which leads to
a discrete degeneracy of the orbits.  Such a degeneracy can be
simply taken into account multiplying the amplitudes in the trace formula
(\ref{dgscl}) by a factor 3. The cut along $u=0$ (right) shows a barrier 
at the saddle $e=1$ with two turning points $v_1 \leq v_2$ at $0<e< 1$;
$v_n$ are the real solutions of the cubic equation $e=3 v^2- 2 v^3\leq 1$
(\ref{yn}).

In order to simplify the amplitude function $\mathcal{A}$ in the ISPM 
trace formula (\ref{dgscl}), we note that sufficiently far from the 
symmetry breaking at $\vareps=0$, the integration limits in (\ref{amphh})
can be extended to $\pm \infty$ (convergence being guaranteed by the
finite fourth-order terms). Then, the amplitudes $\mathcal{A}$ (\ref{amphh})
can be expressed through integral representations of the Bessel functions
$J_{\pm 1/4}(x)$: 
\bea\l{amphhpp}
\hspace{-0.5cm}\mathcal{A}(\xi)\!=\!&\;
\frac{\pi}{2} \sqrt{\xi}
\left\{\exp\!\left[-i \left(\frac{\xi^2}{8}\!-\!\frac{\pi}{8}\right)\right]
\!J_{-1/4}\left(\frac{\xi^2}{8}\right) \!-\,\right. \\
&-\; \frac{\xi}{|\xi|}\,\left.\exp\left[-i \left(\xi +
\frac{\pi}{8}\right)\right]
J_{1/4}\left(\frac{\xi^2}{8}\right) \right\}. \nonumber
\eea
Here we took into account
a time-reversal symmetry by inclusion of the factor 2 where necessary. 
Note that more exact trace formulae (with additional terms
proportional to Bessel functions with indices $\pm\, 3/4$ etc.) 
can be derived by taking into account higher-order terms in the phase
and amplitude factors, respectively. This gives results similar to those
obtained in 
\cite{ssun} using the normal forms for pitchfork
bifurcations. 

Using asymptotic forms of the Bessel functions for large arguments $\xi$
in (\ref{amphhpp}), one obtains from (\ref{dgscl}) (with $\phi=0$) the 
standard Gutzwiller trace formula 
\cite{gutz,sclbook} valid for isolated POs:
\bea\l{gutz}
\hspace{-0.5cm}\delta g_{\rm scl}(\vareps)\!\rightarrow \!
\!\sum_{\rm PO}\! A_{\rm PO}^{\rm G}(\vareps)\!
\cos\!\left[\!\frac{S_{\rm PO}(E)}{\hbar}\!-\!\frac{i\pi}{2} 
\!\sigma_{{}_{{\rm PO}}}\!\right], \\
A_{\rm PO}^{\rm G}(E)=T_{\rm PO}\left(\pi \hbar
\sqrt{\Big|2 - {\Tr\mathcal{M}}_{\rm PO}\Big|} \right)^{-1}, \nonumber
\eea
where  
$\mathcal{M}_{\rm PO}$ is the stability matrix for the PO (Appendix A7).
Numerical and
analytical calculations and the remarkable 
``fan'' structure of the pitchfork bifurcations of the straight-line
orbits $A_\sigma$ were analyzed in the case of the HH potential
\cite{bmtjpa2001,fmbpre2008,bkwf2006}.  
Several analytical approximations for $\Tr M_A$ can be derived in terms of the
simplest Mathieu functions for smaller energies $e \siml 0.8$,
and in terms of the improved Legendre solutions for the whole region
from the zero energy to the saddle, $0 \leq e \leq 1$, in good agreement 
with the numerical results
\cite{kkmabPS2015,bmtjpa2001}.

The trace formula (\ref{dgscl}) also has the correct harmonic-oscillator (HO) 
limit for $\vareps \rightarrow 0$, 
where $\Tr\mathcal{M}_{\rm PO} \rightarrow 2$, and 
all coefficients in the expansion of the action phase in $Q$ and $p$ go to zero 
(and $\int d Q d p \rightarrow 2 \pi \vareps$). The Poincar\'e variables 
$Q$ and $p$ become cyclic in this HO limit. In the spirit of the uniform 
approximations 
(\cite{sclbook} and 
\cite{bracktanaka}) 
within the ISPM, we may use a canonical transformation from the variables 
($Q,p$) to new variables ($\widetilde{Q},\widetilde{p}$) 
in which one has a simple 
analytical expression for the PO amplitude
\be\l{uniform} 
A_{\rm PO}^{\rm G}\left[1-\exp(-\vareps/A_{\rm PO}^{\rm G})\right]\;, 
\ee
instead of 
$A_{\rm PO}^{\rm G}$ in (\ref{gutz}), with the two correct limits 
to the HO trace formula 
\cite{sclbook} for $E\to 0$ and to the Gutzwiller trace formula (\ref{gutz})
for large $E$. We should note that this procedure is not unique, 
see \cite{kkmabPS2015}. 
On the other hand, within 
the ISPM, we use as ``normal forms'' equation (\ref{genfunact})  
for the generating function with expansion (\ref{genfunexp}) near the 
stationary points rather than near the 
bifurcations. A similarity to the normal form theory is manifested
if we put formally $Q^*=0$, $p^*=0$ in (\ref{genfunexp}) in the system
of coordinates related to the bifurcation point reducing the non-local ISPM 
to its local approximation valid nearly the bifurcation points. 
Moreover, from a  more
pragmatic point of view, the details of the required canonical transformation 
do not matter for the SPM approximation in narrow regions of phase space around 
the critical points:  
The limit $\hbar \rightarrow 0$  
in practice
corresponds to large particle numbers $N$ through the Fermi energy 
$\vareps^{}_F$
at a rather small parameter $\alpha$ and larger 
averaging width $\gamma$ of the gross shell structure. 
We emphasize also a chaos-to-order transition of the PSS in the limit to the 
symmetry breaking point $e \rightarrow 0$ (Fig.\ \ref{fig1} 
).  
In this limit, the isolated trajectories 
are transformed into 
the degenerate PO families.

Expressions found from (\ref{dgscl}) locally for the separate bifurcations
of the rotating ($R$) or librating ($L$) orbits are in agreement 
with the results
 \cite{sie97,ssun,schomerus}
obtained using the standard normal forms for
the pitchfork bifurcations. However, for the full cascade of bifurcations near 
the saddle energy of the HH potential, our result (\ref{dgscl}) goes beyond the 
normal-form theory. It is a continuous function through all bifurcation points
near the saddle energy and also down to the limit to the 
symmetry-breaking point 
at $E=0$. The coefficients $\epsilon(\vareps)$ and $a(\vareps)$ 
in  (\ref{amphh}) are 
also continuous functions of the energy $\vareps$ 
through all stationary points (POs).
Note also that our ISPM expression (\ref{dgscl}) for the shell correction to
the level density is a sum of separate contributions of all involved POs,
and a coarse-graining over the energy $\vareps$ (cf.\ below) may therefore be 
performed analytically. Thus, one has 
a possibility to study analytically both gross and fine shell structures.
This is in contrast to the results \cite{kaidelbrack,bracktanaka} using uniform 
approximations based on the normal-form 
theory 
\cite{sie97,ssun,schomerus}, where at each critical point all
involved POs give one common contribution.

\subsubsection{Discussion of results}

For the purpose of studying the improved level density around
the bifurcation points, we consider a slightly averaged 
level density, thus avoiding the convergence problems that usually arise
when one is interested in a full semiclassical quantization.
Such a ``coarse-graining'' can be done by folding the level density over
a Gaussian of width $\gamma$ \cite{sclbook,migdalrev}.
(The particular choice of a Gaussian form of the averaging 
function is immaterial
and guided only by mathematical simplicity.) 
Applying this procedure to the semiclassical
level density (\ref{dgscl}), one obtains
(\ref{avdeltadentot}) for the
averaged level-density shell correction $\delta g^{}_{\Gamma,scl}(\vareps)$
\cite{strumag,sclbook,migdalrev}.

The averaging of the oscillating level density yields an exponential decrease
of the amplitudes with increasing periods $t_{\rm PO}$ and/or $\Gamma $.  As
shown in \cite{MAFptp2006}, for $\gamma$ about 1/3 
(in $\hbar \omega$ units), all
large-action paths are strongly damped and only the time-shortest POs
contribute to the oscillating part of the level density, yielding its
gross-shell structure.  For a study of the bifurcation phenomenon,
however, we need smaller values of $\gamma$. In Fig.\ \ref{fig3} 
we used the coarse-grained Gutzwiller trace formula 
(\ref{avdeltadentot}) with
(\ref{gutz})
including the simplest primitive orbits 
$A, B=L_4$ and $C=R_3$. 

It is interesting that the gross-shell structure manifests itself
for the HH parameter $\alpha=0.04$ even for a relatively 
small averaging parameter 
$\gamma=0.25 \hbar \omega$. Therefore, we should expect also a good agreement
between semiclassical and quantum results for the shell-correction energy
$\delta U$ as function of the particle numbers 
$N^{1/2}$ for the same $\alpha=0.04$ 
for larger energies (but still far enough from the bifurcation 
points, cf. Fig.\ \ref{fig4}
). 

The shell-correction energy $\delta U$,
i.e., the oscillating part of the total energy $U$ of a system of $N$
fermions occupying the lowest quantum levels in a given potential,
can be expressed in terms of the oscillating components 
$\delta g_{\rm PO}^{}(\vareps)$ at the Fermi energy 
$\vareps=\vareps^{}_F$ of the semiclassical level density
(\ref{dgscl})
and (\ref{deltadenstot}),
 as in \cite{strumag,smod,sclbook,migdalrev}, see also (\ref{escscl})
for the shell-correction energy $\delta U$.
We are taking into account the spin degeneracy factor 2  
in (\ref{escscl}).
The semiclassical representation of the shell-correction energy 
(\ref{escscl}) 
differs from that of 
$\delta g_{\rm scl}(\vareps)$ 
(\ref{dgsc})
(at $\vareps=\vareps^{}_F$) only by a factor
$(\hbar/t^{}_{\rm PO})^2$ under the sum, which suppresses
contributions from orbits of larger time periods (actions). Thus the 
periodic orbits with smaller periods play a dominant role in determining 
the shell-correction energy \cite{strumag,smod}.
Finally, we should note that the higher the degeneracy of an orbit,
the larger the volume occupied by the orbit family in the phase space,
and also the smaller its time period (action), the more important is its
contribution to the shell-correction energy (\ref{escscl}).

Fig.\ \ref{fig3} and \ref{fig4} 
show a good agreement 
between the semiclassical and quantum results, in spite of 
using only the three shortest orbits $A$, $L_4$ ($B$), and $R_3$ (C). 
These are seen to yield the correct gross-shell 
structure for the parameter $\alpha=0.04$ (and widths for the 
Gaussian averaging of the level density shell corrections 
$\gamma=0.25\hbar\omega$ or, similarly, for 
$\gamma=0.6\hbar\omega$) in the energy region below the saddle ($E=E_{barr}$) 
and above the bottom ($E=0$).
The discrepancies at smaller energies are related to the symmetry breaking 
at $E=0$, as discussed above, and will be removed when using our full ISPM 
trace formula (\ref{dgscl}). In the quantum-mechanical determination 
of $\delta U$ (see \cite{sclbook,fuhi} 
for discussions of the Strutinsky averaging method), the plateau condition 
for the averaged energy was satisfied for a
 Gaussian width $\tilde{\gamma} \simeq 
1.75 \hbar \omega$ and a curvature correction parameter $M=6$.

\section{ FISSION-CAVITY MODEL AND SHAPE ISOMERS}

In this section, we shall present some applications of the 
POT to nuclear deformation energies and discuss in more detail the relation 
of the bifurcations of periodic orbits with the pronounced shell effects and 
fission isomers.

According to the SCM, the oscillating part of the total energy of a finite 
fermion system, the shell-correction energy $\delta U$, 
is associated 
with an inhomogeneity of the s.p. energy levels near the Fermi 
surface. Its existence in dense fermion systems is a basic point of Landau's 
quasi-particle theory of infinite Fermi liquids, 
as extended to self-consistent finite fermion systems by Migdal and  
collaborators 
\cite{migdal,khodsap}.
Depending on the level density at the Fermi energy -- and with 
it the shell-correction energy $\delta U$ -- being a maximum or a minimum, 
the nucleus is particularly unstable or stable in the case of dense and
sparse
s.p.\ spectra, respectively. This
situation varies with particle numbers and deformations of the 
nucleus. In consequence, the shapes of stable nuclei depend strongly 
on the particle numbers and deformations. This is illustrated in 
\cite{smod,sclbook}.
The shell correction $\delta U$ of neutrons is 
shown as a function of the neutron number $N$ and the deformation parameter 
$\eta$ of a Woods-Saxon potential  
with spheroidal shape,
$\eta$ being the ratio of the semi-axes. If we fix the neutron number 
$N$, e.g.\ $N=150$, and increase the deformation $\eta$, we meet the 
first minimum (ground state) at about $\eta \sim 1.25$ and the next one 
(isomeric state) at much larger deformations $\eta \sim 1.9-2.1$. 
The experimental data corresponding to these deformations are 
in good agreement with the semiclassical slopes.   

The SCM was successfully used to describe nuclear masses and deformation
energies and, in particular, fission barriers of heavy nuclei, see
an 
early review by Strutinsky's group \cite{fuhi}, 
in which also the microscopic
foundations of the SCM are discussed. 
As shown in \cite{smod}, the predictions of the POT for a loci of the 
ground-state minima, using the shortest POs in a spheroidal cavity are
basically in agreement with experimental data.
Bifurcations of POs under the variation of a deformation parameter or the
(Fermi) energy can have noticeable effects for the shell structure 
\cite{smod,spheroidpre,spheroidptp,MAFptp2006,ellipseptp}.
In this section, we review the semiclassical description (see also 
\cite{brreisie}) of a 
typical nuclear fission barrier in terms of the shortest periodic orbits, 
employing a cavity model with the realistic shape parametrization developed 
in \cite{fuhi}. In particular, the effect of the 
left-right asymmetric deformations 
on a height of the outer fission barrier will be discussed. Isochronous 
bifurcations of the shortest orbits are treated in 
\cite{brreisie} by using
the
 uniform approximation 
employing a suitable normal form for the action function. 
The relation of the bifurcations of POs to the foundation of the
local second minima at large isomer deformations will be discussed 
for the cavity model
with the realistic parametrization \cite{fuhi}.

One prominent feature in the fission of actinide nuclei (isotopes of
U, Pu, etc.) is that their fragment distributions are asymmetric with a
most probable ratio of fragment masses of $\sim$ 1.3 - 1.5 (cf.\ \cite{fuhi}). 
This is an effect that cannot be described within the LDM which always 
favors the highest possible symmetries. It was one of the big successes 
of the SCM to explain the mass asymmetry of fission fragments.
The fragment distribution is, of course, a result of nuclear dynamics.
However, already in static calculations of fission barriers, the onset
of the mass (or left-right) asymmetry at an outer fission barrier
was found in SCM calculations with realistic nuclear shell models. 
On the l.h.s.\ of Fig.\ \ref{fig5},  
we show a 
schematic picture of the deformation energy of a typical actinide nucleus, 
as plotted versus suitably a chosen deformation parameter (see below for
a specific choice of deformations). The heavy dashed line is the average 
deformation energy obtained in the LDM; the thin lines are the results 
obtained when the shell-correction energy $\delta U$ is included. They 
exhibit the characteristic deformation effects of the shell structure 
in these nuclei: a deformed ground state and the characteristic double-humped 
fission barrier, split by a second minimum corresponding to the fission
isomer. The solid line is obtained when only the left-right symmetric
deformations are used; the dashed thin line is obtained when one allows
for the left-right asymmetric shapes. As we see, the asymmetric shapes  
are displayed  considerably lower the outer fission barrier. All shapes here are
taken to be axially symmetric.

The mass asymmetry in nuclear fission was therefore understood as a quantum 
shell effect. In a detailed microscopical study \cite{gumni} 
of the Lund 
group using the Nilsson model, it was shown that those s.p.\ states 
which are most sensitive to the left-right asymmetric deformations are 
pairs of states with opposite parity, having the nodes and extrema of 
their wave functions on parallel planes perpendicular to
the symmetry axis at and near the waist-line of the fissioning nucleus, 
as shown on the r.h.s.\ of Fig.\ \ref{fig5}.  
Under the effect of the 
neck constriction one of these s.p.\ levels, which for actinides is just 
lying below the Fermi energy, is further lowered when the mass asymmetry 
is turned on. As a consequence, the asymmetry leads to a lowering of the 
total shell-correction energy, and hence 
of the outer fission barrier,
the LDM part of the energy being much less sensitive to the mass asymmetry.

In this section, we want to show that the POT is able to reproduce this 
quantum shell effect, at least qualitatively, in the semiclassical description
using the POT. We will focus here only on the gross-shell structure, like 
that seen in the qualitative picture of a  fission barrier in 
Fig.\ \ref{fig5}. 

The spheroidal cavity model used in \cite{smod} and discussed in 
\cite{smod,migdalrev,spheroidptp} 
allows one to describe only qualitatively a 
nuclear fission, since an ellipsoidal 
deformation is not sufficient to yield a finite barrier towards fission. In 
\cite{migdalrev,brreisie}, a 
simple but more realistic ``fission cavity 
model'' was used. It consists of a cavity with the $(c,h,\alpha)$ shape 
parametrization that was used both for the LDM and for the deformed 
Woods-Saxon type shell-model potentials in the SCM calculations of 
\cite{fuhi}. These axially symmetric shapes are shown in Fig.\ \ref{fig6}.
The parameter $c$ describes the elongation of the nucleus (in units of the 
radius $R_0$ 
of a sphere containing $N=\rho_0 4\pi R_0^3/3$ particles, where $\rho_0$ is 
the bulk particle density), $h$ is a necking parameter, and $\alpha\neq 0$ 
describes the left-right asymmetric shapes shown by the dotted lines. The 
sequence of shapes with $h=\alpha=0$ reproduces the optimized shapes  of 
the LDM \cite{myswann69,cosw} 
(see \cite{fuhi} for details). As in \cite{smod}, the
spin-orbit and pairing interactions were 
neglected in  
\cite{migdalrev,brreisie} and, for simplicity, only one kind 
of nucleons (without Coulomb interaction) was used. The only parameter in 
the fission cavity model, the Fermi wave number $k_F=12.1/R_0$, was adjusted 
to yield the second minimum at the deformation $h=\alpha=0,\,c=1.42$ which 
is that of the fission isomer obtained in \cite{fuhi} for the nucleus 
$^{240}$Pu. This corresponds here to a particle number $N\simeq 180$, i.e., 
to $N^{1/3}\simeq 5.65$ when the spin-orbit interaction is neglected. 

This procedure is justified by the observation that, to a first 
approximation, the spin-orbit and Coulomb interactions essentially lead to 
a shift of the magic numbers, preserving the relative shell structures in 
the energy shell-correction. This shift can be simulated by a shift of the
Fermi energy as in \cite{smod}. The procedure works, however, only locally 
in a limited region of deformations and particle numbers. The results shown
below suggest that it is successful in the region $1.3 \siml c \siml 1.65$; the 
ground-state deformations would, e.g., not be reproduced correctly with the 
same Fermi energy. [Note that, in principle, spin-orbit effects can be 
included in the POT, 
see \cite{migdalrev} and references cited therein. 
However, in non-integrable systems one is met with lots of 
bifurcations under the variation of the spin-orbit strength,  
which makes the POT with spin-orbit interactions very cumbersome. Similarly,
the pairing interactions can also be included in the POT, 
but this has not been done for the nuclear-deformation energies so far.]

In \cite{brreisie}, the shortest POs in the $(c,h,\alpha)$ cavity were 
found to dominate the gross-shell features of the double-humped fission 
barrier. For the deformations around the barriers ($c \simg 1.3$), the 
shortest POs are the primitive diagonal and regular polygonal orbits in 
planes perpendicular to the nuclear symmetry axis, situated at the extrema 
of the cavity shape function (seen in Fig.\ \ref{fig6} 
).  
At the onset of the neck ($c=1.49$ 
for $h=\alpha=0$), the orbits in the 
central equatorial plane become unstable with respect to small perturbations
perpendicular to the equatorial plane and give birth to new stable orbits 
lying in planes parallel to the equatorial plane. In the restricted 
deformation space with $\alpha=0$, these bifurcations are of pitchfork
type; they are isochronous from the reflection symmetry with respect
to the equatorial plane.  
When the asymmetry $\alpha\neq 0$ is turned
on in the presence of a neck, the bifurcation is of a more complicated
type. These bifurcations were treated in the uniform approximation employing a 
suitable normal form for the action function \cite{brreisie}. 
(Note that with respect to small perturbations within the equatorial plane,
all these orbits are marginally stable, forming degenerate families with the
degeneracy $\cK=1$ due to the axial symmetry of the cavity.)
Before summarizing the results of \cite{brreisie}, let us study the general 
trends of the shell effects obtained in the fission cavity model and try to 
understand them in terms of the leading POs.

In Fig.\ \ref{fig7}  
we show a contour plot
of the quantum-mechanical 
shell-correction energy $\delta U$ calculated from the s.p.\ energy
spectrum of the fission cavity model, shown versus the cube-root 
of the particle number $N^{1/3}$ and the elongation parameter $c$ along
$h=\alpha=0$ (white: positive values, gray to black: negative values,
see \cite{migdalrev}). 
The horizontal dotted line for $N\simeq 180$ (i.e., $k_F=12.1/R_0$)
corresponds to the situation where the isomer minimum lies at $c\simeq 
1.4$ and the outer (symmetric) barrier is peaked around $c\simeq 1.55$, 
as shown explicitly below. The heavy lines 
give the loci of constant actions of the leading POs (3,1,1)s:
meridian
triangles (triangle orbits in the meridian plane, i.e.,
the plane containing
the symmetry $z$ axis); (2,1)EQ: equatorial diameters; (2,1)AQ: diameter orbits in  
planes parallel to the equator plane. As seen, these lines follow the 
valleys of the minimal shell-correction energy.
As shown in Section II, these periodic orbits are dominating in the PO
expansion (\ref{escscl}) of the shell-correction energy $\delta U$.
Assuming that a certain PO yields the dominant contribution into the PO
sum (\ref{escscl}), one can approximate 
the shell-correction energy by its main term: 
\bea\l{dgpoEeta}
& \!\delta U_{\rm scl}(N,c) \!\approx\! 2\left(\frac{\hbar}{t^{}_{\rm PO}(E_F)}
 \right)^2\delta g^{}_{\rm PO}(E_F,c) \!= \\
& =2\left(\frac{\hbar}{t^{}_{\rm PO}(E_F)}\right)^2
 A_{\rm PO}\;\cos\left[\frac{1}{\hbar}S_{\rm PO}(E_F,c)
 -\frac{\pi}{2}\mu^{}_{\rm PO}\right] \nonumber\\
& \qquad\mbox{with}\quad E_F=E_F(N). \nonumber
\eea
Then, the minima of the shell-correction energy should be distributed along
the lines where the phase takes the values $(2n+1)\pi$ with 
an integer $n$.  These conditions satisfying along the constant-action lines
in the particle number-deformation, $N$-$c$, plane take the form as a 
generalized multi-dimensional quantization rule \cite{strumag},
\bea \l{eq:cac}
S_{\rm PO}(E_F,c)&=&2\pi\hbar\left(n+\frac12+\frac{\mu^{}_{\rm PO}}{4}\right),
\\
&&\qquad n=0,1,2,\cdots\;.\nonumber
\eea
For the valleys corresponding to 
the ground-state deformations, the situation is like in \cite{smod}
obtained for the 
spheroidal models, but here for the more realistic 
fission-cavity model; 
in all cases the meridian orbits dominate the
ground-state valleys. The valleys corresponding to the fission isomers, 
are starting around $c \sim 1.3$. They are determined by 
the shortest POs in 
planes perpendicular to the symmetry axis: up to $c \sim 1.5$, these are 
the equatorial orbits EQ; after their bifurcation at $c=1.49$, the 
valleys are seen to curve down towards smaller values of $N^{1/3}$, 
following the constant-action lines of the stable POs in planes 
parallel to the equator plane (dashed lines, AQ). The bifurcating AQ POs
have larger 
semiclassical amplitudes than the equatorial orbits (EQ) that for $c>1.49$ 
have become unstable. The fact that the quantum-mechanically obtained 
stability valleys follow the (dashed) lines AQ after their branching 
from the lines EQ is a remarkable quantum signature of the classical 
bifurcation effect
as the level-density amplitude enhancement (Section IIB).

The most striking feature of the gross-shell structure
(Fig.\ \ref{fig7}), namely the opposite slopes of the
ground-state valleys ($1 \siml c \siml 1.25$) and the isomer
valleys ($c \simg 1.3$), 
are thus understood semiclassically in terms of the opposite deformation
dependence of the dominating meridian POs in the former valleys and the 
POs in planes perpendicular to the symmetry axis in the latter valleys, 
respectively.

These results can be further elucidated by looking at the Fourier spectra 
in Fig.\ \ref{fig8} for the five values (from top to bottom) $c=1.1,\,
1.2,\,1.4,\,1.5$, and 1.6 (all for symmetric shapes with $h=\alpha=0$). 
The short
arrows underneath the Fourier peaks indicate the lengths of the equatorial
orbits: diameter (2,1)EQ and its second repetition 2(2,1)EQ, triangle (3,1)EQ,
etc., and (for $c=1.6$) the corresponding orbits AQ in the planes parallel
to the equator plane. The long arrows correspond to the meridian
orbits: triangle (3,1,1)s and quadrangle (4,1,1)s.  For
small deformations $c=1.1$ and 1.2, the meridian orbits have the 
strongest amplitudes, and hence, dominate 
the shell structure in yielding
the ground-state deformation valleys (Fig.\ \ref{fig7}).  The 
equatorial orbits EQ and their bifurcated partners AQ have the largest 
amplitudes for $c=1.4$ -- 1.6, which explains their dominance in yielding 
the isomer valleys.

As we discussed in Section IID, the factor $(\hbar/t_{\rm PO})^2$ in
the trace formula (\ref{escscl}) brings about
a natural suppression 
of longer orbits contributing to $\delta U$. This ensures the convergence 
of the PO sum, particularly in non-integrable systems (like the one 
considered here) where the PO sum for the level density 
(\ref{deltadenstot}) 
usually does not converge \cite{gutz}.
This suppression is particularly effective
amongst orbits with comparable amplitudes $A_{\rm PO}$. 
It explains why 
already at $c=1.4$, where the meridian orbits (3,1,1)s and (4,1,1)s still 
have similar amplitudes as the EQ orbits,\footnote{The strong Fourier peak 
near $L/R_0\sim 6.3$ for $c=1.4$ in Fig.\ \ref{fig8} 
contains the 
combined amplitudes of the meridian quadrangle (4,1,1)s and the second 
repetition of the equatorial diameter orbit, 2(2,1)EQ. Although the two 
cannot be disentangled, we estimate that both these orbits have comparable
amplitudes (Section II).} 
the latter dominate the shell structure (by a factor 
$\sim 4$ in the case of the EQ2 orbit), as suggested by Fig.\ \ref{fig7}.

In Fig.\ \ref{fig8} 
we have marked some of the peaks around
$7 \siml L \siml 8.5$ for $c=1.4$ and around $6.5 \siml L \siml 8$ for
$c=1.5$ and 1.6. They correspond to orbits born from the equatorial 
orbits in the period-doubling bifurcations (similar as discussed 
for the spheroidal cavity in
\cite{migdalrev,spheroidptp}); some of them are 3-dimensional 
orbits.  Similarly, there are many other peaks at $L \simg 8$, some of 
which correspond to orbits born in the high $m$-tupling ($m\geq 3$) 
bifurcations. The contributions of all these orbits 
to the gross-shell structure is, however, practically negligible due 
to their long periods. They have therefore not been included in the 
results presented below. They might, however, become noticeable in POT 
calculations with higher resolution of the shell structure. 

We should also recall the fact that in realistic SCM calculations, the 
pairing interactions are known to reduce the amplitude of $\delta U$ by 
up to $\sim$ 30\% (see, e.g., \cite{fuhi}). In the POT, the pairing 
effects  
yield, indeed, an extra smoothing factor in
the semiclassical amplitudes, which further suppresses the contributions 
of longer orbits \cite{migdalrev}.

Let us now look at the influence of the left-right asymmetric shapes with
$\alpha \neq 0$ on the shell-correction energy and, in particular, on 
the height of the second fission barrier. 
In Fig.\ \ref{fig9} 
, the 
semiclassical result of $\delta U$ is shown in a perspective view as 
a function of the elongation $c$ and left-right asymmetry $\alpha$, taken
along $h=0$ in the region of the isomer minimum and the outer fission 
barrier in \cite{brreisie}, see also \cite{migdalrev}. 
We see how the outer fission barrier 
is lowered for the left-right 
asymmetric shapes. Instead of the higher barrier obtained for these 
shapes with $\alpha=0$ (arrow labeled ``symm.''), the nucleus 
can go towards fission over a lower saddle when asymmetric shapes are 
allowed (arrow labeled ``asymm.''). To the left, we see the shapes 
corresponding to the three points A (fission isomer), B and C (along the 
asymmetric fission path) in the deformation energy surface. The vertical 
lines indicate the planes in which the POs are situated (solid lines for 
stable and dashed line for unstable POs). 

The instability of the outer 
fission barrier towards the left-right asymmetric deformations, known from the 
quantum-mechanical SCM calculations, 
can thus be 
described semiclassically using the POT, indeed \cite{migdalrev}. 
Hereby only the shortest
primitive POs are relevant from the fast convergence of the PO sum for
the semiclassical shell-correction energy $\delta U$, as discussed above.

The old quantum-mechanical
results of SCM calculations \cite{fuhi} with some realistic deformed 
Woods-Saxon potentials are compared to  
the semiclassical POT results 
using the present simple fission-cavity model. 
Shown are contour plots 
of $\delta U$ versus $c$ and $\alpha$ for two values of the neck
parameter $h$.  
The semiclassical results (r.h.s.)
reproduce the gross-shell structure of the quantum results (l.h.s.) 
very well (Fig.\ \ref{fig10}). 
The correct topology is obtained, displaying the lowering 
of the outer barrier for several left-right asymmetric shapes. Also, the
amplitudes of the shell effects on both sides are comparable, which 
justifies our calculations of only the gross-shell structure by using the 
shortest periods on a semiclassical level. Of course, a detailed 
quantitative agreement cannot be expected for the two calculations 
using such different potentials as the sophisticated smooth Woods-Saxon 
potential including pairing, spin-orbit, and Coulomb interactions 
on one side (left), and the simple fission-cavity model without 
these extra interactions on the other side (right). The more gratifying 
is the overall good qualitative agreement of the gross-shell 
structure. This agreement demonstrates, by the way, an experience made 
from the quantum-mechanical SCM calculations using a realistic nuclear 
shell-model potential: 
The gross-shell features of the fission barriers are 
much less sensitive to the radial dependence of the potential than to 
its deformation. Hence, the success of a simple cavity model that is 
very schematic, but uses the realistic $c,h,\alpha$ deformations.

The white dashed lines in the r.h.s.\ panels of 
Fig.\ \ref{fig10} 
shows the loci of constant classical actions $S_{\rm PO}$ of the leading POs. 
They follow exactly the valleys of minimal energy in the $(c,\alpha)$ planes 
which define the adiabatic fission paths. Thus, as it was already observed in 
\cite{smod} and seen in Fig.\ \ref{fig7}, 
the condition for minimizing 
the shell-correction energy is semiclassically given by a least-action 
principle: $\delta S_{\rm PO} = 0$. 

We should emphasize that in Figs.\ \ref{fig9} and \ref{fig10} 
only the shell-correction energy $\delta U$ is shown. 
The complete fission
barrier is obtained by adding its smooth LDM part (within the SCM) 
which for $^{240}$Pu in the $(c,h,\alpha)$ parametrization occurs \cite{fuhi} 
at $c\simeq 1.45 - 1.5,\, h=\alpha=0$. Since the LDM barrier is rather 
smooth around its maximum, the relative heights of the isomer minimum 
and the outer barrier are not affected much by it. However, for $c 
\simg 1.6$ the LDM barrier is already going steeply down. Therefore, in 
the total energy, the minimum around $c\simeq 1.65 - 1.7$ (Figs.\ 
 \ref{fig9} and \ref{fig10}  
)
along $h=0=\alpha=0$ 
vanishes in the steep slope of the total fission barrier, 
as shown schematically in Fig.\ \ref{fig5}. 

It is also interesting to note  
that the quantum-mechanical probability maxima of those s.p.\ states which 
microscopically are responsible 
for the asymmetry effect in 
the SCM approach (see the schematic plot on the r.h.s.\ of 
Fig.\ \ref{fig5}) 
lie exactly in the planes perpendicular to the symmetry
axis that contain the classical POs. This constitutes a nice 
quantum-to-classical relationship.  
The classical dynamics of the nucleons with small angular momenta
$L_z$ is more than 90\% chaotic in the region of the outer barrier
\cite{migdalrev}. A very 
small phase-space region of regular motion is thus sufficient to create 
the shell effect that leads towards the asymmetric fission of the nucleus.

We emphasize once more that the fission cavity model, 
in its present form without spin-orbit and Coulomb interactions, 
is not suitable for predicting fission barriers for a larger range of 
nuclear isotopes and deformations. The present 
semiclassical calculation should be taken as a model study of a typical 
actinide fission barrier, demonstrating that the POT in principle is 
capable of explaining the existence of a double-humped barrier, and also 
the onset of mass asymmetry around the outer barrier, in terms of a few
short classical POs. It was in no way meant as
a substitute for the quantum-mechanical SCM calculations of 
static-fission barriers. Its aim was rather to provide, as 
suggested by the late Strutinsky, a qualitative physical understanding
of a sophisticated quantum shell effect by means of simple classical 
pictures.

\section{RADIAL POWER-LAW POTENTIALS}

This Section is devoted to  
the analytical POT derivations for the radial power-low potential
(RPLP). 
The main scaling properties
and classical dynamics in the RPLP will be dealt with in Section IVA. 
The trace formulae for different PO families in this Hamiltonian will be 
derived (Sections IVB-IVD). 
In Section IVE , they will be summarized in 
terms of the total POT sums for
the level-density and energy shell corrections, and 
the Fourier transforms of the quantum level densities and their relation
to the level-density amplitudes for different POs will be obtained.
The semiclassical POT and quantum-mechanical results 
for the RPLP will be compared in Section IVF.

\subsection{Scaling and classical dynamics}

The idea of \cite{arita2012,aritapap} is that the spherical WS 
potential, known as a realistic mean-field potential model for spherical
nuclei and metallic clusters, is nicely approximated (up to a constant shift
and without the spin-orbit term) by a significantly simpler RPLP
which is proportional to a power of the radial coordinate $r^\alpha$,
\footnote{%
In the following, the parameter $\alpha$ is used for the power
parameter and should not be confused with that for reflection-asymmetry 
in the previous section.}
\be\l{potenra}
V(r)=V_0(r/R)^\alpha.
\ee
With a suitable choice of the parameters $V_0$ and $\alpha$, for the realistic
WS potential $V_{\rm WS}(r)$ the approximate 
equality,
\be\l{ramod}
V_{\rm WS}(r) \approx V_{WS}(0) + V_0 (r/R)^\alpha\;,
\ee
holds up to $r\siml R$, where $V_{WS}(0)$ is a WS depth constant.
The mean nuclear radius $R$ for a definite mass number $N$ is given by 
$R=r_0 N^{1/3}$ with $r_0 \approx 1.2$~fm.
Thus, one finds a nice agreement of the quantum 
spectra for the approximation (\ref{ramod}) to the WS potential
 up to and around the Fermi energy $\vareps^{}_F$.

In the RPLP well (\ref{potenra}) [or (\ref{ramod})], there are 
the obvious two-parametric ($\mathcal{K}=2$) diameter families, and
the specific $\mathcal{K}=2$ 
family of circle orbits at edges $L=0$ and $L_C$ of the energy surface 
$I_r=I_r(E,L)$ as for any spherical potentials, respectively,
see Section IIE  and 
Appendix A1.  Again, in a fixed plane crossing the center 
with radius
$r=r_{min}=r_{max}=r_c$, one has the isolated circle
 periodic orbit.  The first equation in (\ref{turneq}) determines 
the turning points $r_{min}(L)$ and $r_{max}(L)$ 
as functions of $L$ at the fixed energy $\vareps$, and their cross 
gives the specific $L=L_{\rm C}=p(r^{}_{\rm C}) r^{}_{\rm C}$. 
As mentioned in Section IIE2,
 according
to the expression for the radial momentum $p_r$ from (\ref{pr})
which must be real, 
the maximal angular momentum $L$  is namely this $L_{\rm C}$ which is
related to the zero $p_r$, and the zero minimal value of the angular
momentum $L~$ 
($0 \leq   L \leq L_{\rm C}$).
As for general spherical potentials (Section IIE), 
the maximal value 
of the angular momentum 
$L=L_{\rm C}$ is related to
a circular orbit and the minimal one $L=0$ corresponds to a diameter.
The critical values $r=r^{}_{\rm C}$ and  $L=L_{\rm C}$ are determined
as the solution of system of the two equations (\ref{rc})
with respect to $r$
and $L$
[${\rm d}^2 p_r^2(r,L) /{\rm d}r^2\neq 0\;$],
as assumed to be the case for the $r^\alpha$ model.
The first equation claims that there is no radial velocity, $\dot{r}=0$,
and the next equation is that the radial force is equilibrated by the
centrifugal force.
For instance, for the potential (\ref{ramod}) the solution of the two  
equations (\ref{rc}) is given by  
\cite{arita2012,aritapap,MVApre2013},
\be\l{rcLcd2Fcra}
\hspace{-0.5cm}r^{}_{\rm C}\!=\!R\left(\frac{2\vareps}{(2\!+\!\alpha) V_0}\right)^{1/\alpha},\,\,\, 
L_{\rm C}\!=\!p(r^{}_{\rm C})r^{}_{\rm C}\;.  
\ee
Another key quantity in the RPLP POT
is the curvature $K$  
of the energy surface $I_r=I_r(\vareps,L)$, (\ref{curv}).

Using the scale invariance valid for the RPLP,
\bea\l{scaling}
&&{\bf r}\to s^{1/\alpha}{\bf r}, \quad
{\bf p}\to s^{1/2}{\bf p}, \\
&&t \to s^{1/2-1/\alpha}t \quad
\mbox{for} \quad \vareps\to s\vareps\;, \nonumber
\eea
one may factorize 
the action integral
$S_{\rm PO}(\vareps)$ along the PO as
\bea\l{actionsc}
&S_{\rm PO}(\vareps)=\oint_{{\rm PO}(\vareps)} \p \cdot \d \r = \\
&\quad =\left(\frac{\vareps}{V_0}\right)^{\frac12+\frac{1}{\alpha}}
\oint_{{\rm PO}(\vareps=V_0)} \p \cdot \d \r \equiv \nonumber\\
&\quad \equiv\hbar\epsi\tau^{}_{\rm PO}\;. \nonumber
\eea
In the last equation, we define the dimensionless
variables $\epsi$ and $\tau_{\rm PO}$, which we call
\textit{scaled energy} and \textit{scaled period}, respectively;
\bea\l{eq:scaledentau}
\epsi&=&\left(\vareps/V_0\right)^{\frac12+\frac{1}{\alpha}},
\\
\tau^{}_{\rm PO}&=&\frac{1}{\hbar}\oint_{{\rm PO}(\vareps=V_0)}
\p \cdot \d \r\;,\nonumber
\eea
as classical characteristics of the particle motion.
To realize the advantage of the scaling
invariance (\ref{scaling}), it is helpful to use
$\epsi$ and $\tau_{\rm PO}$ in place of the
energy $E$ and the period $t_{\rm PO}$, respectively.
In the HO limit ($\alpha\to 2$), $\epsi$
and $\tau^{}_{\rm PO}$ are 
proportional to 
$\vareps$ and $t^{}_{\rm PO}$;
while in the cavity limit
($\alpha \rightarrow \infty$), they are proportional 
to the momentum $p$ and length $\mathcal{L}_{\rm PO}$,
respectively.

The PO (resonance) condition (\ref{percond}) determines 
several PO families in the RPLP well, namely the 
polygonal-like  ($\mathcal{K}=3$), the circular and diametric
($\mathcal{K}=2$)
POs. Fig.\ \ref{fig11} 
shows these POs in the RPLP (\ref{potenra}) in the 
$\{\tau,\alpha\}$ plane, where $\tau(\alpha,L)$ is the
scaled period
and $\tau^{}_{\rm PO}=\tau(\alpha,L_{\rm PO})$ at the angular 
momentum $L=L_{\rm PO}=|{\bf r}\times {\bf p}|^{}_{{\rm PO}}$.
It is clearly seen from this Figure that at $\alpha \geq \alpha_{\rm bif}$
the polygonal-like orbit $M(n_r,n_\theta)$ 
appears, and exists, after the bifurcations,
from the parent circle 
orbit $M$C ($M$-th repetition of the primitive circle orbit C).
The diameter orbits $M$(2,1) are exclusion because their
birth arise exactly at the harmonic oscillator (HO) symmetry-breaking point
$\alpha=2$ and exist for all larger values, $\alpha>2$.

\subsection{Three-parametric PO families}

\subsubsection{ISPM derivations of the trace formula}

For the contribution of the  three-parametric   ($\mathcal{K}=3$) 
{\it families} 
into the trace formula (\ref{pstraceactang}) for the shell correction, 
after the exact integration
over $L_z$, having $2L$; and $\varphi$, $2\pi$,  
for $\mathcal{D}=3$ one obtains
\bea\l{pstraceactang1}
&\hspace{-0.2cm}\delta g_{\rm scl}(\vareps)
\!=\!\frac{4\pi}{(2\pi\hbar)^3}\! \Re\sum_{\rm CT}
\!\int \d \theta''\,\d \Theta_r''
\, \d I_r
\!\times \\
&\times\int \d L\,L\;
\delta\left(\vareps-H(I_r,L)\right)\left|\mathcal{J}_{\rm CT}
\left(\p''_\perp,\p'_\perp\right)\right|^{1/2} \times \nonumber\\
&\quad\times\exp\left(\frac{i}{\hbar}
\Phi_{\rm CT} - \frac{i \pi}{2}
\mu_{\rm CT}\right). \nonumber
\eea
Taking the integral
over $I_r$ exactly by using the $\delta$-function which ensures the 
energy conservation, one has
\bea\l{pstraceactang2}
&\hspace{-0.4cm}\delta g_{\rm scl}(\vareps)
=\frac{4\pi}{(2\pi\hbar)^3}\, \Re\sum_{\rm CT}
\int \d \theta''\, \frac{\d \Theta_r''}{\omega_r}
\, \d L\,L \times \\
&\quad\times\;\left|\mathcal{J}_{\rm CT}
\left(\p''_\perp,\p'_\perp\right)\right|^{1/2}
\exp\left(\frac{i}{\hbar}
\Phi_{\rm CT} -\frac{i \pi}{2}
\mu^{}_{\rm CT}\right)\;. \nonumber
\eea
Then, we integrate over the angle variable $\Theta_r''$ 
accounting for independence
of the integrand, in particular, of the
action phase $\Phi_{\rm CT}$, on any variations of this angle.
With the corresponding time variable, $\d \Theta_r''/\omega_r=
\d t$, along the
POs, one finds
\be\l{periodint}
\int_0^{2\pi} \frac{{\rm d}\Theta_r''}{|\omega_r|}=
\int _0^{T_{r\;{\rm CT}}} {\rm d}t = T_{r\;{\rm CT}}\;,
\ee
where $T_{r\;{\rm CT}}$ is the time duration for a primitive (without
repetitions) particle motion along the CT, one obtains
\bea\l{pstraceactang3}
&\hspace{-0.2cm}\delta g_{\rm scl}(E)
\!=\!\frac{4\pi}{(2\pi\hbar)^3}\! \Re\sum_{\rm CT}
\int \d \theta'' 
\d L L\, T_{r\;{\rm CT}}\!\times \\
&\quad\times \left|\mathcal{J}_{\rm CT}({\bf p}''_\perp,{\bf p}'_\perp)\right|^{1/2}
\exp\left(\frac{i}{\hbar}
\Phi_{\rm CT} - \frac{i\pi}{2}
\mu^{}_{\rm CT}\right). \nonumber
\eea
All quantities in the integrand are taken at the energy surface
$I_r=I_r(E,L)$, defined by (\ref{actionvar}).

Applying the SPM conditions (Sections
IIA and IIE \cite{MAFptp2006})
for the perpendicular angle $\theta'$, 
one notes that there is the continuum of the stationary points 
$\theta''=\theta^{\prime\prime\;*}=\theta^*$ within
$0\leq \theta^* \leq 2\pi$ as solutions of the
SPM equations \cite{MAFptp2006}. Therefore, 
the phase $\Phi_{\rm CT}$
in exponent does not depend on this angle,
\be\l{phase3}
\Phi_{\rm CT}=2\pi \left[M_r\,I_r(E,L) + M_{\theta} \,L\right]\;,
\ee
where $M_r,M_\theta$ are integers,
$M_r=M n_r$, $M_\theta=M n_\theta$, $n_r$ and $n_\theta$ 
are the {\it positive} co-primitive
integers, $M$ is nonzero integer. 
So, writing exactly $2 \pi$ for the integral over $\theta''$ 
in (\ref{pstraceactang3}), one
obtains the semiclassical Poisson summation trace formula
which can be derived alternatively from the quantum Poisson summation trace
formula by using the EBK quantization rules \cite{bt76}
for the case of the spherical symmetry of the Hamiltonian,
\bea\l{poissonsum}
&g_{\rm scl}(\vareps) 
= \frac{2}{\hbar^3}\Re\sum_{M,n_r,n_\theta}\, \int
\d L\:\frac{L}{\omega_r} \times \\ 
&\quad\;\times\exp\left\{\frac{2\pi i}{\hbar}
M\,\left[n_r\,I_r(\vareps,L)
+n_\theta L\right] -\right.\nonumber\\
&\qquad\qquad\left. -\frac{i\pi}{2}\mu^{}_{M,n_r,n_\theta}
\right\}\;, \nonumber
\eea
Formally, before taking the trace integral over the angular
momentum $L$ by the SPM in 
(\ref{poissonsum}),  one can 
consider positive and negative $M$ as related to the two
opposite directions of motion along a classical trajectory CT. This yields
the equivalent contributions into the trace formula 
due to a time-reversibility invariance of 
the Hamiltonian. Therefore, we may write simply the additional factor 2 in
(\ref{poissonsum}) but with replacing the summation over 
$M$ by  positive integers
($M>0$). 

We emphasize that for $\mathcal{K}=3$ families 
the generating function
(\ref{legendtrans}) 
becomes independent of the perpendicular
angle variable $\theta''$  for the integrable Hamiltonian, see
(\ref{phase3}),
in contrast to the Hamiltonian $H(I_r,L)$ itself which 
always does not depend on the angle variables because of  
integrability of the system \cite{MAFptp2006}. 
Exceptions are the 
complete degeneracy as the HO, see below.

The integration range in (\ref{poissonsum}) 
taken from the minimal, $L=0$, to the maximal value, 
$L=L_{+}$, 
covers in the integration variable $L$
the contributions of whole manifold of closed and unclosed
trajectories of the tori in the phase space around
the stationary point $L^*$, which 
corresponds to the periodic
orbit. The maximal angular momentum $L_{+}$ is restricted
by the energy conservation for a given energy $\vareps$. 
By the finite limits for contributions of different orbits they
are approximately independent, except for some exclusions
\cite{MAFptp2006}. 
Here, in the case of the
spherical Hamiltonians such a relationship between the
finite limits for different kind of orbits takes place too
and will be discussed later in relation to the HO
limit $\alpha \rightarrow 2$. In this limit, the sum of the
the trace formulas for the 
two different kinds of
families, with the maximal degeneracy $\mathcal{K}=3$, and 
smaller for $\mathcal{K}=2$ circle and diameter orbits, 
turns into the spherical HO trace formula.
 They are assumed naturally to be in
different parts of the four parametric
($\mathcal{K}=4$) continuum of the HO periodic-orbit tori in such a 
symmetry-breaking limit \cite{MAFptp2006}.
 The latter
is an exclusion because of the bifurcation at $\alpha=2$ where
we meet the maximally degenerated spherical HO.
We shall specify the integration limits $L_{+}$ for the contribution
of the ${\cal K}=3$ families into (\ref{poissonsum}) 
in relation to the corresponding integration limits 
for the circular orbits and HO limit below.

We apply then the stationary phase condition
with respect to the variable $L$ for the exponent phase $\Phi_{\rm CT}$
(\ref{phase3}) in the integrands of (\ref{poissonsum}),
\be\l{statcond3L}
\left(\frac{\partial \Phi_{\rm CT}}{\partial L}\right)^*=0\;,
\ee
which is exactly the resonance condition
(\ref{percond}) [see (\ref{phase3})].  
This condition
determines the stationary phase point $L=L^*$
related to the POs $M(n_r,n_\theta)$ of $\mathcal{K}=3$
families. Note that the semiclassical
Poisson-summation formula (\ref{poissonsum}) obtained from the quantum
Poisson-summation trace formula with further using the EBK
quantization \cite{sclbook,kaidelbrack} contains
the sum over integers for $M_r$ from $-\infty $ to $+\infty$ and
independently for  $M_\theta$  from $-\infty $ to $+\infty$, where
$M_r=0,M_\theta=0$ is related to TF smooth density.
For the derivation of the Gutzwiller trace formula for isolated
orbits \cite{richens} 
it was important that $M_\theta/M_r$ can be also negative
and they can take also zero values, except for simultaneous zeros 
$M_r=0,M_\theta=0$ of the TF component. Another assumption
is that the end points are not the stationary points, in contrast to
our derivations within the extended Gutzwiller approach (EGA).

We expand now the exponent phase $\Phi_{\rm CT}$ 
(\ref{phase3}) in the variable $L$ near the stationary point $L^*$ 
to second
order assuming that there
is no singularities in the curvature (\ref{curv}) for the
contribution of all 
${\cal K}=3$ families,
\bea\l{phaseexp}
\Phi_{\rm CT} &\equiv & 2\pi\left[M_rI_r(\vareps,L) + 
M_\theta L\right] \approx \\ 
&\approx& 
S_{M {\rm P}}(\vareps) +\frac12 J_{M{\rm P}}^{({\tt L})}(L- L^*)^2\;, \nonumber
\eea
where $S_{M{\rm P}}(\vareps)$ is the action along the 
polygon-like PO 
families specified by the two
integers $n_\theta$ and $n_r>2 n_\theta$ from the resonance (PO) 
condition (\ref{percond}) for such a primitive PO, P$=(n_r,n_\theta)$, 
and the number of its repetitions
$M$,
\begin{equation}
S_{M{\rm P}}(\vareps) = 2\pi M \left[n_r\, I_r(\vareps,L^*)+n_\theta \,L^*\right]\;.
\label{actionpo}
\end{equation}
In (\ref{phaseexp})  and (\ref{actionpo}),
 $L^*=L^*(n_r,n_\theta)$ is the solution of the PO
equation (\ref{percond}). The Jacobian $J_{M{\rm P}}^{({\tt L})}$ in 
(\ref{phaseexp}) is the 
stability of the PO with respect to the
variation of the angular momentum $L$ 
at the same energy surface,
\bea\l{jacobpar}
J_{M{\rm P}}^{({\tt L})} &=&\left(\frac{\partial^2 S_{\rm CT}}{\partial L^2} 
\right)_{L=L^*} =2\pi M n_r K_{\rm P}\;, \\  
K_{\rm P} &=&  \left(\frac{\partial^2 I_r}{\partial L^2}\right)_{L=L^*}\;, 
\nonumber
\eea
where $K_{\rm P}$ 
is the curvature (\ref{curv}) of the energy surface $I_r(\vareps,L)$, see the
first equation in (\ref{actionvar}).

We substitute now the expansion (\ref{phaseexp}) into the last equation
of (\ref{poissonsum}) and take there the
pre-exponential factor off the integral at $L=L^*$. 
For the sake of
simplicity, we shall discuss the lowest, i.e., second-order expansion 
of the exponent phase, and the zero-order expansion of the 
pre-exponent factor with respect to the
$L$ variable in (\ref{poissonsum}). Thus, we are left with the integral over 
$L$ of a Gaussian-type integrand within the finite limits 
mentioned above for contributions of
the three-parametric polygon-like families, including 
the contribution of the boundaries for $0<M_\theta/M_r <1/2$.
The integer numbers $M_r=2\,M_\theta=2\,M$ and $M_r=M_\theta=M$ are 
related to the diameter and circle POs.

When the stationary point $L^*$ is far away from the
physical integration ends, one can extend the integration limits to the infinity
region from $-\infty$  to $\infty$, 
and we arrive asymptotically at the Berry\&Tabor result of the
standard POT extended to continuous symmetries 
for the contribution of the 
families ($\mathcal{K}=3$)\cite{bt76} as applied to spherical
potentials. If
the stationary point is close to these ends of the physical tori,
one has to use the finite limits,
i.e.\ , within the ISPM of second order for the phase expansion,
as the simplest approach. For instance, it is the case near $L=L_{\rm C}$
where $I_r=0$, and one has bifurcations of the 
${\cal K}=3$ polygon-like 
from the corresponding $\mathcal{K}=2$ circle family.

Taking the integral over $L$ within the finite limits, we obtain
the trace formula in terms of the error functions. Thus, for the 
ISPM contributions of families of
the three-parametric ($\mathcal{K}=3$) orbits 
$\delta g^{(3)}(E)$, one obtains
\bea\l{deltag3isp}
&&\delta g^{(3)}(\vareps) = \sum_{M{\rm P}}
\mathcal{A}_{M{\rm P}}^{(3)}(\vareps)\times \\ 
&&\times\exp\left[\frac{i}{\hbar}\,
S_{M{\rm P}}(\vareps)-i \frac{\pi}{2} \mu^{}_{M{\rm P}}\right]\;. \nonumber
\eea
Here, the sum is taken over the 
families of the periodic
orbits, $M$P (with accounting for the repetition number $M$), in the
spherical potential, $S_{M{\rm P}}(\vareps)$ is the action
(\ref{actionpo}) along the PO, $M(n_r,n_\theta)$ (with $n_r >
2\,n_\theta$).  For the amplitude $\mathcal{A}_{M{\rm P}}^{(3)}$, one
finds
\be\l{amp3isp}
{\cal A}_{M{\rm P}}^{(3)}=\frac{L_{\rm P} \,T_{\rm P}}{\pi\hbar^{5/2} \sqrt{M n_r\,
K_{\rm P}}}\, \mbox{erf}\left(\mathcal{Z}_{M{\rm P}}^{+},\mathcal{
  Z}_{M{\rm P}}^{-}\right)
\,{\rm e}^{i \pi/4}\;,
\ee
where
$T_{\rm P}=T_{n_r,n_\theta}$ is the period of the
primitive (M=1) periodic orbit P, $(n_\theta,n_r)$, of the 
three-parametric families
for the stationary point $L=L^*$ determined by 
the PO equation (\ref{percond}),
\be\l{period3}
T_{\rm P}=\frac{2\pi n_r}{\omega_r}=\frac{2\pi n_\theta}{\omega_\theta}\;,
\ee
$L_{\rm P}=L^*$ is the classical angular momentum for the particle motion
along a P PO.
The function $\mbox{erf}(v,u)$ in (\ref{amp3isp})  is
expressed through the standard error
function  of a complex argument,
\be\l{errorf}
\mbox{erf}(v,u)=\frac{2}{\sqrt{\pi}}\int_{u}^{v} dz e^{-z^2}
=\mbox{erf}(v)-\mbox{erf}(u)\;, 
\ee
The complex arguments $\mathcal{Z}_{M{\rm P}}^{\pm}$  of the error
functions in (\ref{amp3isp}) are expressed in terms of the curvature
$K_{\rm P}$, see (\ref{jacobpar}) at $L=L^*$,
through the Jacobian $J_{M{\rm P}}^{({\tt L})}$  (\ref{jacobpar}) with
the explicit stationary points $L^*=L_{\rm P}$,
\bea\l{argerrorpar}
\mathcal{Z}_{M{\rm P}}^{-} &=& \sqrt{-i \pi\, M\,n_r\, K_{\rm P}/\hbar}\,\, 
\left(L_{-}-L_{\rm P}\right)\;, \\  
{\cal Z}_{M{\rm P}}^{+} &=& \sqrt{-i\pi \,M\,n_r\, K_{\rm P}/\hbar}
 \,\left(L_{+}-L_{\rm P}\right)\;. \nonumber
\eea
We used here the simplest approximation
for the finite integration limits within the tori,
minimal $L_{-}=0$ and maximal $L_{+}=L_{\rm C}$ values
of the angular-momentum integration variable for the ${\cal K}=3$ family 
contribution. The phase $\mu^{}_{M{\rm P}}$ in (\ref{deltag3isp}) 
is related to the Maslov index as
in the asymptotic Berry\&Tabor trace formula.

Note that the family amplitude ${\cal A}_{M{\rm P}}^{(3)}$
(\ref{amp3isp})
is continuous in the HO limit where $K_{\rm P} \rightarrow 0,$ 
due to cancellation of the singularities in the denominator
proportional to $\sqrt{K_{\rm P}}~$, with the same coming
from the finite limits (\ref{argerrorpar}) (see below in this 
Section for the total trace formula including the circle orbits). There is no
singularity also coming from the separatrix (the potential barrier,
for instance) where $K_{\rm P} \rightarrow \infty$. In this limit
one has obviously zero limit as for the integrable
H\'enon-Heiles potential \cite{MAFptp2006}.

\subsubsection{The SSPM limit}
 
In order to get the standard Berry\&Tabor trace formula limit we 
consider the stationary point  being inside of the integration
interval asymptotically far from the bifurcation points $L=L^*=L_{\rm C}$.
In this case for the contribution 
of the ($\mathcal{K}=3$) families (\ref{deltag3isp}), one can transform the
error functions to the complex Fresnel functions 
with the real limits. 
As noted above, in this case one can extend the upper limit to
$\infty$ and the lower one to $-\infty$ far from the bifurcations
of a circular orbit. 
In this way we arrive at
the result (\ref{deltag3isp}) with the amplitude $A_{M{\rm P}}^{(3)}$
of the standard SPM (SSPM) identical to the 
Berry\&Tabor trace formula 
\cite{bt76}, 
\begin{equation}
A_{M{\rm P}}^{(3)}= 
\frac{2\,L_{\rm P} \,T_{\rm P}}{\pi\hbar^{5/2} \sqrt{M n_r\,
K_{\rm P}}}\, e^{i\pi/4}.
\label{amp3ssp}
\end{equation}

The Maslov phase   
$\mu^{}_{M{\rm P}}$ (\ref{deltag3isp}) 
is determined in terms of the number of
turning and caustic points by using the Maslov\&Fedoriuk theory
\cite{bablo,fedoryuk_pr,maslov,fedoryuk_book1,MAFptp2006}.
It is different for the smoothed,
\bea\l{masl3Dra}
\mu^{}_{M{\rm P}} 
&=& \sigma^{}_{M{\rm P}}-1 \;, \\
\sigma^{}_{M{\rm P}}&=&3\,M\,n_r + 4\,M\,n_\theta\;, \nonumber
\eea
and billiard,
\bea\l{masl3Dbill}
\mu^{}_{M{\rm P}}&=& \sigma^{}_{M{\rm P}}\;, \\
\sigma^{}_{M{\rm P}}&=&2\,(M\,n_r + M\,n_\theta)\;, \nonumber
\eea
spherical potentials due to a difference in the quantum boundary conditions.
The total Maslov phase $\mu_{M{\rm P}}^{\rm (tot)}$ defined as
a sum of the two terms, this asymptotic part (\ref{masl3Dra})
and the argument of the complex density amplitude (\ref{amp3isp}),
which is additional to the
asymptotic one, depends on the energy $\vareps$ and parameters of
the spherical potential as $\alpha$ in (\ref{potenra}).
This total Maslov phase is changed through the bifurcation points 
smoothly due to the second term.

The amplitude (\ref{amp3isp}) of our solution (\ref{deltag3isp}) is
regular at the bifurcations which are the end points $L=L^*=L_{\rm C}$ 
of the action ($L$) part of a tori.
The essential difference of the ISPM from the Berry\&Tabor 
theory \cite{bt76}
is that the equation (\ref{deltag3isp}) for the
orbits with the highest degeneracy
is one of terms of the total solution of
the breaking-of-symmetry problem.
Thus, within the SPM of the EGA 
we have to consider separately the derivations of the other 
families in the spherical potentials, namely the circle and diameter 
$\mathcal{K}=2$ orbits
beyond the  semiclassical Poisson summation trace formula 
(\ref{poissonsum}).

The ISPM trace formula (\ref{deltag3isp}) for the contribution of
the three-parametric $M$P
families contains the {\it end} contributions
related to the finite limits of the integrations in the error functions. 
This trace formula yields the contribution of
the isolated $\mathcal{K}=3$ families. This essentially was used 
in the derivation of (\ref{deltag3isp}) from the
initial trace formula (\ref{pstraceactang2}) taking $2\pi$ 
for the integral over 
the perpendicular angle $\Theta''$. Therefore,  in (\ref{deltag3isp}),
there is no 
contributions of the both circular and diametric orbits
which correspond to the {\it end} stationary-phase points $L=L_{\rm C}$
and $L=0$, respectively.

\subsubsection{The spherical billiard limit}

In this limit [$\alpha \rightarrow \infty$ for the RPLP 
(\ref{potenra}), and $V_0 \rightarrow \infty$ 
for the WS potential (\ref{ramod})] 
the action $S_{M{\rm P}}(\vareps)$
of the general spherical trace formula 
(\ref{poissonsum}) is given by 
\begin{equation}
S_{M{\rm P}}(\vareps)= p\,\mathcal{L}_{M{\rm P}},\qquad p=\sqrt{2\,m\,\vareps}\;,
\end{equation}
where $p$ is the momentum modulus [$V(r)=0$ for $r\leq R$ and $V(r)=\infty$
at $r>R$ for $\alpha \rightarrow \infty$], 
$\mathcal{L}_{M{\rm P}}=2 \,M \,n_r \,R\,\sin \phi$ is
the length of the 
polygonal orbit $M$P, 
 $M(n_\theta,n_r)$, $R=r_{\rm max}$, the radius
of the billiard, $\phi=\pi\, n_\theta/n_r$.
$L_{\rm P}=p \,r_{\rm min}=p\,R\,\cos \phi$ the angular momentum, and
$t^{}_{M{\rm P}}=2\,M R\,m\,n_r\,\sin \phi/p$  
the period of the periodic orbit
$M(n_\theta,n_r)$. 
The curvature $K_{\rm P}$ (\ref{curv}) of (\ref{jacobpar})
and (\ref{amp3ssp})
can be calculated explicitly in this limit,
\begin{equation}
K_{\rm P}=\left(\pi\,p\,R\,\sin \phi\right)^{-1}\;.
\label{curvbb}
\end{equation}
Substituting all of these quantities into (\ref{amp3ssp}),
one obtains the Balian\&Bloch trace formula for
spherical billiards \cite{bablo}:
\bea\l{bb3}
\hspace{-0.5cm}\delta g^{(3)}(\vareps)&=&
\frac{2mR^2}{\hbar^2}\,{\rm Re}\,\sum_{M_\theta} \sum_{M_r > 2\,M_\theta}
\sin\left(2\phi\right)\times \\
&\times&\sqrt{\frac{p\,R\,\sin \phi}{\pi \,\hbar\,M_r}}\,
\exp\left\{\frac{i}{\hbar}\,p\,\mathcal{L}_{\rm PO}\; -\right.\nonumber\\ 
&-&\left. 
\frac{i\pi}{2}\,\mu^{}_{M_r,M_\theta}-
\frac{3i\pi}{4}\right\}\;, \nonumber
\eea
where $M_r=M n_r$ and $M_\theta=M n_\theta$.   
However, the Maslov phase $\mu^{}_{M_r,M_\theta}$ is different 
for a slightly smoothed diffuse edge  and strictly cavity potential,
according to the Maslov\& Fedoriuk catastrophe theory 
\cite{fedoryuk_pr,maslov,fedoryuk_book1,chester,MAFptp2006}, see 
(\ref{masl3Dra}) and (\ref{masl3Dbill}),
respectively.

Note that all roots $L=L^*=L_{\rm P}$ of the stationary phase
equation (\ref{percond}) for $\mathcal{K}=3$ families
{\rm PO}, $M(n_\theta,n_r)$, 
are in between minimal $L=L^*=0$ for the 
diameters and maximal $L=L_{\rm C}$ for the circular
orbits, $0 < L_{\rm P} <L_{\rm C}$. The boundary stationary
points $L=0$ and $L=L_{\rm C}$ are exclusion cases 
in relation to the derivations of their contributions
into the phase space trace formula.
The HO limit of the ISPM trace formula (\ref{deltag3isp}) 
is considered below together with the ISPM trace formula for the
circular and diametric orbits.

\subsection{ TWO-PARAMETRIC  
CIRCLE FAMILIES}

\subsubsection{ISPM trace formulae for circle orbits}

Within the EGA, for the contribution of the $\mathcal{K}=2$
families of the circle ($M$C) orbits
into the trace formula (\ref{pstraceactang}), 
we first integrate over $\varphi''$ and $L_z$ as in 
Section IVB, see (\ref{pstraceactang1}).
In contrast to the derivations of contribution of the maximally
degenerate ${\cal K}=3$ orbits, for the circular orbits
we  now take into account existence of the isolated
stationary point of the phase integral $\Phi_{\rm CT}$ 
(\ref{legendtrans}) in the perpendicular spherical 
phase-space variables $r''=r^{\prime\prime\;*}=r^{}_{\rm C}~~$, 
$~~p_r'=p_r^{\prime\;*}=0$
in the center plane using the spherical phase-space variables 
$~\{r'',\theta'',\varphi''; p_r', p_\theta', p_\varphi'\} ~$.
Integrating exactly over the angular-momentum
projection, $p_{\varphi}=L_z$, in (\ref{pstraceactang}), where the integrand 
is independent of  $L_z$, and also, over $p_\theta$ by using the 
$\delta$-function of the energy conservation, and 
taking
then the integrals over $\varphi''$ and $\theta''$, 
 one obtains
 \bea\l{pstracerpC}
&&\hspace{-0.5cm}\delta g_{\rm scl}(\vareps)
\!=\!\frac{1}{2\pi^2 \hbar^3}\!\Re\sum_{\rm CT}
\!\int_{r_{-}}^{r_{+}}\! {\rm d} r^{\prime\prime}
\int_{p_r^{-}}^{p_r^{+}} {\rm d} p_r^{\prime}L\times \\
\hspace{-1.0cm}&\times&T_{\theta\;{\rm CT}}
\left|\mathcal{J}_{\rm CT}(p'_r,p''_r)\right|^{1/2}
\exp\left[\frac{i}{\hbar}
\Phi_{\rm CT} - \frac{i \pi}{2}
\mu^{}_{\rm CT}\right]. \nonumber
\eea  
Here we used the identity $\int \d \Theta''/\omega_\theta=T_{\theta,{\rm CT}}~$
[$T_{\theta,{\rm CT}}$ is the time of particle motion along a CT]  
in (\ref{pstraceactang1})] as in Section IVB, see
(\ref{periodint}).  The limits $r_\pm$ and 
$p_r^{\pm}$ for the remaining integrals in (\ref{pstracerpC})
correspond to the interval from the minimal $r_{-}=0$ and 
$p_r^{-}=-p(r)$ 
to the maximal  $r_{+}=r_{\rm max}=R (\vareps/V_0)^{1/\alpha}$ 
and $p_r^{+}=p(r)$ values.

The stationary phase conditions for the SPM integration over 
the radial momentum $p_r'$ and coordinate $r''$  in 
(\ref{pstracerpC}) are given by
\bea\l{statcondCp}
&&\left(\frac{\partial \Phi_{\rm CT}}{\partial p_r'}\right)^\ast
\equiv \left(r' -r''\right)^\ast=0\;, \\ 
&&\left(\frac{\partial \Phi_{\rm CT}}{\partial r''}\right)^\ast
\equiv -\left(p_r' -p_r''\right)^\ast=0\;. \nonumber
\eea
Solutions of these equations are the isolated 
stationary point $p_r^{\prime}=
p_r^{\prime\;\ast}=0$ and $r^{\prime\prime}=r^{\prime\prime\;\ast}=r^{}_{\rm C}$ 
related to the stationary point $L=L^\ast=L_{\rm C}$. in the $L$ variable. 
They are equivalent to  the closing (PO) conditions 
$r'=r''=r^{}_{\rm C}$ and $p_r' = p_r''=0$ in the 
phase space. Here and below in this section
the upper index star means that the corresponding
quantity is taken at
the stationary point $p_r^{\prime}=p_r^{\prime\;*}=0$ 
of the isolated periodic circular orbit in a center plane.
We expand now the phase $\Phi_{\rm CT}$ (\ref{legendtrans})
in the momentum $p_r'$  and radial coordinate $r''$ near this phase-space 
point $\{p_r^{\prime\;\ast}=0~, ~r''=r^{}_{\rm C}\}$.
At second order, one has
\be\l{phipiexpC}
\!\Phi_{\rm CT}\!=\!
S_{M{\rm C}}(\vareps) + \frac{1}{2}{\cal J}_{M{\rm C}}^{(p)}
\left(p_r^{\prime}\right)^2 + \frac{1}{2}{\cal J}_{M{\rm C}}^{(r)}
\left(r''\!-\!r^{}_{\rm C}\right)^2, 
\ee
where $S_{M{\rm C}}(\vareps)$ is the action along the circular PO, 
$M$C with the period number $M$
($\Phi_{\rm CT}^\ast =S_{M{\rm C}}$),
\be\l{actionC}
S_{M{\rm C}}(\vareps)=M \int_{0}^{2\pi} L \,\d\,\theta = 2\pi M\,L_{\rm C}\;,
\end{equation}
$L_{\rm C}$ is the angular momentum of the particle moving along the PO 
$M$C, and $M$ its repetition number. 
For the Jacobians ${\cal J}_{\rm PO}^{(p)}$ and ${\cal J}_{\rm PO}^{(r)}$, one 
finds
\be\l{jacpC}
\mathcal{J}_{M{\rm C}}^{(p)}=
\left(\frac{\partial^2 \Phi_{\rm CT}}{\partial
  p_r^{\prime\,2}}\right)_{M{\rm C}}=
\left(\frac{\partial r^{\prime\prime}}{\partial p_r^{\prime}}\right)_{M{\rm C}}\;, 
\ee 
\bea\l{jacrC}
\!\mathcal{J}_{M{\rm C}}^{(r)}&=& 
\left(\frac{\partial^2 \Phi_{\rm CT}}{\partial
  r^{\prime\prime\;2}}\!+\!2\frac{\partial^2 \Phi_{\rm CT}}{\partial
  r^{\prime}\partial r^{\prime\prime}}\!+\!
\frac{\partial^2 \Phi_{\rm CT}}{\partial
  r^{\prime\,2}}\right)^*\!= \\ 
&=&
\left(\frac{\partial p_r''}{\partial
  r^{\prime\prime}} - 2\frac{\partial p_r'}{\partial r^{\prime\prime}}-
\frac{\partial p_r'}{\partial
  r^{\prime}}\right)_{M{\rm C}} = \nonumber\\
&=& -F_{M{\rm C}} \left[{\cal J}_{M{\rm C}}^{(p)}\right]^{-1}\;, \nonumber
\eea 
$F_{M{\rm C}}$ is the Gutzwiller stability factor,
\be\l{fgutzin}
F_{M{\rm C}}=2-\Tr\mathcal{M}_{M{\rm C}} = 4 \sin^2\left[\frac{\pi M  
\Omega_{\rm C}}{\omega^{}_{\rm C}}\right]\;,
\ee
\be\l{Omrcra}
\Omega_{\rm C} 
=\sqrt{\frac{2\alpha \vareps}{mR^2}}\,
\left[\frac{(2+\alpha) V_0}{2\vareps}\right]^{1/\alpha}>0\;,
\ee
is the radial frequency, $\omega^{}_{\rm C}$
the azimuthal one, 
\bea\l{omtcra}
\omega^{}_{\rm C}&=&\omega_\theta(L=L_{\rm C})=L_{\rm C}/(m\,r_{\rm C}^2) = \\
&=&\sqrt{\frac{\alpha V_0}{mR^2}}\,
\left(\frac{2\vareps}{(2+\alpha) V_0}\right)^{1/2-1/\alpha}\;. \nonumber
\eea
$r^{}_{\rm C}$  is the radius of the C orbit, and 
$L_{\rm C}$ the angular momentum for a particle motion along the C PO,
see (\ref{rcLcd2Fcra}).
Therefore, for the bifurcation values of $\alpha$ 
which yield zeros of the Gutzwiller
stability factor (\ref{fgutzin}), one has
\be\l{bifeq}
\frac{\Omega_{\rm C}}{\omega_{\rm C}} \equiv \sqrt{\alpha+2}
=\frac{n_r}{n_\theta}\;.
\ee
With the identity in (\ref{bifeq}), from (\ref{fgutzin})
one explicitly obtains
\be\l{fgutz}
F_{M{\rm C}}= 4 \sin^2\left(\pi M\sqrt{\alpha+2}\right)\;.
\ee
For the bifurcation point $\alpha_{\rm bif}$ which turns the stability
factor (\ref{fgutz}) into zero, 
from (\ref{bifeq})  
one finds  
\cite{arita2012,aritapap,MVApre2013}
\be\l{bifeqra}
\alpha_{\rm bif}=n_r^2/n_\theta^2-2\;.
\ee
At this point, the P families appear and exist
at $\alpha \geq \alpha_{\rm bif}$ with the stationary point 
$L=L^* \leq L_{\rm C}$ 
as the solution $L^*$ of the PO equations (\ref{percond}).
There is one specific bifurcation point $\alpha=2$ 
which corresponds to the spherical harmonic oscillator (HO) with the
frequency $\omega_0=\sqrt{2V_0/R^2}$. For this $\alpha$,  the ratio 
$\omega_r/\omega_\theta$ found from (\ref{freq}) is identical 
to 2 at all of accessible $L$, i.e.\ , 
one has 4 parametric families with $n_r=2, n_\theta=1$ which exist
at any $L$ within continuum $0 \leq L\leq \vareps/\omega_0$.
In this limit the $(\mathcal{K}=2)$ $M$C
family is not isolated 
but belongs to the four parametric
family mentioned above along with diameters (see below).
For spherical billiard limit, $\alpha \rightarrow \infty$, 
the circle orbit $(\mathcal{K}=2)$
family  
disappears as degenerated
into the center of billiard  ($r^{}_{\rm C} \rightarrow 0$,
$L_{\rm C} \rightarrow 0$).

In the last equation of (\ref{jacrC}) we used the general
definitions of the stability matrix (\ref{fgutzin}) and 
 properties of the action phase
as a generating function, see also very right of 
(\ref{jacpC}) for the Jacobian
${\cal J}_{M{\rm C}}^{(p)}$.
Thus, the Jacobian calculations are reduced to those of the Jacobian 
$\mathcal{J}_{M{\rm C}}^{(p)}$ (Appendix A4), 
\be\l{jacpCres}
\mathcal{J}_{M{\rm C}}^{(p)}= 2 \pi (\alpha+2)\;
M\; K_{\rm C}\;r_{\rm C}^{2}\;,
\ee
where $K_{\rm C}$ is the curvature
for C orbits \cite{MVApre2013},
\be\l{curvraC}
K_{\rm C} = -\frac{(\alpha+1)(\alpha-2)}{12 \,(\sqrt{\alpha+2})^3\,
  L_{\rm C}}\;. 
\ee

Substituting the expansion (\ref{phipiexpC}) into (\ref{pstracerpC})
and taking the pre-exponent amplitude factor 
off the integrals at the stationary point
$p_r^{\prime}=p_r^{\prime\;\ast}=0$ and $r^{\prime\prime}= 
r^{\prime\prime\;\ast}=r^{}_{\rm C}$, one finally arrives at
\bea\l{dengenC}
&&\delta g_{{\rm scl,C}}^{(2)}(\vareps)=
\Re\sum_{M}\,A_{M{\rm C}}^{(2)}\times \\ 
&\times&\exp\left[\frac{i}{\hbar}
\,S_{M{\rm C}}\left(\vareps\right) - \frac{i \pi}{2}
\mu^{}_{M{\rm C}}\right]\;. \nonumber
\eea
The sum is taken over  the repetition number $M$ for the circle 
PO, $M=1,2,...$;
$S_{M{\rm C}}\left(\vareps\right)$ is the action along the orbit 
$M$C (\ref{actionC}).  For amplitudes of the $M$C orbit
contributions, one obtains 
\bea\l{amp2ispC}
&&\!{\cal A}_{M{\rm C}}^{(2)}(\vareps)
= \frac{2 i L_{\rm C}\,T_{C}}{ \pi\,\hbar^2\sqrt{F_{M{\rm C}}}}\times\\
&\times& \mbox{erf}\left(\mathcal{Z}_{p\;{M{\rm C}}}^{+}\right)\,
 \mbox{erf}\left(\mathcal{Z}_{r\;{M{\rm C}}}^{-},
\mathcal{Z}_{r\;{M{\rm C} }}^{+}\right),\nonumber
\eea  
where $T_{\rm C}$ is the period of a particle motion along
the primitive (one-repeated, $M=1$) orbit C, 
$T_{\rm C}=2 \pi/\omega^{}_{\rm C}$
(\ref{omtcra}), and
$F_{M{\rm C}}$ is the Gutzwiller stability 
factor (\ref{fgutz}).
The error functions are defined in (\ref{errorf}).
In these derivations, we used transformations to the
new integration variables,
\bea\l{zpvarC}
z_p &=&p_r^{\prime}\,\sqrt{-\frac{i}{2\hbar}\,{\cal J}_{M{\rm C}}^{(p)}}\,,
\\ 
z_r&=&\left(r^{\prime\prime}-r^{}_{\rm C}\right)\;
\sqrt{\frac{i F_{M{\rm C}}}{2\hbar \mathcal{J}_{M{\rm C}}^{(p)}}}\,. \nonumber
\eea
The finite limits in the error functions 
of (\ref{amp2ispC}), ${\mathcal Z}_{p\;M{\rm C}}^{\pm}$ and 
${\mathcal Z}_{r\;M{\rm C}}^{\pm}$,
are given by
\bea\l{ZplimC}
\!{\mathcal Z}_{p\;M{\rm C}}^{-}&=&0,\,\,\, 
{\mathcal Z}_{p\;M{\rm C}}^{+}=\frac{L_{\rm C}}{r^{}_{\rm C}}\, 
\sqrt{-\frac{i}{2\,\hbar}\,
{\cal J}_{M{\rm C}}^{(p)}}\;, \\
{\mathcal Z}_{r\; M{\rm C}}^{-}&=&
-r^{}_{\rm C}\;\sqrt{\frac{F_{M{\rm C}}}{2\,i \hbar\; \mathcal{J}_{M{\rm C}}^{(p)}}}\,
\nonumber\\
{\mathcal Z}_{r\; M{\rm C}}^{+}&=&
\left(r^{}_{\rm max}-r_{\rm C}\right)
\sqrt{\frac{iF_{M{\rm C}}}{2\,\hbar\; \mathcal{J}_{M{\rm C}}^{(p)}}}\,, \nonumber
\eea
where $\mathcal{J}_{M{\rm C}}^{(p)}$ 
 is the Jacobian (\ref{jacpCres}).
We took into account that the upper limit for the radial 
momentum $p_r^{\prime}$ is the momentum modulus
$p\left(r^{}_{\rm C}\right)=L_{\rm C}/r^{}_{\rm C}$ 
(lower limit is zero) and the upper limit
for the radial coordinate $r^{\prime\prime}$ is $r^{}_{\rm max}$ of 
(\ref{actionvar}).  
Substituting \ (\ref{jacpCres}) 
into (\ref{ZplimC}), one finally arrives at the 
following arguments of error functions,
\bea\l{ZplimCfin}
\!{\mathcal Z}_{p\;M{\rm C}}^{-}\!&=&\!0,\,\,\, 
{\mathcal Z}_{p\;M{\rm C}}^{+}\!=\!L_{\rm C} \sqrt{-\frac{i \pi}{\hbar}
(\alpha+2) M K_{\rm C}}\;,\qquad\\
{\mathcal Z}_{r\;M{\rm C}}^{-}&=&
-\;\sqrt{\frac{i F_{M{\rm C}}}{4 \pi 
(\alpha+2)\,\hbar M K_{\rm C}}}\;,\nonumber\\
{\mathcal Z}_{r\;M{\rm C}}^{+}&=&
\left(\frac{r_{\rm max}}{r^{}_{r_{\rm C}}}-1\right) \sqrt{\frac{i F_{M{\rm C}}}{4 \pi 
(\alpha+2)\,\hbar M K_{\rm C}}}\,. \nonumber
\eea

\subsubsection{The SSPM limit} 

Asymptotically far 
from the bifurcations determined by roots
of (\ref{bifeq}) [or (\ref{bifeqra}), also far from the symmetry 
breaking point $\alpha=2$ ] for the RPLP
(\ref{potenra}), we derive from (\ref{amp2ispC})
\begin{equation}
\mathcal{A}_{M{\rm C}}^{(2)}(\vareps)
= 
\frac{2 L_{\rm C}\,T_{C}}{\pi \hbar^2\;\sqrt{F_{M{\rm C}}}}\;.
\label{amp2sspC}
\end{equation} 
The same result can be obtained directly from the initial
trace formula (\ref{pstracerpC}) far from the symmetry breaking and bifurcation
points by applying the SSPM.
Indeed, assuming the isolated stationary point,  $p_r^{\prime\;\ast}=0$
and $r^{\prime\prime\;\ast}=r^{}_{\rm C}$, in a plane
crossing the center of the spherical symmetry, 
 and expanding the action phase
$\Phi$, according to (\ref{phipiexpC}), one may
extend the upper integration 
limit $p_{r}^{+}$ to $\infty$. The lower integration limit 
$p_r^{-}=0$ can be extended to $-\infty$
if we consider the repetition numbers
$M=\pm 1$, $\pm 2$ and so on assuming the motion of particle in both direction.
Then, one may reduce the trace formula to the same result
above mentioned  multiplying  by factor 2 due to the time reversibility symmetry
of the Hamiltonian independent of time as in subsection IVA.
Taking the
pre-exponent amplitudes off the integrals, we calculate explicitly 
the both integrals over $p_r$ and $r$, and 
arrive at (\ref{dengenC}) 
with the amplitude 
$A_{M{\rm C}}^{(2)}$, equal exactly to (\ref{amp2sspC}).
Note also that there is a difference from the
standard Gutzwiller trace formula
for isolated orbits because  (\ref{dengenC}) with
the amplitude (\ref{amp2sspC})
is the SSPM trace formula for a ${\mathcal K}=2$ family of the circle orbits
which are isolated in a fixed plane crossing the center.

The Maslov phase $\mu^{}_{M{\rm C}}$
in (\ref{dengenC}) is given by
\begin{equation}
\mu^{}_{M{\rm C}}=\sigma^{}_{M{\rm C}} + 1\;,\qquad \sigma^{}_{M{\rm C}}=2\,M\;.
\label{masl2}
\end{equation}
For the calculation of this asymptotic 
Maslov index $\sigma^{}_{M{\rm C}}$ through
the turning and caustic points [see the trace formula 
(\ref{dengenC}) with the ISPM 
(\ref{amp2ispC}) and the SSPM  
(\ref{amp2sspC}) amplitudes], one can use the
Maslov\&Fedoriuk catastrophe theory  
(\cite{fedoryuk_pr,maslov,fedoryuk_book1,chester,MVApre2013}).
As usual, there is the two components of the Maslov phase in the ISPM
trace formula (\ref{dengenC}) for the circular orbit families
as in (\ref{deltag3isp}) for $\mathcal{K}=3$ family. 
One of them is the
asymptotic constant part (\ref{masl2})  independent of the 
energy. Another part changes smoothly (continuously, with no jumps) 
through the bifurcation points.
The total Maslov phase $\mu_{M{\rm C}}^{({\rm tot})}$ for the 
 family of $M$C orbits is then 
given by the sum of these two contributions
mentioned above.

\subsubsection{The harmonic oscillator limit}

 In the harmonic oscillator limit to the 
symmetry breaking point ($\alpha \rightarrow 2$) the stability factor 
$F_{M{\rm C}}$ (\ref{fgutz})
and curvature $K_{\rm C}$ , see (\ref{curvraC}), tend both to zero. 
However, there 
is a finite limit of the
amplitude $A_{M{\rm C}}^{(2)}$  
(\ref{amp2ispC}) at $\alpha \rightarrow 2$
because of the exact cancellation of singularities for $K_{\rm C} \rightarrow 0$ 
and $F_{M{\rm C}}/K_{\rm C} \rightarrow 0$ 
in the expansion of the product of 
error functions
($\propto \mathcal{Z}_{p\,M{\rm C}}^{+} \mathcal{Z}_{r\,M{\rm C}}^{-}$) 
with the denominator $\sqrt{F_{M{\rm C}}}$ of 
(\ref{amp2ispC}). 
Taking into account the relation between the angular momentum $L_{\rm C}$ and 
energy $\vareps$ for this HO limit, $\vareps=\omega^{}_{\rm C}L_{\rm C}$, 
one obtains
\be\l{HOlim}
\mathcal{A}_{M{\rm C}}^{(2)}(\vareps)
\rightarrow \frac{\vareps^2}{2(\hbar \omega^{}_{\rm C})^3}\;.
\ee
This is a half of the spherical HO density amplitude \cite{sclbook},
in addition to another one half coming from the diameters, 
as shown in
the next Section IVD  
(the repetition number counts
rotations along the circular orbit in one direction, $M=1,2, ...$). 
The limit (\ref{HOlim}) 
can be found also directly from the trace formula 
(\ref{pstracerpC}) accounting for constant action phase and finite integration
limits $r_{-}=0$, $~r_{+}=r_{\rm max}$, $~p_r^{-}=0$, and
$~p_r^{+}=p(r^{}_{C})=L_{\rm C}/r^{}_{\rm C}~$
($~L_{\rm C}=\vareps/\omega^{}_{C}$, $~T_{\rm C}=2 \pi/\omega^{}_{\rm C}$), 
\bea\l{ampholimdirect}
&&A_{M{\rm C}}^{(2)} \rightarrow \frac{L_{\rm C}\;T_{\rm C}}{4 \pi \hbar^3}\;
\int_{r_{-}}^{r_{+}} \d r\; \int_{p_r^{-}}^{p_r^{+}} \d p_r~= \\ 
&=& 
\frac{L_{\rm C}^2\;T_{\rm C}}{4 \pi \hbar^3}=
\frac{\vareps^2}{2\left(\hbar \omega^{}_{\rm C}\right)^3}\;. \nonumber
\eea
Thus, one has a half of the HO trace formula amplitude
(independent of $M$),
up to relatively small second order terms in $\hbar$,
\begin{equation}
\delta g_{M{\rm C}}^{(2)}(\vareps) \rightarrow \frac{1}{2}\,
\delta g_{\rm HO}^{(4)}(\vareps),
\label{holimC}
\end{equation}
where 
\be\l{denhod}
\delta g_{\rm HO}^{(4)}(\vareps) =\frac{\vareps^2}{(\hbar \omega)^3}
\sum_{{\tt M}=1}^{\infty}\cos\left[\frac{1}{\hbar}
S_{\rm HO}\left(\vareps\right) - \frac{\pi}{2}
\mu^{}_{\rm HO}\right],
\ee
$S_{\rm HO}=2 \pi M \vareps/(\hbar \omega)$ is the action
along POs of the 4 parametric family in the HO potential , 
$\omega=\omega_\theta=\omega_r/2$ is the HO frequency,
$\mu^{}_{\rm HO} =\sigma_{\rm HO}= 2M$ \cite{magosc,sclbook}.

\subsubsection{The bifurcation limit} 

Taking the limit  to the bifurcations 
$F_{M{\rm C}} \rightarrow 0$,
 where
$\alpha \rightarrow \alpha_{\rm bif}=n_r^2/n_\theta^2-2$ for 
the potential (\ref{potenra}), see (\ref{bifeqra}), but far from 
the HO limit $\alpha=2$, 
one finds that the argument of the second  error function
in (\ref{amp2ispC}) coming from the radial-coordinate integration 
tends to zero as $\sqrt{|F_{M{\rm C}}|}$  
(\ref{ZplimCfin}).
Thus, the singular  
stability 
factor $F_{M{\rm C}}$ of 
the denominator (\ref{amp2ispC}) is exactly canceled by 
the same from the numerator, and we arrive at the finite result:
\bea\l{amp2ispCbif}
&&\hspace{-0.5cm}{\mathcal A}_{M{\rm C}}^{(2)}(\vareps)
= \frac{i L_{\rm C}}{4 \hbar^2\; \omega^{}_{\rm C}}\;
\sqrt{(\alpha+2)\,M\,K_{\rm C}}\times \\
 &\times&\mbox{erf}\left({\cal Z}_{p\;{M{\rm C}}}^{(+)}\right)\;. \nonumber
\eea  

\subsubsection{The separatrix} 

At the potential barrier (separatrix) when
$K_{\rm C} \rightarrow \infty$ one has the continuous zero
limit. Thus, one finds the continuous
transition  of the ISPM amplitude
$\mathcal{A}_{M{\rm C}}^{(2)}$ through 
all bifurcation points and separatrix,
including potential barriers. The HO limit will be 
discussed after derivations of the diameter family contributions.

\subsection{TWO-PARAMETRIC 
DIAMETER FAMILIES}

\subsubsection{ISPM trace formulae for the diameters }

 For the diameter-orbit $\mathcal{K}=2$ family contribution
into the trace formula (\ref{pstrace}) for a spherical potential, we
take first the integral over $p_{\parallel}^\prime$ of the momentum integration 
by using the energy conservation
$\delta$-function, 
\bea\l{pstraceppar1}
&\delta g_{\rm scl}(\vareps)=
\frac{m}{(2\pi\hbar)^3} \Re\sum^{}_{\rm CT}
\int \d {\bf r}^{\prime\prime}
\int \frac{\d {\bf p}_\perp^\prime}{p_\parallel^{\prime}}
\times \\
&\quad\times\left|\mathcal{J}_{\rm CT}({\bf p}^\prime_\perp,
{\bf p}^{\prime\prime}_\perp)\right|^{1/2}
\exp\left[\frac{i}{\hbar}
\Phi_{\rm CT} - \frac{i\pi}{2}
\mu^{}_{\rm CT}\right]\;. \nonumber
\eea
We used a local
coordinate system with $x$ directed to this diametric orbit, 
$x^{}_\parallel=x$, and
${\bf r}_\perp=\{y,z\}$ perpendicular to it, and similarly, for 
Cartesian momenta, $p^{}_\parallel=p_x$ and
${\bf p}^{}_\perp=\{p_y,p_z\}$. 
The phase integral $\Phi_{\rm CT}$ (\ref{legendtrans})
in the integrand over the perpendicular momentum 
${{\bf p}_\perp}$,
has obviously the isolated stationary point,
${\bf p}_\perp^{\prime}={\bf p}_\perp^{\prime\;\ast}\equiv 
{\bf p}_\perp^{\ast}=0$, in the subspace
of the perpendicular variables in the local Cartesian coordinate
system introduced above. This point is the solution of the stationary 
phase condition:
\be\l{statcondDp}
\left(\frac{\partial \Phi_{\rm CT}}{\partial {\bf p}_\perp^{\prime}}\right)^\ast
\equiv \left({\bf r}_\perp^{\prime} - {\bf r}_\perp^{\prime\prime}\right)^\ast=0\;.
\ee
The SPM condition determining 
the isolated stationary point is the closing condition:
\be\l{rclosing}
{\bf r}_\perp^{\prime\;\ast}={\bf r}_\perp^{\prime\prime\; \ast} \equiv {\bf r}_\perp,
\ee
according to the definition of the phase $\Phi_{\rm CT}$
(\ref{legendtrans}) with
(\ref{actionp}). As the stationary phase conditions for the
next integration over the spacial coordinates are identities
due to the spherical symmetry, the star means that CT is the periodic
orbit (PO) which is now the diameter orbit, $M$D$=M$(2,1) 
(with the $M$ number of periods).
Expanding the phase integral $\Phi_{\rm CT}$ (\ref{legendtrans})
near the isolated stationary point at second order, one has
\bea\l{phiexppD}
&&\Phi_{\rm CT} = S_{M{\rm D}}(\vareps) +\frac12\,J_{y\;{\rm D}}^{(p)}\,
\left(p_y^{\prime}\right)^2 + \\ 
&+&\frac12 J_{z\;{\rm D}}^{(p)}\,
\left(p_z^\prime\right)^2=
S_{M{\rm D}}(\vareps) +\frac12\,J_{\perp\;{M{\rm D}}}^{(p)}\,
\left(p_\perp^{\prime}\right)^2\;, \nonumber
\eea
where $S_{M{\rm D}}(\vareps)$
is the diameter action accounting for the
period number $M$, $S_{M{\rm D}}=M S_{\rm D}$,
$S_{\rm D}$ is 
the action along the primitive ($M=1$) diameter (2,1) 
($\Phi_{\rm CT}^\ast=S_{M{\rm D}}$). 
As shown in Appendix A5 
by using the symmetry of Jacobians 
[taken at the diameter orbit (D), $\mathcal{J}_{y\;{\rm D}}^{(p)}=
\mathcal{J}_{z\;{\rm D}}^{(p)}=
\mathcal{J}_{\perp\;{\rm D}}^{(p)}$] with respect to rotations around
the diameter, one obtains
\be\l{jacpD}
\mathcal{J}_{\perp\;M{\rm D}}^{(p)} 
=\left(\frac{\partial^2 \Phi_{\rm CT}}{\partial
  p_y^{\prime\,2}}\right)^\ast
=-2\pi M K_{\rm D}r^2,
\ee
where the star means 
${\bf p}_\perp^{\prime}=
{\bf p}_\perp^{\ast}=0\;,$
at the $M$D PO, and $K_{\rm D}$ 
is the diameter curvature \cite{MVApre2013},
 \be\l{curvraD}
K_{\rm D}=
\frac{\Gamma\left(1-1/\alpha\right)}{\Gamma\left(1/2-1/\alpha\right)\epsi 
\;\sqrt{2 \pi m R^2 V_0}}\;,
\ee
where $\Gamma(x)$ is the Gamma function.
The last expression was obtained 
by using the explicit expression of the
solution for the trajectory $\theta(r)$ 
in an azimuthal plane crossing the diameter PO  
\cite{nishioka}. 
In this plane, one considers a small variation 
of the initial momentum $\delta p_y'$ perpendicular to the diameter orbit (D)
of a particle moving along a trajectory perturbed 
near the diameter at its given point $r'=r$ through variations
 of the  perpendicular coordinate $\delta y'$  and the 
corresponding change of angular momentum $\delta L$, see 
(\ref{jacrelation1}). We used also  
the standard Jacobian relations (\ref{jacrelation2}) to reduce the calculation
of the Jacobian $\mathcal{J}_{\perp\;{\rm D}}^{(p)}$ to 
another simpler Jacobian
$(\delta \theta^{\prime\prime}/\delta L)^{}_{\rm D}$.
As shown in Appendix A5, due to the azimuthal symmetry 
for rotations around the diameter, this intermediate Jacobian 
$(\delta \theta^{\prime\prime}/\delta L)^{}_{D}$ taken at the D
is invariant \cite{annphysik97} as expressed through
the invariant diameter curvature (\ref{curv}) [see (\ref{invjac})].

We substitute now the expansion (\ref{phiexppD}) for the phase
integral $\Phi_{\rm CT}$ and (\ref{jacpD}) for the Jacobian $J_\perp^{(p)}$ into 
(\ref{pstraceppar1}). We use also the cylindrical  
coordinates 
for the integration over the perpendicular momentum $\p_\perp^{\prime}$,
$\d \p_\perp = p_\perp \d  p_\perp \d \varphi_{p}$,  and spherical
coordinate system for the integration over $\r^{\prime\prime}$,
$\d \r = r^2 \d r\; \sin\theta \d \theta\;  \d \varphi$. Taking the
integrals over the azimuthal angles $\varphi_{p}$ and  spherical 
angles $\theta, \varphi$ ($2 \pi$ and $4\pi$, respectively,
 because of independence of
the integrand on these angles from the spherical symmetry), one obtains
\bea\l{dengenD}
&\delta g_{{\rm scl}\; {\rm D}}^{(2)}(\vareps)=
\Re\sum_{M} A_{M{\rm D}}^{(2)}\times\\
&\times\exp\left[\frac{i}{\hbar}
\,S_{M{\rm D}}\left(\vareps\right) - \frac{i\pi}{2}
\mu^{}_{M{\rm D}}\right]\;.\nonumber
\eea
where
\bea\l{ampdgen}
\hspace{-0.7cm}A_{M{\rm D}}^{(2)}&=&\frac{m}{\pi \hbar^3}\int_{0}^{r_{\rm max}} 
\!r^2 \d r 
\!\int_0^{p(r)} \frac{p^{}_\perp \!\d p^{}_\perp}{\sqrt{p^2-p_\perp^2}}\times \\ 
&\times&\exp\left(-\frac{i\pi}{\hbar}
M K_{\rm D}r^2 p_\perp^2\right)\;. \nonumber
\eea 
(We omitted primes and double primes
for the integration variables, for simplicity,
$p^{}_\parallel=\sqrt{p^2-p_\perp^2}$
$p(r)=\sqrt{2 m [\vareps - V(r)]}$ for 
arbitrary spherical potential.
Using the dimensionless variables, $u=r/r^{}_{\rm max}$, and 
$t=p_\perp/p$, one can re-write (\ref{ampdgen}) in a more convenient
form:
\be\l{ampd1}
{\cal A}_{M{\rm D}}^{(2)}=\frac{m}{\pi \hbar^3}\; r_{\rm max}^3\sqrt{2m\vareps}\; 
\mathcal{I}(M\zeta),
\ee
where
\be\l{Iintdoubdef}
\mathcal{I}(M\zeta)=\int_0^1 u^2 \d u\;\sqrt{1-u^\alpha}\; \mathcal{F}(Mq),
\ee
\be\l{Fintdef}
\mathcal{F}(q)=\int_0^1 \frac{t \d t}{\sqrt{1-t^2}}\; 
\exp\left( i q t^2\right)\;,
\ee
with
\be\l{parq}
q=\zeta u^2 (1-u^\alpha)\;,
\ee
\bea\l{parz}
\zeta&=& -2 \pi\; K_{\rm D} r_{\rm max}^2 m\vareps/\hbar\,, 
\\
r_{\rm max}&=&R \left(\vareps/V_0\right)^{1/\alpha}\;. \nonumber
\eea
The internal integral $\mathcal{F}(Mq)$ (\ref{Fintdef}) 
can be calculated exactly 
analytically in terms of the error function of a complex 
argument (\ref{errorf}) by using the suitable integration variables 
\be\l{intvar}
w=\sqrt{1-t^2},\qquad z_w=\sqrt{iM q}\;w,
\ee
to arrive at
\be\l{Finterf}
\mathcal{F}(q)=\frac{\sqrt{\pi}}{2}\; \frac{e^{i q}}{\sqrt{iq}}\;
{\rm erf}(\sqrt{i q}).
\ee
Collecting (\ref{Finterf}), (\ref{Iintdoubdef}), (\ref{parz})
and (\ref{parq})
for the RPLP (\ref{potenra}), one obtains
a simple integral representation of the diameter amplitude through the 
error function,
\bea\l{Iintsinglerf}
\mathcal{I}(M\zeta)&=&\frac{\sqrt{\pi}}{2 \sqrt{iM\zeta}}\;
\int_0^1 u \d u\times\\
&\times& \exp\left[i M\zeta u^2 (1-u^\alpha)\right]\times \nonumber\\
&\times&{\rm erf}\left[\sqrt{i M \zeta u^2 (1-u^\alpha)}\right]. \nonumber
\eea

\subsubsection{The HO limit}

Taking the limit $\zeta \rightarrow 0$ to the spherical 
harmonic oscillator  
value $\alpha \rightarrow 2$
($K_{\rm D} \rightarrow 0$), one obtains a half of the HO amplitude
(accounting negative and positive repetition numbers $M$ and time-reversibility
by factor 2 with the summation in positive integers 
$M=1, 2, ... $), one finds for 
(\ref{ampdgen})
\be\l{ampholimd}
A_{M{\rm D}} \rightarrow \frac12 \frac{\vareps^2}{(\hbar \omega)^3}\;.
\ee
This limit is a half of the HO
 shell-correction density amplitude (independent of $M$)
 (\ref{denhod}).

\subsubsection{Simplified diameter trace formulae}

As usual, in these derivations within the ISPM,
the constant $\zeta$, which is proportional to the  curvature 
$K_{\rm D}$ [see (\ref{curvraD})] going to zero, is canceled with that
coming from the expansion of the integrand near the catastrophe 
points. We took also into
account that $V_0=mR^2\omega^2/2$ for the spherical harmonic oscillator
($\alpha \rightarrow 2$).
The radial period can be explicitly calculated,
\bea\l{periodfreq} 
T_r&=&\frac{2 \pi}{\omega_r}=2 m \int_0^{r_{\rm max}}\frac{\d r}{p(r)}
= \\
&=& \frac{2 m r_{\rm \max}}{\sqrt{2 m \vareps}}\;
\frac{\sqrt{\pi} \Gamma\left(1+1/\alpha\right)}{ 
\Gamma\left(1/2 + 1/\alpha\right)}\;\rightarrow \nonumber\\
& \rightarrow & \frac{2 \pi}{\omega} \quad\mbox{for} \quad \alpha 
\rightarrow 2\;. \nonumber
\eea
[$r^{}_{\rm max} \rightarrow (2\vareps/m 
\omega^2)^{1/2}$ for $\alpha \rightarrow 2$ 
but $\omega_r=2 \omega$].

 Within the accuracy
of the simplest version of the ISPM
(second order expansion of the phase
and zero order of the amplitude in the integrand), taking
 the pre-exponent amplitude factor 
$\sqrt{p^2-p_\perp^2} $ in the internal integral in (\ref{pstraceppar})
over the modulus of the perpendicular momentum $p_\perp$
off the integral at the stationary point 
$p^{}_\perp=p_\perp^{\ast}=0$, 
$p_\parallel^\ast=\sqrt{p^2-\left(p^{\ast}_\perp\right)^2} = 
p(r)$,  
one finds that this integral can be calculated analytically with the 
finite integration limits in terms of the elementary functions.
Substituting also the Jacobian
expression (\ref{jacpD}) and using the spherical coordinates 
for the spacial-coordinate integration variable 
${\bf r}$, $\left(r, \theta, \varphi\right)$ as above, 
we calculate the integral over 
angles $\theta,\varphi$ (just $4\pi$ from the 
spherical symmetry). 
Then, the diameter contribution (\ref{pstraceppar}) into the 
trace formula (\ref{pstrace}) for any spherical potentials $V(r)$ 
can be resulted in
 (\ref{dengenD}) but with a much simpler amplitude,
\bea\l{ampd0}
\hspace{-0.5cm}A_{M{\rm D}}^{(2)}&=&\frac{m}{\pi \hbar^3}\int_{0}^{r_{\rm max}} 
\frac{r^2 \d r}{p(r)} 
\int_0^{p(r)} p^{}_\perp \d p^{}_\perp\times \\ 
&\times& \exp\left(-\frac{i\pi}{2 \hbar}
M K_{\rm D}r^2 p_\perp^2\right)\;, \nonumber
\eea
where $p(r)$ is the momentum modulus depending on the
radial coordinate $r$ through the radial potential, e.g., the
RPLP (\ref{potenra}).
Taking the internal integral explicitly analytically, one finds from
 (\ref{ampd0})  for
the diameter amplitude ${\cal A}_{M{\rm D}}^{(2)}$ 
in the case of the RPLP 
(\ref{potenra}) the following 
simple one-dimensional integral representation:
\begin{equation}
A_{M{\rm D}}^{(2)}=
\frac{m r_{\rm max}^3 \sqrt{2m E}}{2 \pi i \hbar^3 M \zeta}\, 
\left[I(M\zeta) -I(0)\right]\;,
\label{ampgenD}
\end{equation}
where
\be\l{intradu}
I(\zeta)=\frac{1}{r_{\rm max}}
\int_{r_{\rm min}}^{r_{\rm max}} \frac{\d r}{\sqrt{1-V(r)/E}}\; 
\exp\left[i \zeta \Phi(r)\right]\;,
\ee
with the same $\zeta$ (\ref{parz}),
\be\l{phaserad}
\Phi(r)=r^2 \left(1-V(r)/\vareps\right)/r^2_{\rm max}\;,
\ee
$r_{\rm min}$ and $r_{\rm max}$ are the turning points in the RPLP 
$V(r)$,  
$r_{\rm min}=0$ for the diameter PO. 

Substituting now the RPLP
$V(r)$ (\ref{potenra}) and using a new radial variable
$u=r/r_{\rm max}$ as above (\ref{parz}), 
we present the 
integral $I(M\zeta)$ in the dimensionless
form (\ref{integrald}) with the                         
amplitude $\mathcal{A}(u)$ (\ref{ampu})
and phase $\Phi(u)$ (\ref{phiu}). For the RPLP
(\ref{potenra}), one explicitly finds 
(\ref{rmaxa}), (\ref{zetasc}), (\ref{curvraD}) and (\ref{freq})
(Appendix A6).

For large $\zeta$ (\ref{parz}), one can use the
SPM evaluation of the integral (\ref{intradu}) in terms of the error functions 
(\ref{erfampu}). Accounting for the two stationary points $u_1$ and $u_2$ 
(\ref{statpointsun}) ($u_2=0$)
in the phase function $\Phi(u)$ (\ref{phasu}), one obtains
\bea\l{erfampu}
\!I(M\zeta)&\approx&
\sqrt{\frac{\pi (\alpha+2)}{4 i M \zeta \alpha^2}} \times \\
&\times& {\rm erf}\left(\mathcal{Z}_{1\;M{\rm D}}^{+},\mathcal{Z}_{1\;M{\rm D}}^{-}\right)\;e^{i M\zeta \Phi_1 } +\nonumber\\ 
&+&\sqrt{\frac{i\pi}{4 M\zeta}}\;
{\rm erf}\left(\mathcal{Z}_{2\;M{\rm D}}^{+}\right)\;. \nonumber
\eea
In the contribution of the second stationary point $u_2$ we have to keep
the amplitude $u$ as it is.
The arguments of the error functions, 
$\mathcal{Z}_{1\;M{\rm D}}^{\pm}$
and $\mathcal{Z}_{2\;M{\rm D}}^{+}$  in 
(\ref{erfampu}), are given by
\bea\l{argerf}
\mathcal{Z}_{1\; M{\rm D}}^{+} 
&=& \sqrt{-i M\zeta \Phi_1''/2}\; (1-u_1)\;, \\
\mathcal{Z}_{1\;M{\rm D}}^{-} &=& -\sqrt{-i M\zeta \Phi_1''/2 }\; u_1\;,
\nonumber\\
\mathcal{Z}_{2\; M{\rm D}}^{+} 
&=& \sqrt{-i M\zeta \Phi_2''/2}\;\;, \nonumber
\eea
where 
\be\l{Phi1}
\Phi_1=\Phi\left(u_1^{}\right)=\frac{\alpha}{\alpha+2}\;
\left(\frac{2}{\alpha+2}\right)^{2/\alpha}\;,
\ee
and $\Phi_1''=\Phi''(u_1^{})$ and 
$\Phi_2''=\Phi''(u_2^{})$,
are curvatures (\ref{curvu});
see also (\ref{statpointsun}) for $u_1$.
For the limit $|\zeta| \rightarrow \infty$, 
from (\ref{erfampu}) 
one has the same as directly from
(\ref{ampgenD}),
\be\l{izetassp}
I(M\zeta) \rightarrow \sqrt{\frac{\pi (\alpha+2)}{i M\zeta \alpha^2}}\;
e^{i M\zeta \Phi_1} + \sqrt{\frac{i \pi}{4 M\zeta}}\;.
\ee

\subsubsection{The SSPM and billiard limit} 

Taking the limit $|\zeta|>>1$, 
from (\ref{ampgenD})
one finally obtains the SSPM 
trace formula with the same amplitude 
$A_{M{\rm D}}^{(2)}$  as
we would get directly from integrations (\ref{pstraceppar}) over $p_y'$ 
and $p_z'$ at the simplest 
 second order in the phase integral
and zero order in the amplitude expansion, 
and extension of the integration limits   
up to $\pm \infty$ (taking also into account the factor 2 
in the trace formula reducing it to the summation over $M=1,2,...$ ).
Finally, from (\ref{ampgenD}) [or (\ref{pstraceppar})] one finds 
the SSPM limit:
\be\l{amp2sspD}
{\cal A}_{M{\rm D}}^{(2)} \rightarrow \frac{1}{i \pi M K_{\rm D}\omega_r \hbar^2}\;.
\ee
With this amplitude for 
$\alpha \rightarrow \infty$, one then obtains from (\ref{dengenD})  
the Balian-Bloch trace formula \cite{bablo}
for the diameter orbits in the spherical billiard,
 \be\l{bb2d}
\delta g_{\rm D,BB}^{(2)}(\vareps) = -\frac{2m R^2}{\hbar^2}
\sum_{M=1}^{\infty} \frac{1}{2\pi M}
 \sin\left(\frac{4}{\hbar}\,M p R\right)\;.
\ee
The Maslov phase $\mu^{}_{M{\rm D}}$
for the diameters in spherical billiard
is given by
\be\l{masldbil}
\mu^{}_{M{\rm D}}= \sigma^{}_{M{\rm D}} - 1\;,\quad \sigma^{}_{M{\rm D}}=4 \,M\;.
\ee
In these derivations of the spherical billiard limit, one should take 
into account the asymptotics
$\alpha \rightarrow \infty$ for the curvature,
\be\l{curvbbD}
K_{\rm D} \rightarrow -\frac{1}{\pi pR}= -\frac{1}{\pi R \sqrt{2 m \vareps }}\;.
\ee
Thus, from the amplitude in the one-integral representation (\ref{Iintsinglerf})
we obtained the two limits $\alpha \rightarrow 2$ (a half of the 
HO trace formula (\ref{denhod}) \cite{sclbook,magosc}) and 
$\alpha \rightarrow \infty$ (the spherical billiard \cite{bablo}), 
and also the standard SSPM approach.

The constant $\zeta$ which is proportional to the  singular curvature 
$K_{\rm D}$ (\ref{curvraD}) is canceled near the catastrophe 
points, as usual within the ISPM. In these derivations, one should take into
account that $V_0=mR^2\omega^2/2$ for the spherical harmonic oscillator
($\alpha \rightarrow 2$), and (\ref{periodfreq}) for the diameter period.

\subsection{TOTAL RPLP TRACE FORMULAE}

\subsubsection{Averaged density}
\label{tottraceesc}

 The total ISPM trace formula for the RPLP is the sum of the contribution of
the $\mathcal{K}=3$ polygon-like (P) families $\delta g_{\rm P}^{(3)}(\vareps)$
(\ref{deltag3isp}) with (\ref{amp3isp}),
the $\mathcal{K}=2$ circular (C) orbits 
$\delta g_{\rm C}^{(2)}(\vareps)$ (\ref{dengenC}) [with (\ref{amp2ispC})], 
 and the $\mathcal{K}=2$ diameters (D) 
$\delta g_{\rm D}^{(2)}(\vareps)$
[(\ref{dengenD}) with (\ref{ampgenD}) 
for larger $\alpha$ and 
(\ref{ampd1}) for smaller $\alpha \rightarrow 2$],
\be\l{deltadenstotprlp}
\delta g_{\rm scl}(\vareps)
=\delta g_{\rm P}^{(3)}(\vareps) + \delta g_{{\rm C}}^{(2)}(\vareps) +
\delta g_{{\rm D}}^{(2)}(\vareps)\;.
\ee
This trace formula has the correct finite asymptotic limits to
the SSPM, the Berry\&Tabor result (\ref{deltag3isp}) 
and (\ref{amp3ssp}) 
for the $\mathcal{K}=3$ polygon-like; and the 
$\mathcal{K}=2$ for the diameter (\ref{dengenD}) and [ 
(\ref{amp2sspD})]; and the
circle [\ref{dengenC}) and (\ref{amp2sspC})]
POs. 

According to the general trace
formulas  (\ref{deltadenstot}) and (\ref{avdeltadentot}) 
\cite{strumag,sclbook,migdalrev,MVApre2013}, for the
averaged density $\delta g^{}_\Gamma(\vareps)$ (with the Gaussian weight 
function specified by the averaging
parameter $\Gamma$ much smaller than
the Fermi energy $\vareps^{}_F$) in terms of the 
scaled-energy level density,
\be\l{scdentot}
\mathcal{G}_\gamma(\epsi)=g^{}_\Gamma(\vareps)\;\frac{\d\vareps}{\d\epsi},\quad
\gamma=\Gamma \frac{\d\epsi}{\d \vareps}\;,
\ee
as a function of the scaled energy $\epsi$ (\ref{eq:scaledentau}), 
one obtains 
\bea\l{scldenstyra1}
&&\hspace{-0.5cm}\mathcal{G}_{\gamma\;{\rm scl}}(\epsi)
=\mathcal{G}_{\rm ETF}(\epsi)+
\delta \mathcal{G}_{\gamma\;{\rm P}}^{(3)}(\epsi) +\\ 
&+&\delta \mathcal{G}_{\gamma\;{\rm C}}^{(2)}(\epsi) +
\delta \mathcal{G}_{\gamma\;{\rm D}}^{(2)}(\epsi)\;.\nonumber
\eea
Here, $\gamma$ is the scaled averaging parameter,
 $\mathcal{G}^{}_{\rm ETF}(\epsi)$ is the scaled 
smooth part obtained within the ETF approximation \cite{sclbook}.
Its TF component $\mathcal{G}_{\rm ETF}(\epsi)$ corresponds to
\be\l{TFden}
g^{}_{\rm TF}(\vareps) 
=
 \frac{2\,m}{\pi \hbar^3}\,\int_{0}^{r^{}_{\rm max}} \,r^2\,\d r\,p(r)\;,
\ee
with $H(p,r)=p^2/2m + V(r)$ and $p(r)$ given by 
(\ref{pr}). For the scaled TF approach, one obtains
explicitly \cite{arita2012},
\be\l{scldenstyraTF}
\mathcal{G}^{}_{\rm TF}(\epsi)
=c_0\epsi^2, \quad c_0=
\frac{2\sqrt{2}}{\pi}B\left(1+\frac{3}{\alpha},\frac32\right)\;,
\ee
with the Euler beta function $B(s,t)$. 
The next-order term \cite{sclbook} can be also given
analytically \cite{arita2012}, and one has the ETF density
\bea\l{eq:getf}
&\mathcal{G}_{\rm ETF}(\epsi)
=c_0\epsi^2+c_1, \\ 
&c_1=-\frac{\alpha+1}{12\sqrt{2}\pi}
B\left(1+\frac{1}{\alpha},\frac12\right)\;.\nonumber
\eea

According to (\ref{avdeltadentot}), 
the oscillating terms of (\ref{scldenstyra1}) take the form:
\be\l{dscldenstyra}
\delta \mathcal{G}^{\mathcal{K}}_{\gamma,{\rm PO}}(\epsi)=
\delta \mathcal{G}^{\mathcal{K}}_{\rm PO}(\epsi)\;
\exp\left(-\tau^2_{\rm PO}\gamma^2/4\right)\;,
\ee
where $\tau^{}_{\rm PO}=t^{}_{\rm PO}\;\d \vareps/\d \epsi$ is the
scaled period (\ref{eq:scaledentau}),
\be\l{dscldenstyradef}
\mathcal{G}^{(\mathcal{K})}_{\rm PO}=
g^{(\mathcal{K})}_{\rm PO}(\vareps)\;\frac{\d\vareps}{\d \epsi}\;,
\ee
see (\ref{deltag3isp}), (\ref{dengenC}) and (\ref{dengenD}).

\subsubsection{Shell-correction energy}

The periodic orbit expansion 
for the semiclassical shell-correction energy
$\delta U$ is shown by (\ref{escscl}) with $\delta g^{}_{\rm PO}(\vareps)$
given in (\ref{deltadenstotprlp}) at $\vareps=\vareps^{}_{F}$
\cite{strumag,sclbook,spheroidptp,ellipseptp,MVApre2013}
where $t^{}_{\rm PO}=M t^{M=1}_{\rm PO}(\vareps^{}_F)$ 
is the time of a particle motion
along the PO in the RPLP (taking into
account its repetition number $M$) at the Fermi energy $\vareps=\vareps^{}_F$.
The Fermi energy $\vareps^{}_F$ is related to the conservation of the particle
number $N$ through the equation (\ref{partnum}),
\be\l{partnumsc}
N=2\int_0^{\epsi^{}_F} \d \epsi\;
\mathcal{G}(\epsi)\;. 
\ee
According to \cite{strumag,sclbook,strut,MVApre2013}, for the corresponding
dimensionless scaled shell-correction energy,
\be\l{scdedef} 
\delta \mathcal{U}_{\rm scl}=(\d \epsi/\d \vareps)\;
\delta U_{\rm scl}\;,
\ee
with (\ref{escscl}) for 
$\delta U_{\rm scl}$, one obtains
\bea\l{scdescra}
&&\delta \mathcal{U}_{\rm scl}
=2\sum_i \epsi^{}_{i}\delta n_i\approx \\
&&\approx
2\sum^{}_{\rm PO} 
\frac{\hbar^2}{\tau^2_{\rm PO}}\; 
\delta \mathcal{G}_{\rm PO}^{(\mathcal{K})}(\epsi^{}_{F})\;. \nonumber
\eea
Here, $\delta n_i$ is the variation of occupation numbers 
defined by the Strutinsky smoothing procedure \cite{strut,fuhi}, and
the sum over PO runs all P, C and D families (with repetitions)
[(\ref{deltag3isp}), (\ref{dengenC}) and (\ref{dengenD}) for 
$\delta g_{\rm PO}^{\mathcal{K}}$ components, relatively].

As mentioned in Section IIA3, the scaled shell-correction energy 
$\delta \mathcal{U}$ (\ref{scdescra}), which is
the observed (dimensionless) physical quantity, does not
contain arbitrary averaging parameter $\gamma$, in contrast to the
oscillating level density $\delta \mathcal{G}^{}_\gamma(\epsi)$. The 
convergence of the PO
sum (\ref{scdescra}) is ensured by the additional factor in front
of the density component $\delta \mathcal{G}^{\mathcal{K}}_{\rm PO}$ 
which is inversely
proportional to the scaled period
$\tau^{}_{\rm PO}$ squared along the PO. 
Therefore, we need (scaled) short-time POs in the RPLP if they
 occupy sufficiently large
phase-space volume in terms of the scaled quantities.

\subsubsection{Fourier transform}

The Fourier transform of the
semiclassical  
level density with respect to the scaled-period variable $\tau$ 
is given by 
\bea\l{fourierpower}
   &&F(\tau)\!=\int \d \epsi\; g(\epsi)e^{i\epsi\tau}
e^{-\gamma^2\epsi^2/2}\approx \\
&&\approx F_0(\tau)+\pi\sum^{}_{\rm PO}F_{\rm PO}\;
\delta_\gamma(\tau-\tau^{}_{\rm PO})\;, \nonumber\\
&&\qquad\delta_\gamma(x)\equiv\frac{1}{\sqrt{2\pi}\gamma}\;
e^{-x^2/2\gamma^2}\;,\nonumber
\eea
which exhibits peaks at periodic orbits $\tau=\tau^{}_{PO}$. 
In (\ref{fourierpower}), $F_0(\tau)$ 
represents the Fourier transform of the smooth ETF level density and 
has a peak at $\tau=0$ related, in the case of the simple TF
approach,  to the direct (zero-action) trajectory
\cite{migdalrev,MVApre2013}. 
Thus, from the Fourier transform of the
quantum-mechanical level density,
\bea\l{fourierpower1}
&&\hspace{-0.5cm}F(\tau)\!=\!\int \left[\sum_i\delta(\epsi-\epsi_{i})\right]
\d\epsi \times\\ &\times& e^{i\epsi \tau}e^{-\gamma^2\epsi^2/2} 
=\sum_i e^{i\epsi_{i}\tau}e^{-\gamma^2\epsi_i^2/2},\nonumber\\ 
&&\psi_{i}=\left(\vareps_i/V_0\right)^{\frac12+\frac{1}{\alpha}}\;,
\nonumber
\eea
one can directly extract information about the amplitudes
$A_{\rm PO}$ of the classical PO contributions into the level density
(\ref{deltadenstot}).

\subsubsection{The harmonic oscillator limit}

In the harmonic oscillator limit [$\alpha \rightarrow 2$  
in the RPLP (\ref{potenra})], the energy surface
is simplified to the linear function in actions,
\begin{equation}
E=\omega_{r}\,I_{r} + \omega_{\theta}\,I_{\theta} =
\omega_{\theta}\,\left(2\,I_{r} + L\right)\;.
\label{Eho}
\end{equation}
In this limit
the curvature $K_{\rm PO}$, see (\ref{curv}), at $L=L_{\rm PO}$
for all POs including the maximal value 
$L=L_{\rm C}=E/\omega_\theta$ for circular orbits  
as well as the Gutzwiller
stability factor $F_{M{\rm C}}$ turn into zero, see also the 
specific expressions for the diameter curvature $K_{\rm D}$ (\ref{curvraD})
and the circle one $K_{\rm C}$ (\ref{curvraC}). However, there is no
singularities in amplitudes of the ISPM trace formula (\ref{dengenC}) 
[with (\ref{amp2ispC})] for 
$F_{M{\rm C}} \rightarrow 0$ and $K_{\rm C} \rightarrow 0$ because
the arguments of all error functions go to zero as explained
above, see also a similar cancelation of the singularities for 
$K_{\rm D}\rightarrow 0$ ($\zeta \rightarrow 0$) in 
(\ref{ampd1}) with (\ref{Iintsinglerf}).

The contribution of the three-parametric polygonal-like orbits 
$\delta g^{(3)}_{\rm P}(E)$ in (\ref{deltadenstotprlp}) 
disappears in the HO limit 
because the action (time
of the particle motion) goes to the infinity for $n_r/n_{\theta} \rightarrow 2$ 
when $n_{\theta} \rightarrow \infty$ and $n_r > 2 n_{\theta} \rightarrow \infty$.
Therefore, for any finite averaging parameter $\Gamma$, 
one can neglect their contributions to the averaged level density
(\ref{avdeltadentot}) [or in (\ref{scldenstyra1}) with (\ref{dscldenstyra})
and (\ref{dscldenstyradef})]. 
These P orbits do not contribute also into
the shell-correction energy (\ref{escscl}) in this limit 
$\alpha \rightarrow 2$ due to the factor $\propto 1/t^{2}_{\rm PO} 
\rightarrow 0$ for time-long POs.

Thus, in the HO limit ($\alpha \rightarrow 2$), 
 one can assume that
the two kind of other families with
$\mathcal{K}=2$
of the diameter and circular orbits 
form together the $\mathcal{K}=4$ family of the HO Hamiltonian. 
As shown above, in this limit 
they give the same contributions and their sum  
is the HO trace formula $\delta g_{\rm HO}^{(4)}(E)$ 
(\ref{denhod}) up to the higher order terms in $\hbar$
which were neglected in our derivations.

As the result, the HO limit of the sum of the circular and 
diameter orbit contributions
is the HO trace formula with the precision of the higher order terms 
in $\hbar$.

\subsection{COMPARISON WITH QUANTUM RESULTS}
\label{comptottrace}

Figure\ \ref{fig12} shows the Fourier transform of the 
quantum-mechanical level density $g(\epsi)$ for the RPLP 
[see (\ref{fourierpower})]. For a smaller 
$\alpha=2.1$, the diameter (2,1) (D),
and circle (1,1) (C) ($M=1$)
orbits yield the dominant contribution to the gross-shell structure
as the shortest POs (see the peak at $\tau\sim 5.0$). These
primitive diameter D and circle C
peaks which appear at almost the same $\tau\approx 5.0$
cannot be distinguished, and they are seen as a common peak of their
sum. Other $M$D and $M$C orbits 
with $M>1$ give smaller contributions
at larger $\tau$.
With increasing $\alpha$ ($\alpha=\alpha_{\rm bif}=4.25$ and $7$), the 
amplitudes of the oscillating level density for these orbits are decreased, 
and one finds a prominent enhancement 
around the
bifurcation points , $\tau \sim 11.2$ at $\alpha^{}_{\rm bif}=4.25$ 
and $\tau\sim 6.2$ at $\alpha^{}_{\rm bif}=7.0$.
However, near these bifurcations, the contribution of the newborns 
star- (5,2) and triangle-like 
(3,1) families, having a higher degeneracy $\mathcal{K}=3$, 
becomes important also  
for larger $\alpha >\alpha^{}_{\rm bif}$. 
The newborn (5,2) and (3,1) peaks cannot be distinguished from the 
parent circular  
2C and C
orbits near the corresponding
bifurcation points $\alpha^{}_{\rm bif}$,
as in the case of the diameter and circular orbits 
at $\alpha$ close to the HO limit.
Notice, in good agreement with the Fourier transforms 
(Fig.\ \ref{fig12}), the remarkable enhancement 
is found in the oscillating 
level density amplitude $A^{(\mathcal{K})}_{\rm PO}$ of
the PO family having different order of $\hbar$
with respect to the Gutzwiller trace formula for the isolated
POs (see \ref{ampenhance}).  However, the phase-space volume occupied
by a circular-orbit family at its bifurcation becomes negligible,  and
all amplitude enhancement becomes inherent to the bifurcating 
$\mathcal{K}=3$ family 
on right of the bifurcation point.

Figs.\ \ref{fig13}--\ref{fig15}  
show a nice agreement of 
the coarse-grained ($\gamma=0.6$) 
and fine-resolved ($\gamma=0.03-0.2$) semiclassical
and the quantum level densities 
$\delta \mathcal{G}^{}_\gamma(\epsi)$ 
(divided by $\epsi$) as functions 
of the scaled energy $\epsi$ at  
$\alpha=6.0$. 
This value of $\alpha$ is remarkable as $\alpha=2$ and $4$ because
all classical critical characteristics of all POs can be found explicitly
analytically, and therefore, one has a very high precision of the calculations
of trace formulas, as shown in \cite{MVApre2013}.
The ISPM results at this value of $\alpha$
are in good agreement with the SSPM ones because $\alpha$ is
far away from the main bifurcations of the short-time POs 
at $\alpha=4.25$ and $7.0$, as well as
from the HO symmetry-breaking point $\alpha=2$. As seen from 
Figs.\
\ref{fig16} -- \ref{fig18}
 for the birth  of 
 triangle-like
(3,1) ($\alpha=7$) POs in a typical bifurcation scenario, one has 
also good agreement
of the ISPM with quantum results. 
Note that the SSPM of these PO contributions show a sharply pronounced
discontinuity of their amplitudes; and for the 
2C and C orbits,
one finds a divergent behavior, in contrast to the continuous ISPM PO 
components. We demonstrate that the ISPM solves successfully these
catastrophe problems of the SSPM.   For the averaged 
semiclassical trace formula  we used (\ref{scldenstyra1}). 
For quantum calculations we employed the standard Strutinsky averaging
(over the scaled energy $\epsi$), finding a good plateau around the 
Gaussian averaging width ${\widetilde \gamma}=2-3$ and curvature-correction 
degree ${\cal M}=6$. 

The C and D
POs with the shortest (scaled) periods $\tau$ are dominating
at large averaging parameter $\gamma=0.6$ (coarse-graining, or 
gross-shell structure) while much more families with a relatively
long period $\tau$ at $\gamma=0.03-0.2$ (fine-resolved shell structure)
become significant in comparison with the quantum results [see the panels
{\it (b)} in Figs.\ \ref{fig13}--\ref{fig15}]. Notice that
for the exemplary bifurcation 
$7.0$ at smaller 
$\gamma \siml 0.2$ the dominating orbits 
become the bifurcating newborn 
(3,1) of the highest
degeneracy $\mathcal{K}=3$ 
along with leading (5,2), (7,3), and (8,3) POs 
which were born at smaller bifurcations. This is in nice agreement with the 
quantum Fourier transforms shown in Fig.\ \ref{fig12}.
These POs yield more 
contributions near the bifurcation values of $\alpha$ and even more on their 
right in a wide region of $\alpha$. 
Moreover, the bifurcating parent
circular-orbit family 
$M$C does not contribute relatively 
in the bifurcation scenario because the lower, and the upper
radial integration limits for the ISPM $M$C  
amplitudes (\ref{amp2ispC}) 
coincide at their
bifurcation point $\alpha=\alpha_{\rm bif}$ with the radius $r=r_{\rm C}$:
They occupy the zero phase-space volume at the bifurcation value 
$\alpha=\alpha_{\rm bif}$.

Figs.\ 
\ref{fig19} and \ref{fig20}
show the semiclassical
shell-correction energy $\delta U_{\rm scl}$ [(\ref{scdescra}) 
in units of ($\epsi^{}\d \vareps/\d \epsi )_{\rm F}$] as function of the
particle (neutron or proton) number variable $N^{1/3}$ (\ref{partnum}).
They were calculated by using the PO sum (\ref{scdescra}) and the 
standard relationship $N=N(\epsi^{}_{\rm F})$, see (\ref{partnumsc}),
after the 
scale transformation (\ref{scdedef})
\cite{MVApre2013}.
The corresponding quantum shell-correction calculations
are performed by using the Strutinsky smoothing procedure 
(see, e.g., \cite{fuhi}). For the sake of convenience, one can use 
averaging of the level density $g(\epsi)$ with a small parameter $\gamma=0.1$
in (\ref{partnumsc}) for the relationship between the particle 
number and the Fermi (scaled)
energy $\epsi_{\rm F}$, $N=N(\epsi_{\rm F})$, as this
integral characteristics is almost insensitive of variations of 
$\gamma \approx 0.02 \div 0.1$, at least.

More precised results for the semiclassical 
shell-correction energy $\delta U$  
as functions of the 
particle number $N^{1/3}$, especially near its
minima, are obtained with using the quantum level density for the
re-calculation of the Fermi energy $\epsi^{}_{F}$ to 
the particle number $N$ through (\ref{partnumsc}). 
The reason is
rather a slow convergence of the semiclassical expansion of the phase integral
in terms of the POs in (\ref{partnumsc}) as compared with the
PO sum for the shell-correction energy (\ref{scdescra})
at a given Fermi energy
$\epsi^{}_{F}$.
A good plateau for the quantum calculations of the scaled 
shell-correction energy
is realized 
near the same averaging parameters ${\widetilde \gamma}$ and ${\cal M}$
 mentioned above.  
We have to point out that the quality of the plateau in the SCM calculations
is much better with using the scaled-energy variable $\epsi$ rather 
than the energy $\vareps$
itself, except for the HO limit $\alpha=2$, where these energies coincide
\cite{MVApre2013}.
For instance, as well-known, for the spherical billiard case 
$\alpha \rightarrow \infty$,
the plateau condition can be obtained in terms of 
the scaled energy $\epsi$ which is the wave number $\sqrt{2 m \vareps}/\hbar$,
instead of the energy $\vareps$.
For the relation $N=N(\epsi^{}_{F})$ we used specifically 
an averaging in small $\gamma=0.1$
because there is almost no sensitivity of this integral 
characteristics within the
interval of smaller $\gamma$ ($\gamma=0.02-0.1$). 
The PO sum (\ref{scldenstyra1}) for the level density converges with the 
averaging width $\gamma=0.2$ of a fine resolution 
of the shell structure as well as the shell-correction energy PO sum
(\ref{scdescra}) by taking into account almost the same major simplest
POs of a smaller action (scaled period $\tau^{}_{\rm PO}$). 
For the shell-correction energy 
at the value 
$6.0$ far from the short-time
bifurcating POs, one finds also a good agreement with the both quantum and SSPM
results, as for density calculations, cf.\ Figs.\ \ref{fig14} and \ref{fig17}, 
with Fig.\ \ref{fig19}.
As seen from Fig.\ \ref{fig20}, 
we obtain a nice 
agreement between the semiclassical (ISPM, dashed) and quantum 
(QM, solid) results at a bifurcation, too. Again,
the dominating contribution in the semiclassical results 
for $\delta U$ [(\ref{scdedef}) and (\ref{scdescra})] 
at the bifurcation point 7.0 
(Fig.\ \ref{fig20})  
give the bifurcating newborn  
triangle-like (3,1) POs at
$\alpha=7.0$ 
together with
other newborns $(5,2)$, $(7,3)$ and $(8,3)$ which appear at smaller
bifurcation values of $\alpha_{\rm bif}$ (and exist for $\alpha \geq 7.0$)
are obviously dominating (cf. lower panels {\it (b)}
in these Figures).

A nice beating seen in these figures [\ref{fig13}--\ref{fig20})] is explained
by the interference of the leading POs. The bifurcating orbits with the 
simple diameters 
 of the same order 
in magnitude but with different phases are responsible for such a beating
at the bifurcation points [\ref{fig16}--\ref{fig18}), and \ref{fig20}].
The ISPM contributions of diameters are close to the SSPM asymptotic ones
near the bifurcation points $\alpha=7$ and $4.25$ and in between (e.g. at 
$\alpha=6.0$), because
they are far away from their symmetry-breaking point at 
the harmonic-oscillator value $\alpha=2$.

\section{DEFORMED SHELL STRUCTURES AND PERIODIC ORBITS
IN THE POWER-LAW POTENTIAL}

\subsection{The power-law potential model} 

In Section III, the deformed shell structure is
investigated by using the nuclear fission-cavity model with the potential having
the sharp infinite walls. As shown in Section IV, this model corresponds
to the $\alpha\to\infty$ limit of the radial power-law potential 
(PLP) model.
In this section, we consider a more realistic 
mean-field PLP by taking the finite values of $\alpha$ 
for the deformed Fermi systems.
We shall investigate the changes of the
shell structures with various types of deformations, and examine
their relation to the periodic orbits, focusing especially on the role of
the periodic-orbit bifurcations.

The general deformed PLP, up to a constant, can be
expressed as
\begin{equation}
V(\r)=V_0\left(\frac{r}{R_0f(\theta,\varphi;\bm{\beta})}
 \right)^\alpha.
\label{PLP}
\end{equation}
The function $f$ describes the shape of the equi-potential surface
$V(\r)=E$ as 
\begin{equation}
r=R(\theta,\varphi;\beta,E) 
\label{effsurf}
\end{equation}
with the effective-surface profile function $R$ in the 
spherical coordinates,
\begin{equation}
R(\theta,\varphi;\beta,E)=(E/V_0)^{1/\alpha}R_0 f(\theta,\varphi;\bm{\beta}),
\label{ESPfun}
\end{equation}
where $\bm{\beta}$ represents the set of deformation parameters.
Imposing the conservation of the volume bounded by 
an equi-potential surface
(\ref{effsurf}) at a given energy $E$ with increasing the deformation
from the spherical ($f=1$) to the deformed
[$f=f(\theta,\varphi;\bm{\beta})$] shape
with the deformation parameters $\bm{\beta}$,
the shape profile function $f$ should be normalized to
satisfy the equation:
\begin{equation}
\frac{1}{4\pi}\int_0^{2\pi}d\varphi \int_0^{\pi}d\theta\sin\theta
 f^3(\theta,\varphi;\bm{\beta})=1\;.
\label{conscond}
\end{equation}
The scaling relation discussed in Section IV is 
valid independently 
of the function $f$.  In the following, we shall
investigate the deformed shell
structures in the PLP 
versus those in spheroidal cavity.

For the Fermi PLP system with a tetrahedral-like deformation,
as a simple and non-trivial exemplary case, 
the emergence of an unexpectedly 
strong shell effect at a large deformation
with a suitable diffuseness 
[power parameter $\alpha$ in the PLP (\ref{PLP})]
will be shown too in this Section.
A study of the anomalous shell effects, and their semiclassical origin with
focusing on the role of the periodic-orbit bifurcations
will be presented in two next sections.

\subsection{Prolate-oblate deformations}

First, we examine the effect of 
spheroidal deformations in the PLP, and discuss the
origin of the asymmetry 
between the deformed shell structure
in the prolate (cigar-like) and the oblate (pan-cake-like)
sides.

The deformed shell structures are
quite different in the HO potential having a soft surface, and 
the cavity potential
with a sharp surface.  The valley lines of the
shell-correction energy minima
in these 
potentials have the
opposite slopes in the plane of the deformation and
particle number for the prolate case at small deformations.
As explained in Section III, any valley can be associated with
a few constant-action lines of the dominant classical periodic
orbits with close (slightly different) actions.
Using the scaling rule for the action integral,
$S_{\rm PO}=\hbar\tau_{\rm PO}\epsi$ [see, e.g., (\ref{actionsc})], 
one can present the constant-action condition
(\ref{eq:cac}) in a more explicit form:
\begin{equation}
\epsi_F
=\frac{2\pi(n+\frac12+\frac{\mu^{}_{\rm PO}}{4})}{\tau^{}_{\rm PO}(\beta)}
\label{epsif}
\end{equation}
with
\begin{equation}
N\approx \int_0^{\epsi_F}\mathcal{G}^{}_{\rm ETF}(\epsi)d\epsi
 =\frac13{c_0}\epsi_F^3+c_1\epsi_F\;.
\label{partepsif}
\end{equation}
In the second equation, the ETF level density 
$\mathcal{G}^{}_{\rm ETF}$ was approximated by
that for the spherical shape (\ref{eq:getf}).%
\footnote{The coefficient $c_0$ in the leading term is
independent of the shape
under the volume conservation condition.  The coefficient
$c_1$ in the next-order term depends on the deformation,
but in practice, its influence on the basic shell-structure 
properties is almost negligable at leading order terms.}
Equations\ (\ref{epsif}) and (\ref{partepsif}) give the parametric
representation for the constant-action lines in terms
of the particle number $N$ as function of the deformation
parameters $\beta$ ($\eta$ for spheroidal cavity or $c$ for more realistic
parametrization of a cavity in Section III) through 
the Fermi energy $\epsi_F$.
Strutinsky et al. have shown that the 
valleys under the consideration (Section III)
are successfully explained by the constant-action
lines of the dominant classical periodic orbits 
within the spheroidal-cavity model of the nucleus
\cite{smod}.
In the HO model,
a two-parametric family of the shortest POs in the equatorial plane
makes the dominant contribution to the periodic-orbit 
sum (\ref{escscl}) for the shell-correction energy $\delta U$.
Their periods become smaller with increasing prolate 
deformation, and
the valley lines have a positive slope while
the two-parametric 
meridian-orbit family (triangles, quadrangles, ...) play the
dominant role in the spheroidal cavity potential for small
deformations. 
Their lengths become larger with
increasing prolate deformation and the slopes of the valleys are
negative \cite{spheroidptp,migdalrev}.

As shown in Section III for the case of a
fission-cavity model, the valleys of the constant-action minima change
their slope sign
to be positive with increasing deformation (at $c \approx 1.3$ 
in Fig.\ \ref{fig7})
because the dominant contributors
become shorter equatorial orbits
as approaching closer to the necking deformation
where they encounter the bifurcation.
At these deformations, such bifurcations are responsible for the 
enhancement of the shell structure through the additional local minima along
the basic growing valleys. As noted in Section III and discussed in 
\cite{migdalrev}, to some extent, it is similar to the 
shell structure enhancement in the  spheroidal cavity.
In case of spheroidal cavity, with increasing
deformation, the dominant 
contributors become the 3D POs which bifurcate from 
the second repetitions of equatorial orbits 
around the shape with the axis ratio
nearly 2:1, and form pronounced superdeformed shell
structures \cite{spheroidpre,spheroidptp,migdalrev}.

On the other hand, the issue of prolate-oblate
asymmetry is relevant to the shell structures in normal
(small-to-medium) deformation regions.
In the HO model, slopes of the shell-energy 
valleys are similar in both
prolate and oblate sides.  For the spheroidal cavity model, 
however,
Frisk noticed \cite{Frisk90} that the above-mentioned
valleys along
the constant-action lines of meridian orbits
are found to be approximately
flat in the oblate-deformation side. 
Due to this flatness, the gross-shell effects at the oblate deformations
is similar to those
for the spherical state, and systems 
will find favorable shapes in the
prolate-deformation side when the particle numbers depart from 
their spherical magic values.
This can be regarded as the origin of the
famous prolate-shape predominance in the nuclear ground-state
deformations.

Here, we are going to generalize Frisk's idea to a more
realistic  PLP model
with spheroidal deformations.
In this model, surface diffuseness is controlled
by the power parameter $\alpha$, and one can study
the dependence of the deformed shell structure on the surface 
diffuseness by taking different values of $\alpha$.  
The shape function $f$ for the spheroidal deformation is given by
\begin{equation}
f(\theta,\varphi;\delta)
\!=\!\frac{1}{\sqrt{e^{-4\delta/3}\cos^2\theta+e^{2\delta/3}\sin^2\theta}}\;,
\label{fspheroid}
\end{equation}
where the deformation parameter $\delta$ is related to the axis ratio
$\eta=R_z/R_\perp$ by $\eta=e^\delta$.  This definition of the
deformation parameter is useful because
one finds the same set of meridian orbits with the same absolute 
value $|\delta|$ in  both  
 the prolate ($\delta>0$, or $\eta>1$) and the
oblate ($\delta<0$, $\eta<1$ ) shapes.

The spheroidal shapes does not seem to be realistic 
in the description of nuclei
with very large deformations while
the gross-shell structures in the 
normal quadrupole deformation region are not much
sensitive to the details of the shape parametrization.  For the
comparison, one can take
a popular axially-symmetric quadrupole
deformation with the shape function
\bea\l{fquadrupole}
&f(\theta,\varphi;\beta_2)
=\frac{1}{(1+\frac35{\beta_2}^2+\frac{2}{35}{\beta_2}^3)^{1/3}}\times\\
&\times(1+\beta_2P_2(\cos\theta)). \nonumber
\eea
Figure~\ref{fig21} displays the difference of quadrupole and
spheroidal shapes.
These shapes are close
at small deformations while, at large ones, they are essentially
different because of 
the formation of the neck 
in the case of a
quadrupole shape.  The necking plays important role in shell structures
at large deformations, e.g., in nuclear fission \cite{fuhi}
whereas,
as usual, it is not so critical in the case of normal
deformations.
In Fig.~\ref{fig22}, we compare the level diagrams for
the spheroidal and axially-symmetric quadrupole deformations.
 These two diagrams have the
resemblance and difference with each other, which
can be understood from the semiclassical point of view
as we shall discuss below.
  
Classical dynamics in a
quadrupole-shape potential at a finite deformation
is much more chaotic than  
that in the spheroidal-shape potential because of
the negative curvature of the
potential surface, although these two PLP 
Hamiltonians
are both non-integrable.
Let us look first at the
Poincar\'{e} surface
of Section  
to observe 
the chaoticity of the classical motion.   
Figure~\ref{fig23} shows the PSS for the shapes given in
Fig.~\ref{fig21}.  Phase space is mostly covered with a 
tori for the spheroidal shape, 
while the large part of the phase space is filled
with a chaotic sea for the quadrupole shape.  
Reflecting a strongly chaotic
nature of the classical motion, 
the single-particle levels show 
their remarkable avoided  crossings (level repulsions)
at a large quadrupole deformation.
However, the gross-shell structures in the 
two level diagrams look quite similar.
This might be related to the short periodic orbits
which have the dominant contribution to the gross-shell
structure.
Figure~\ref{fig24} shows some short meridian orbits in the 
potential
(\ref{PLP}) with both (\ref{fspheroid}) and (\ref{fquadrupole})
for the profile function $f$.  One may find a quite similar 
set of POs
in these potentials.  Hence, we adopt the spheroidal
shape parametrization in order to understand the
gross-shell structures 
in deformed nuclei through the 
quantum-classical correspondence in a simple way.

In the axially-symmetric deformed HO limit $(\alpha=2)$ with
a generic irrational frequency ratio, all
the equatorial orbits are periodic, and they form the two-parametric
degenerate families.
Another periodic orbit is the isolated diameter
along the symmetry axis.  At deformations with a rational 
(resonance) ratio, all the three-dimensional (3D) 
orbits become periodic, and they are created from the corresponding 
EQ POs as the four-parametric degenerate families.
Let us now
consider the system with the power parameter $\alpha$
at a value slightly larger than 2.  
At the spherical shape, only the diameter
and circular POs
are remaining to be periodic.
With increasing spheroidal deformation, the equatorial and
meridian branches cross with each other.
In the HO limit, it corresponds to the resonance ratios
and families of POs with higher degeneracy are found there.
For $\alpha>2$, one finds no such families but new \textit{bridge orbits}
in place of them\cite{bridge}.

Figure~\ref{fig25} shows the
scaled period $\tau$ of some
shortest periodic orbits as function  of the deformation parameter
$\delta$ for the power parameter $\alpha=3.0$.
With increasing $\delta $ deformation through the spherical shape
($\delta=0$),  one of the diameter orbits in the equatorial 
plane, say $M$X ($M$ stands for the repetition number), 
bifurcates into itself (long-dashed), and a meridian (thick
solid and dotted curves) PO
$(M,1,M')$ at one of the 
successive bifurcation points marked by heavy dots.\footnote{%
The 3D orbit $(M_R,M_\varphi,M_z)$ is labeled by the numbers of
oscillations (rotations) in the directions of cylindrical
coordinates $(R,\varphi,z)$ with the symmetry $z$ axis.}
Equatorial polygon-like orbits can also encounter
a bifurcation at a certain deformation
where they are parents for the newborn
3D bridge orbits.  (These orbits are disregarded for simplicity
from Fig.~\ref{fig25}.)
Then, e.g., the meridian PO exists up to
the deformation where it submerge into a diameter $M'$Z
(in general, $M'$ can be different from $M$).
This might be
also considered as a bifurcation
in the opposite direction
of the deformation change,
namely, when the diameter 
$M'$Z bifurcates
 into itself and the meridian PO
$(M,1,M')$ with decreasing the deformation.
Therefore, such a meridian (or 3D) PO, which exists only between the
two deformations (bifurcation points)  $\delta$ and connects 
for instance the equatorial X diameter (at a smaller deformation) 
and the $M$Z diameter along the symmetry axis
(at a larger one) PO, can be transparently called as
the {\it ``bridge orbit''} \cite{bridge}.
In the PLP model, all the equatorial polygon-like POs 
encounter a bifurcation with increasing 
the deformation $\delta$, and
the new-born meridian, or 3D orbits,
make bridges between the
 equatorial
and $M$Z POs.
According to the general POT (Section II), the shortest
POs give the most important contributions into the averaged level density
(\ref{avdeltadentot}) and shell-correction energies (\ref{escscl}). Moreover,
according to the general arguments of the ISPM (Section IIB, III and IV 
for the specific examples), the amplitudes of the
oscillating level density are usually
enhanced by
the shortest bridge orbits due to their bifurcations \cite{migdalrev}.
These bridge orbits play a significant role in the deformed shell structure.  
Superdeformed and hyperdeformed shell
structures are associated with the 2:1 and 3:1 bridge-orbits between
the 2nd and 3rd repetitions of equatorial orbits and a 
primitive
symmetry-axis orbit $Z$, respectively.
With increasing oblate deformation $|\delta|$,
each repeated symmetry-axis
orbit encounters the bifurcation with emerging
a new bridge orbit, in addition to the parent one, and this
bridge orbit is submerged into a certain repeated equatorial
orbit at larger $|\delta|$.
For the normal
deformation region, the most
important orbit is the meridian family C which makes a bridge
between the primitive equatorial diameter family X 
in the oblate, and the
isolated primitive symmetry-axis diameter orbit Z
in the prolate deformation side.

Figure~\ref{fig26} shows the scaled periods
of the orbits C, X and Z, which participate in a  
bridge-orbit
bifurcation scenario for some
three different values of the power
parameter $\alpha$.  For $\alpha$ close to 2, the bridge orbits 
exist in a very
small range of deformations. 
They travel from the orbit X to Z with
only a small change of the deformation, 
and along their paths in the phase space a family of approximately
periodic (quasi-periodic) orbits should be formed.
It yields a coherent contribution into the
trace integral because of a local effective increase of the 
phase-space region of more exact integration, that leads, 
 in the semiclassical approximation 
through the ISPM, to an enhancement of shell effect (Section II, 
see also Sections III and IV
for the specific examples).
In the case of the bridge-orbit bifurcations, 
such a quasi-periodic family
acquires approximately an extra degeneracy along the 
trail of a bridge
orbit in between its emerge and subemerge
deformations, and one will
observe in between a more pronounced shell structure enhancement.
For the spherical PLP with $\alpha$ slightly 
larger than 2,
the existence of such a quasi-periodic 
family  with the increased degeneracy between the two bifurcation
points, 
in contrast to a local
family which is localized at the single bifurcation 
for a non-integrable Hamiltonian system in the symmetry $z$ axis plane
(or in the HO potential with commensurable frequencies)
(Section III and \cite{ozoriobook}), is associated with
a weak breaking of the SU(3) symmetry.
This is also much in contrast to
the PO family which appears at the single bifurcation point and exists
for all larger deformations in the integrable (e.g., spheroid cavity 
and RPLP) 
systems, see \cite{spheroidptp,migdalrev} and Section IV.
Generally speaking, the emergence of such bridge orbits
might be considered as a deep classical sign of the restoration of
a certain dynamical symmetry, increased 
in a finite deformation region
in comparison with the local
family case discussed in Section III.
As $\alpha$ increasing apart from 2, the
range of the bridge orbit becomes wider, and the 
quasi-periodic orbits
only exist around both ends of the bridge, where the
pronounced shell effect might be observed.

In order to find a
quantum-classical correspondence in the
deformed shell structures, we compare
the Fourier transform of the
quantum scaled-energy level density 
(\ref{fourierpower1})
with the scaled periods of the classical POs.  
Figure~\ref{fig27} displays the Fourier amplitude
$|F(\tau;\delta)|$ plotted in the
$(\delta,\tau)$ plane.  The scaled
periods $\tau^{}_{\rm PO}$ of the classical periodic orbits are also
plotted as functions of $\delta$.  As shown in this Figure,
the Fourier amplitude (\ref{fourierpower1})
is concentrated along the classical POs and takes especially
large values around the bifurcation points of the bridge orbits.
Since the gross-shell structure 
is dominated by the shortest periodic orbits,
the primitive bridge orbit C is expected to play the dominant 
role in the shell structure  of nuclei at normal deformations.

Figure~\ref{fig28} shows
the contour plot of the shell-correction energies 
as functions of the deformation $\delta$, and the
particle number $N$.  
The regular-energy valleys running along
the constant-action lines of the bridge orbit C can be observed
for $|\delta|\siml 0.4$.  As clearly seen from this Figure, 
the contribution of the bridge orbit
to the semiclassical shell-correction energies 
is dominant in 
a normal deformation region.
Since the slopes of the constant-action lines are steep in the 
right prolate
side while they are rather flat in the left oblate one
(in accordance with the valleys),
the nucleus 
apart from the spherical
magic numbers cannot find the energetically favorable shape
in the oblate
side. Thus, this isomer shape
will
tend to be deformed 
towards the
prolate deformation.
Obviously, this 
clearly explains the origin why the 
prolate-shape is dominant in the nuclear ground-state deformations,
in line of  
the spheroidal-cavity case 
analyzed by Frisk \cite{Frisk90}.

The origin of the prolate-shape dominance is also considered by
Hamamoto and Mottelson from another point of view \cite{HM2009}.  They
are focusing on the asymmetric manners of the level 
splittings in the prolate and oblate deformation cases,
and discuss the reason of such an asymmetry.  
In their model, the spin-orbit coupling was neglected.
Tajima et al.  
have shown that the feature of the prolate-shape dominance 
is quite
sensitive to the spin-orbit coupling
\cite{Tajima2001,Takahara2011}.
In this sense, our present
explanation of basic features of the prolate-shape 
dominance, without the spin-orbit coupling, is 
preliminary.
The final analysis should be based on a
more realistic
model with accounting for the spin-orbit 
coupling
for a deep understanding
of this long-standing
problem in nuclear structure physics
whose semiclassical study is in
progress \cite{aritapap,AritaXXX}.

\subsection{Tetrahedral deformed shell structures}

Breaking the reflection symmetry, 
one can also find an exciting subject
useful for the nuclear-structure physics \cite{BN1996}.
As shown in \cite{BN1996},
such a symmetry breaking is
experimentally observed through, for instance, the 
collective low-lying negative-parity states or
parity-doublet rotational spectra.
It is also significant for describing the asymmetric fission
process as discussed in Section III.  In the octupole deformation space, 
there are the 4 types of the octupole deformations.
In this subsection, the change of shell
structures with respect to octupole shapes and surface diffuseness
will be discussed from the semiclassical viewpoint.

With a large octupole deformation,
the equi-potential surface has
negative curvatures as for a large quadrupole deformation 
(Section VB),
and the classical
orbits become chaotic.
Significance of the flexible deformed shapes for a deep
understanding of the nuclear structure 
and dynamics were emphasized 
many times by Strutinsky et al. \cite{fuhi,smod,migdalrev} and Swiatecki
et al. \cite{wall,blocki,BMYpr} in different aspects: The nature of
the fission isomer shapes and
order-chaos transitions from spheroidal (or quadrupole) to the
Legendre polynomial high-multipole shapes 
of the cavity models versus the same but within the 
quantum-mechanical approach using the
realistic deformed Woods-Saxon (WS) potential with 
the finite surface diffuseness, respectively.
Taking their good ideas to use
simple analytical approaches (e.g., without the spin-orbit interaction 
but within the mean-field approximation for understanding the major 
properties of the 
shell structure, and chaoticity of the adiabatic 
to the non-adiabatic nucleon 
dynamics with increasing the shape multipolarity in heavy deformed nuclei),  
one should emphasize the diffuseness of the mean-field potentials.
Together with other additional degrees of 
freedom (spin-orbit interaction, for instance), they
lead to some new symmetry-breaking and  bifurcation phenomena. 
According to the general semiclassical ISPM arguments (Section IIB),
these phenomena are associated with the enhancement of the shell structure,
even in the case of a very chaotic nucleon motion inside of a
complicate non-integrable shape.
This seems to be in contrast to 
the results expected from \cite{blocki,BMYpr}, 
however, for the adiabatic approximation.  The diffuseness
of the mean-field potentials was underestimated in the
works \cite{smod,migdalrev} supporting the cavity models with
realistic surface shapes, too.  But, in some another sense that
there is no increasing much chaoticity with complexity of the 
deformed shapes, our results turn to be more
in line of \cite{fuhi,smod,migdalrev,bridge}.

In spite of the chaoticity in internal classical dynamics,
it is suggested in several
works that the system with $Y_{32}$-type shape provides
a remarkable shell effect at finite deformations
\cite{HMXZ91,RKHL97,TYM98,DGSM02}.
To see the case where such enhancement of gross shell effect is especially
pronounced, we shall examine now the special-shape parametrization which
smoothly connects the sphere and tetrahedron under variations of
a single deformation parameter
$\beta_{\rm td}$.  The shape function $f$ of (\ref{PLP}),
\bea\l{fftilde}
&f(\theta,\varphi)\!=\!\widetilde{f}(\theta,\varphi)\!\times \\
&\times \left[\frac{1}{4\pi}\int_0^{2\pi}d\varphi'\int_0^\pi d\theta'\sin\theta'
 \widetilde{f}^3(\theta',\varphi')\right]^{-1/3},\nonumber
\eea
can be expressed in terms of an auxiliary function $\widetilde{f}$.
This function is given by the largest positive
root $\widetilde{f}$ of the following quartic equation \cite{AM2014}:
\begin{equation}
\widetilde{f}^2+\frac{\beta_{\rm td}}{2}
 \left\{1+u_3(\theta,\varphi)\tilde{f}^3
 -u_4(\theta,\varphi)\widetilde{f}^4\right\}=1\;, 
\label{eqftilde}
\end{equation}
where
\bea
u_3&=&\frac{4}{15}P_{32}(\cos\theta)\;\sin2\varphi\;, \\ 
u_4&=&\frac15+\frac45P_4(\cos\theta)+\frac{1}{210}P_{44}(\cos\theta)
\;\cos4\varphi\;. \nonumber
\eea
Both functions $u_3$ and $u_4$ are invariants
under all the symmetry transformations which define
a tetrahedral $T_d$ group.  
The solution $\widetilde{f}$ of the equation (\ref{eqftilde}) 
possesses the same symmetry for any value of $\beta_{\rm td}$.  
Equation (\ref{fftilde}) for the shape function $f$ satisfies
the volume conservation condition (\ref{conscond}), as a scaling
of the function $\widetilde{f}$. 
For $\beta_{\rm td}=0$, one obtains the 
equi-potential surfaces of the spherical shape
($f=1$) while, at $\beta_{\rm td}=1$, one finds
those of the shape of tetrahedron.
Hence, the transition of the
shell structures from the sphere to the tetrahedron can be examined by
changing the single tetrahedral-deformation parameter
$\beta_{\rm td}$
continuously from 0 to 1.  The shapes of the equi-potential surface at
several values of $\beta_{\rm td}$ are displayed in
Fig.\ \ref{fig29}.  The $T_d$ symmetry group
consists of the
24 symmetry
transformations: The three-fold rotations around
 the four $C_3$ axes
(there are 8 such transformations); the four-fold rotatory
reflections around
the three $S_4$ axes (6 transformations); the two-fold rotations
around 
the $C_2$ axes, equivalent to the $S_4$ axes, (3 ones); the
reflections about the six symmetry planes $\sigma_d$ (6 ones); and 
finally, in addition, the identity (one). 
This group has five irreps (irreducible representations);
the two 1D irreps (A$_1$, A$_2$), one 2D irrep
(E), and two 3D irreps (F$_2$, F$_1$). 
The quantum levels can be
classified in terms of  
these five irreps.  
The levels belonging 
the $n$-dimensional irrep
have the $n$-fold degeneracy, and thus, the
quantum energy spectrum
in the potential with the $T_d$ symmetry generally
contains levels with the three-fold degeneracy.

In addition to these 
geometric degeneracy properties,
Hamamoto et al. have
found that the $Y_{32}$ deformed system 
shows a quite strong 
gross-shell effect as compared to
other types of the octupole shapes
\cite{HMXZ91}.  With our shape parametrization which connects the
 sphere and the tetrahedron, one obtains 
a significantly pronounced shell effect arising
at a large tetrahedral deformation.  
This effect might be related to the
stabilities of
classical periodic orbits.  As shown in
Fig.~\ref{fig29}, the equi-potential surface for the current
shape parametrization (\ref{fftilde}) is convex everywhere, 
and the classical
trajectory is more stable than in the case of a pure $Y_{32}$ 
shape.

Figure~\ref{fig30} shows the single-particle level diagram for
two values of the 
power parameter $\alpha=4.0$ and 6.0, where the
scaled-energy levels $\epsi_i=(E_i/V_0)^{1/2+1/\alpha}$ 
are
plotted as functions of the tetrahedral 
parameter $\beta_{\rm td}$.  One
finds the prominent shell effect 
at  
a large tetrahedral-like
deformation $\beta_{\rm td}\approx 0.6$ and $ 0.7$ for 
$\alpha=4.0$ and $6.0$, respectively.
It is extremely interesting that the 
deformed magic
numbers are exactly found to be equivalent to 
those of the spherical
harmonic oscillator at such large tetrahedral-like deformations.
According to the POT (Section IIB),
these HO magic numbers might be explained by the approximate symmetry
restoration,
specific to the bridge-orbit bifurcation.
In a simple bifurcation, the formation of the local quasi-periodic
family is 
limited to the vicinity of the single bifurcating PO,
while, in the bridge-orbit bifurcation, 
the bridge PO family 
is traveling between the two POs which are
close in the deformation but widely
separated from each other in the phase space.
Therefore, 
because of increasing the phase space volume occupied by the 
bridge 
PO family, one can find  
 a remarkable
enhancement 
of the  bridge
PO amplitudes of the oscillating level 
density,
in contrast to the case of
simple local 
bifurcations.

Let us investigate the origin of the
unexpectedly strong and regular shell effects
which emerge at certain combinations of the surface diffuseness
and large tetrahedral deformation.
Figure~\ref{fig31} shows some short classical
periodic orbits for $\alpha=6.0$.  There are 
the diameter- and the circle-orbit
families in the spherical power-law potential $(\beta_{\rm td}=0)$.
With increasing $\beta_{\rm td}$, each of these 
two families triplicate into the
three branches.  The diameter family triplicate
into the linear orbit
DA along the $C_{2v}$ axis, the linear orbit DB 
along the $C_3$ axis, and
the self-retracing planar orbit PA in the $\sigma_d$ plane.
The circle orbit triplicate into the 3D
orbit TA having the $S_4$ and $\sigma_d$
symmetries with respect to the two symmetry planes containing
the common $S_4$ axis,
the 3D TB having the $C_3$
symmetry, and the planar 
PB (in the $\sigma_d$ plane)
having the $C_2$ and 
$\sigma_d$ symmetries as related 
to the two symmetry planes
containing the common $C_2$ axis.

Figure~\ref{fig32} shows the scaled periods of some short
periodic orbits as functions of the tetrahedral 
deformation parameter
$\beta_{\rm td}$.  With increasing $\beta_{\rm rd}$, the scaled
periods of the major four orbits gather 
around $\beta_{\rm td}=0.6$, and
bifurcations take place almost at the
same values of $\beta_{\rm td}$.
Some of the new-born periodic orbits 
are bridges between the two crossing orbits.
As we discussed above, the emergence of these
bridges is a good signature of
the restoration of
an 
approximate relatively high
dynamical symmetry.
A family of 
the quasi-periodic orbits are formed along
the bridge orbits connecting some two distinct 
periodic orbits in the
phase space, and hence, the dynamical 
symmetry is not localized to a
single PO, in contrast to
the single local 
bifurcation \cite{AM2014}.

Figure~\ref{fig33} shows the Fourier amplitudes
of the quantum scaled-energy level density, which are defined in
(\ref{fourierpower1}), 
in the $(\beta_{\rm td},\tau)$ plane.  Scaled periods
$\tau^{}_{\rm PO}$ of some shortest
 classical POs
are also plotted as
functions of $\beta_{\rm td}$ in the same plane.
As seen from this map, the
Fourier amplitudes exhibit peaks along the classical periodic orbits,
and show remarkable enhancement around the bifurcation points
$\beta_{\rm td}=0.6\sim 0.8$, where a strong shell effect
appears 
in the
quantum spectrum.
Again, the gathering of 
POs 
and almost
the  
isochronous occurrence of the 
bridge bifurcations
among them 
can be associated with
the 
restoration of an approximate high dynamical
symmetry.
As a  conclusion, the latter is 
the origin of the
anomalous properties of the shell structure for a large
tetrahedral-like deformation.
Recalling also the
magic numbers equivalent to the spherical harmonic oscillator, one may
relate the restored symmetry 
to that of the SU(3) nature.

Since the tetrahedral-like deformed potential 
has no continuous symmetries, the
Gutzwiller trace formula for the isolated orbits can be applied if all the
significant
periodic orbits are sufficiently apart from the bifurcation
points.  In order to compare with the quantum
calculations, one can
apply the Gutzwiller
trace formula for the level density to the semiclassical
shell-correction energy (\ref{escscl}) for a given deformation 
$\beta_{\rm td}$,
\bea\label{eq:gutz1}
&\delta U(\epsi_F)=2\left.\frac{dE}{d\epsi}\right|_{\epsi=\epsi_F}\times\\
&\times \sum_{\rm PO}\frac{\mathcal{A}_{\rm PO}}{\tau_{\rm PO}^2}
\cos\left(\tau^{}_{\rm PO}\epsi_F - \frac{\pi\mu^{}_{\rm PO}}{2}\right),
\nonumber\\
&N(\epsi_F)=2\int_0^{\epsi_F}d\epsi\left[\mathcal{G}_{\rm ETF}(\epsi)+
\right.\\
&+\left.\sum_{\rm PO} \mathcal{A}_{\rm PO}\;\cos\left(\tau^{}_{\rm PO}\epsi
-\frac{\pi\mu^{}_{\rm PO}}{2}\right)\right],
\label{eq:gutz2} \nonumber\\
&\mathcal{A}_{\rm PO}
\approx \mathcal{A}_{\rm PO}^{\rm G}
=\frac{\tau_{\rm PO}}{\pi\sqrt{|\det(\mathcal{M}_{\rm PO}-I)|}},
\label{eq:gutz3}
\eea
where $\mathcal{G}_{\rm ETF}$ is approximated by the 
spherical expression (\ref{eq:getf}) (Section IVE1).
Equations (\ref{eq:gutz1}) and
(\ref{eq:gutz2}) provide the parametric
representation for the function
$\delta U$ of $N$ through 
the Fermi scaled energy 
variable $\epsi_F=(\vareps_F/U_0)^{1/2+1/\alpha}$, 
where $E_F$ is the Fermi energy.

The two bottom panels in Fig.~\ref{fig34} show the comparison
between the quantum and semiclassical shell-correction
energies for $\beta_{\rm td}=0.1$
and $0.5$.  For $\beta_{\rm td}=0.1$, we take the main 6 shortest
periodic orbits DA, DB, PA, PB, TA and TB.  The 
gross structures
are successfully described 
by the Gutzwiller trace formula.  Some
fine structures might be well approximated semiclassically
by the proper treatment of 
longer periodic orbits.  For $\beta_{\rm td}=0.5$, we 
include the
contribution of the orbit TC, which 
emerges through the bifurcation of PB
at $\beta_{\rm td}=0.28$,
in addition to the six orbits mentioned above.
The main oscillating
pattern is reproduced well
but the semiclassical formula generally overestimate
the amplitude of the oscillations.  A most probable reason of this 
discrepancy is that
the deformation  $\beta_{\rm td}$ becomes too   
close to one of the bifurcation points where the lost of
accuracy of 
the standard
stationary-phase approximation takes place because of the 
divergence of the Gutzwiller
trace formula. 

In the top panel of Fig.~\ref{fig34}, we display the quantum
shell-correction energies for $\beta_{\rm td}=0.7$. 
At this deformation, one finds the most pronounced shell effect
with
a very strong and regular oscillations in the shell-correction
energy, which should be dominated by the bifurcating orbits having
almost the same periods.  Here, the standard 
stationary-phase method completely breaks down
due to the
dense bifurcations, and one cannot certainly
apply the Gutzwiller 
trace formula.
As shown in Section IIE, 
the ISPM is useful also to 
solve the
bifurcation problem in a non-integrable system.  
Its application to the description of
the
anomalous behavior in the amplitudes of the
oscillating level density and energy  
at large oblate-prolate and tetrahedral-type
deformations
should be an interesting and
challenging future subject.

\section{Conclusions}

In Section II we have given a short review of 
the semiclassical theory (POT)  accounting for the PO bifurcations 
and symmetry breaking  in 
different potential wells.
A general trace formula for the oscillating level density
was derived in the phase space variables (see also Appendix A2).
Extensions of the POT to the treatment of the bifurcations and symmetry breaking
was presented as the improved stationary phase method 
in close analogy with
the catastrophe theory of Fedoriuk and Maslov, and hereby,
overcoming the divergence of the semiclassical amplitudes of the 
Gutzwiller theory and discontinuity of them in the Berry\&Tabor approach
at bifurcations. The improved semiclassical amplitudes
within the ISPM typically exhibit an enhancement
of the shell structure locally near a bifurcation and on right 
side of it where new orbits 
emerge which is of high order in the inverse
semiclassical parameter, $1/\hbar$, with respect to the Gutzwiller
trace formula amplitude for the isolated orbits.
The PO expansions for the averaged level density and 
the shell-correction energy $\delta U$, and their PO convergence were shown
too.
The semiclassical
trace formulae for $\delta U$ exhibit a 
rapid convergence of the PO sum, due to an
inverse dependence of the individual orbit 
contributions on the squares of their
periods (actions), in addition to  the phase-space 
volume and degeneracy symmetry factors 
of the iscillating level density.
This allows one often to express significant 
features of the shell structure in terms of a 
few short periodic orbits. In many cases, 
the shortest POs are sufficient to describe 
the gross-shell features in $\delta U$.
We have obtained an analytical trace 
formula for the oscillating part 
of the level
density in the H\'enon-Helies Hamiltonian 
as a sum over periodic orbits 
in a non-integrable potential. 
It is continuous through all critical points, in particular here the 
harmonic oscillator 
limit at zero energy and the cascade of pitchfork bifurcations near 
the saddle energy.
We find an enhancement of the semiclassical amplitudes near the most 
critical points. 
The numerical agreement with quantum results is good,
in spite of a simple uniform ISPM approximation including only
the simplest primitive periodic orbits. The quantum-classical correspondence
for the chaos-order transitions is shown through the Poincar\'e surface of
sections in the limit from the non-integrable region of the energies to
the symmetry breaking point.

In Section III, we have presented a semiclassical calculation of a 
typical actinide 
fission barrier using the POT, employing a fission cavity model 
that uses a realistic 
description of the three principal axially-symmetric 
deformations (elongation, neck 
formation and left/right asymmetry) occurring 
in the (adiabatic) fission process.
The characteristic double-humped barrier 
and, in particular, the sensitivity of the outer barrier to 
left/right-asymmetric deformations 
can be qualitatively well described by the POT. 
The loci of minimal 
quantum shell-correction energies 
$\delta U$, both in particle number vs.\  
deformation space  
and in
two-dimensional deformation space, 
are correctly followed
by the constant-action loci of the shortest POs. Hereby we observe a clear
signature of period-one bifurcations of the shortest equatorial orbits
which were
treated semiclassically using normal forms and uniform approximations.

We found that the local minima of the shell-correction energy calculated
for the non-integrable (in the symmetry $z$ axis plane)
cavity potential with the realistic
parametrization of the shapes of the fission cavity model 
\cite{fuhi} can be associated with
the bifurcations of the POs at large deformations 
as for the 
spheroidal cavity \cite{migdalrev,spheroidptp}. 
The quantum-mechanical Fourier spectra of the corresponding 
Hamiltonian exhibit a nice quantum-to-classical correspondence, in that 
the enhanced
Fourier signals follow exactly the PO lengths of 
their semiclassically enhanced 
amplitudes.
This correspondence appears also in the correct 
description of the loci of large deformations in particle number vs.\ 
deformation space by the constant-action lines of the bifurcated 
period-two and -three orbits.
An important reason for their strong enhancement at large deformations, 
in addition to the general argument 
given above (and explained in Sec.\ 2), 
is also the fact that the new bifurcated orbits have locally a larger classical
degeneracy than their parent orbits and the period-one  
orbits ($\cK=1$) (except near the bifurcations).

In Section IV we presented a class of the radial power-law
potentials, $V(r) \propto r^\alpha$, 
which up to a constant turn out as good approximations to the
 popular WS potential
in the spatial region where the particles are bound. 
The advantage of the RPLP is that it is
capable of controlling surface diffuseness, and in the same time,
the classical dynamics scales
with simple powers of the energy, that
makes the POT calculations greatly
easy. It can be done sometimes even 
explicitly analytically for $\alpha= 4$
and $6$ (besides of the well-known HO case $\alpha=2$)
in terms of the simple special functions. 
The quantum Fourier spectra yield directly the amplitudes of the
quantum level density in terms of periods (actions) of the
leading classical POs.

We described the main PO properties of the 
classical dynamics in the
RPLP as the key quantities of the POT.  
We developed semiclassical trace formulae for any power $\alpha$
of this
potential and studied various limits of $\alpha$ (the harmonic oscillator
potential for $\alpha=2$ and the cavity potential for $\alpha\to\infty$).
We presented a semiclassical theory of quantum
oscillations of the level density and shell-correction energy for
RPLPs. It is based upon extended Gutzwiller's trace formula in the 
convenient phase space variables which 
connects the oscillating component of the level density for a 
quantum system to a sum over POs of the corresponding 
classical system. This POT was applied 
to express the shell-correction
energy of a finite fermion system in terms of POs. 
We obtained good agreement between the 
ISPM semiclassical and quantum results for the 
level densities and shell-correction
at several powers of the PRLPs. 
For the power $6$ we found also good agreement
of the ISPM trace formulas with the SSPM ones far from the
bifurcations of the leading  short-time POs of a maximal degeneracy of the 
classical POs. 
The strong amplitude
enhancement phenomena at a bifurcation point $\alpha=7$
 in the oscillating (shell) components of the level density and 
energy, due to the bifurcations, and on their right side,
were remarkably agreed with the peaks of 
the Fourier spectra. We found a significant influence 
of bifurcations of periodic orbits on the main characteristics 
(oscillating components of the level densities and energy-shell corrections)
 of a fermionic 
quantum system. They leave signatures in its energy spectrum (visualized, e.g., 
by its Fourier transform), and hence, its shell structure.

The trace formulae are in good agreement with the
quantum-mechanical level-density oscillations for gross- (larger 
averaging parameter $\gamma$ and a few shortest POs) 
and fine-resolved (smaller $\gamma$ and time-longer bifurcating POs)
shell structures, also for the shell-correction 
energies independent of $\gamma$. We found a similar gross shell structure
in the shell-correction 
energies with the corresponding densities at $\gamma \approx 0.2$
for $\alpha \simg 6$ . 
The fine-resolved shell structures were
found with the corresponding densities at 
$\gamma=0.03 - 0.2$ for $\alpha \simg 6$.

Section V is devoted to the extension of the  
semiclassical POT
to a more realistic deformed Fermi systems with their surface diffuseness
by using scaling of the power law potentials (PLP) from the RPLP to 
the superdeformed shapes.
We may state that the POT is
well capable of explaining the main features of quantum shell structure 
in terms of a few short classical POs for such 
deformed PLPs.
Bifurcations of POs are
not simply an obstacle of the semiclassical theory, but they leave
clear signatures both in the quantum Fourier spectra and the locations
of minima of the shell-correction energy $\delta U$ plotted versus
particle number and potential parameters
such as 
the power (potential surface diffuseness) 
and the deformation parameter.
As examples of the deformed PLP with the surface diffuseness,
we have discussed in Sec.~5  the prolate-oblate
asymmetry in shell structures for the
quadrupole-type deformations, and
anomalous enhancement of  shell effects at 
reflection-asymmetric deformations
with the tetrahedral symmetry.  In the both cases, 
the bifurcations of bridge orbits,
play the significant role in enhancement 
of the shell effect through
approximate restorations of the dynamical symmetry  which depends
essentially on the certain combinations of the diffuseness (power) and
deformation parameters.  Especially, in the latter case of the
tetrahedral deformation, one finds unexpectedly remarkable enhancement
of the shell effect.  It might be explained 
by the formations of the short
bridge PO families: If the two bifurcation  
points, corresponding to
the two ends of the bridge, are close to each other, 
such PO
families are formed along the path of the bridge connecting the
 two widely
separated  (in the phase space) POs. These bridge families
 should occupy  larger phase-space 
volumes as compared
to the cases of a simple local bifurcation.
 Thus, these specific
bridge-orbit bifurcations  might be responsible
for the 
significant influence
on the shell structure in
a quantum deformed Fermi system 
with a finite diffuse edge.

In our future work we intend to further study finer shell structures
due to longer orbits, applying both the ISPM and uniform approximations 
to treat their bifurcations \cite{sclbook,brreisie,trap}. 
In particular, we want to understand the
``quantization'' into isolated minima that occurs along the valleys of
both the ground-state and the secondary (isomer) minima and, in 
particular, we want to better understand the superdeformed shell 
structure taking into account the diffuseness 
of the mean-potential edge.  In the
region of superdeformation in the spheroidal cavity, 
we found clearly
that this is due to the bifurcated
period-two and -three orbits. It will be interesting to study the
analogue situation in the fission cavity model by investigating
the higher-period bifurcations occurring there. One complication
in that model is that we have to vary three deformation parameters
($c,h,\alpha$), instead of only one ($\eta=c^{3/2}$) for the spheroid, in 
order to fully cover the fission barrier landscape.  This will 
naturally lead to a much larger variety of possible bifurcations.
Another object of our studies will be to understand semiclassically
the transition regions between the down-going ground-state valleys
and the initially up-going secondary-minimum valleys. 
They are well described
separately by the shortest meridional and equatorial orbits,
respectively, but in the transition regions ($1.2\siml c\siml 1.35$
and $1.2\siml \eta \siml 1.5$, respectively) we expect further
bifurcations and corresponding new orbits to play a role.
We want to apply the ISPM for deeper studying the enhancement 
of the shell structure in the PLP due to the new bridge-orbit bifurcations.

On a longer time scale, we intend also to include pairing and spin-orbit 
interactions into the POT calculations, in order to come closer
to a realistic description of nuclei, and to extend the POT
towards the collective dynamics
\cite{friskguhr1993,AB_jphys2002,BAPZ_ijmpe2004}. 
We would like to study
within the POT the shell corrections
to the transport coefficients, such as the  
inertia, friction
\cite{magvvhof,GMFprc2007,MGFyaf2007,BMYijmpe2012,BMps2013};
 and moments of inertia \cite{mskbPRC2010,belyaev,GMBBPS2015,GMBBPS2016}
for the nuclear collective dynamics,
see also \cite{bartelnp}
for the  self-consistency and spin effects included into the ETF component 
of the moments of inertia. 
We would like also
to extend the POT to the two-component (neutron-proton) Fermion systems
as atomic nuclei, see 
\cite{strtyap,strmagbr,strmagden,magsangzh,BMRV,BMRPhysScr2014,BMR_PS2015,BMR_PRC2015} for the ETF approach. 
Our semiclassical analysis may therefore lead 
to a deeper understanding of the shell structure effects in 
finite fermionic systems such as nuclei, metal clusters or 
semiconductor quantum dots whose conductance and magnetic 
susceptibilities are significantly modified by shell effects. 

\bigskip
\centerline{{\bf ACKNOWLEDGMENTS}}
\medskip

Authors gratefully acknowledge
Ch.\ Amann, M.\ Brack, S.\ N.\ Fedotkin, F.\ A.\ Ivanyuk,
Ya.\ D.\ Krivenko-Emetov, M.\ Matsuo,
K.\ Matsuyanagi, A.\ I.\ Sanzhur, K.\ Tanaka 
and E.\ E.\ Saperstein for fruitful collaborations and 
many useful discussions.  
One of us (A.G.M.) is
also very gratitude for a nice hospitality during his working visits of the
University
of Regensburg in Germany,
and Physical Department of the Nagoya Institute of Technology,
also the Japanese Society of Promotion of Sciences
for the financial support, ID No. S-14130.

\appendix

\setcounter{equation}{0}
\renewcommand{\theequation}{A.\arabic{equation}}
\renewcommand{\thesubsection}{A\arabic{subsection}}

%
\section{}
\l{appA}
%

\subsection{TO CLASSICAL DYNAMICS \\ FOR THE RPLP}
\l{appA1}

For any spherical potentials $V(r)$ the Hamiltonian $H(\r,\p)$ (\ref{Hsph})
in the spherical canonical phase-space variables 
$\{r,\theta,\varphi,p_r,p_\theta,p_\varphi\}$, one has the dynamical
equations: 
\bea\l{prteq}
p_r(r) 
&\equiv& \sqrt{2m\left[\vareps-V(r)\right] - 
\frac{L^2}{r^2}} 
= m \dot{r}\;, \\ 
\dot{\theta} &\equiv& \frac{L}{mr^2} \equiv \frac{L}{r^2p_r}\;\dot{r}\;.
\nonumber
\eea
As the angular moment $L$ is conserved for a motion of the
particle in a central field $V(r)$, all CTs are
lying in a plane crossing the center $r=0$.
Therefore, the third equation for $\varphi(t)$, additional to 
(\ref{prteq}), is identity
$\varphi=\theta$, and the two other equations (\ref{prteq}) 
can be considered in an azimuthal plane $\varphi=const$. 
As the Hamiltonian (\ref{Hsph}) expressed through the spherical 
action-angle variables does not depend on the angles, i.e., they are
cyclic, the frequencies are constants given by
\bea\l{freq}
&\omega_r=\partial H/\partial I_r=
\left(\partial I_r/\partial E\right)^{-1} = \\
&=\left(\frac{m}{\pi} \,\int_{r_{min}}^{r_{max}}\frac{{\rm d}r}
{\sqrt{2m\left[E-V(r)\right] - L^2/r^2}}\right)^{-1},\nonumber\\
&\omega_\theta \equiv \omega_\varphi=
-\left(\partial I_r/\partial L\right)/\left(\partial I_r/\partial E\right)\;,
 \nonumber
\eea
where $I_r=I_r(E,L)$ is the surface energy in (\ref{actionvar}).
Thus, the PO equation takes the form of the resonance condition 
(\ref{percond}). The energy surface
$I_r=I_r(\vareps,L)$ is simplified to a function of one variable $L$ for a 
given energy $\vareps$ of the particle  because the second
equation for the ratio of frequencies $\omega_\theta/\omega_\varphi=1$ 
is identity
[$I_\theta=L$
for $L_z=0$, see (\ref{actionvar})].

Integrating the differential equations in (\ref{prteq}), 
one obtains the radial trajectory $r=r(t)$ and the angle 
$\theta(r)$ in an azimuthal plane.
For one period $t=T$ along the PO
at $\Theta=\Theta''$ ($r=r''$), one has (\ref{tetar1}).
In order to derive the relation (\ref{tetar1}), one can use
(\ref{freq}) for the definition of frequencies $\omega_\theta$
and $\omega_r$ through the classical Hamiltonian $H$, and
the Jacobian properties for the frequency ratio 
$\omega_\theta/\omega_r$ in an azimuthal plane ($I_\theta=L$ for 
$I_\varphi=L_z=0$):
\bea\l{jacprop}
\frac{\omega_\theta}{\omega_r}&=&
\frac{D(H,I_r)}{D(I_\theta,I_r)}\;\frac{D(I_r,I_\theta)}{D(H,I_\theta)}
= \\
&=&-\left(\frac{\partial I_r}{\partial L}\right)_{H=\vareps}\;. \nonumber
\eea
%

\subsection{SEMICLASSICAL DERIVATION \\ 
OF THE PHASE SPACE TRACE FORMULA}
\l{appA2}

For the derivation of the phase-space trace formula (\ref{pstrace})
we start with the definition of the level density $g(\vareps)$ as the trace of 
Green's function $G(\r',\r'',\vareps)$ in the Cartesian coordinates
 \cite{sie97,ssun,schomerus},
\begin{eqnarray}
&\hspace{-0.5cm}g(\vareps)= - \frac{1}{\pi}\,\Im \int \d \r'' \int \d\r'\,
G\left(\r',\r'',\vareps\right)\times\\
&\times\delta\left(\r''-\r'\right)= - 
\frac{1}{\pi(2 \pi \hbar)^{\mathcal{D}}}\,\Im \int \d\r''\, 
\int \d\r'\, \int \d \widetilde{\p}\times \nonumber\\
&\times G\left(\r',\r'',\vareps\right)\,\exp\left[\frac{i}{\hbar}
  \,\widetilde{\p}
\left(\r'-\r''\right)\right]\;. \nonumber
\label{trace}
\end{eqnarray}
The Fourier representation of $\delta(\r'-\r'')$ 
was used in these derivations.
For any fixed integration variables, one has the single $CT$, 
and therefore, according to 
\cite{strumag,ellipseptp}, we can apply the standard Gutzwiller's
expression for the Green's function related to the isolated paths
CT \cite{gutz,sclbook},
\bea\l{greenfunscl}
&&\hspace{-0.5cm}G_{\rm scl}\left(\r',\r'',\vareps \right)=2 \pi \left(2 \pi i
\hbar\right)^{-(\mathcal{D}+1)/2}\times \\
&\times&\sum^{}_{\rm CT} 
\left|\mathcal{J}_{\rm CT}\left(\p' \, t^{}_{\rm CT},
      \r''\,\vareps\right)\right|^{1/2}\times \nonumber\\
&\times&\exp\left[\frac{i}{\hbar} S_{\rm CT}\left(\r',\r'',\vareps\right)
 - \frac{i \pi}{2}\,\mu^{}_{\rm CT}\right]. \nonumber
\eea
Substituting this approximate expression for the  Green's function into
the second equation of (\ref{trace}) we calculate the integral
over $\r'$ by the SPM 
extended to the continuous symmetries. 
The stationary phase
conditions for the integration over the parallel $y'$ 
and perpendicular
$\r_\perp'=\{x',z'\}$ components of $\r'$ 
in the 3D case
($x'$ in the 2D Fermi system) are given by
\begin{eqnarray}
&&\left(\frac{\partial S_{\rm CT}}{\partial {y}'}\right)^* =
-\widetilde{p}_y,\quad \rightarrow \quad {p_y}^{\prime *}=\widetilde{p}_y\;,
\\
&&\left(\frac{\partial S_{\rm CT}}{\partial {\r_\perp}'}\right)^* 
= \widetilde{\p}_\perp,
\quad \rightarrow \quad \p_\perp^*=
\widetilde{\p}_\perp\;. \nonumber
\label{statcondrfin}
\end{eqnarray}
Taking the integral over the perpendicular coordinates $\r_\perp'$
by the SSPM with the second equation in (\ref{statcondrfin})
for the stationary points $(\r'_\perp)^*$, one notes that the first
equation for the stationary values of the parallel component
$y'$ is the identity. Using the expansion of the action in 
$y'-y''$ up to the main
linear term, and  keeping
the zero order terms of the pre-exponent factors, we obtain the Fourier 
integral, which leads to the energy-conservation $\delta$-function,
\bea\l{enconserv}
&\frac{1}{2 \pi \hbar}\, \int \d y'\;
\exp\left[\frac{i}{\hbar}\left(p-p_y'\right)
\left(y'-y''\right)\right]=\\
&=\delta (p-p_y')=\frac{p}{m}\,\delta \left(\vareps-H(\r',\p')\right)\;. 
\nonumber
\eea
Applying the Hamilton-Jacobi equations,
\begin{equation}
H(\r',\p')=\vareps, \qquad\qquad H(\r'',\p'')=\vareps\;,
\label{hjeq}
\end{equation}
for the calculation of derivatives in the Jacobian
$\mathcal{J}_{\rm CT}\left(\p',; t^{}_{\rm CT},\r''\,\vareps\right)$
of the semiclassical Green's function (\ref{greenfunscl})
and using the standard Jacobian transformations, one has \cite{gutz,strumag}
\begin{equation}
\hspace{-0.3cm}\mathcal{J}_{\rm CT}\left(\p',t^{}_{\rm CT};\r'',\vareps\right)=
-\left(\frac{m}{p}\right)^2\!
\mathcal{J}_{\rm CT}\left(\p'_{\perp},\r''_{\perp}\right).
\label{jacrelgutz}
\end{equation}
With further applying the standard
Jacobian transformations,
\begin{equation}
\frac{\mathcal{J}\left(\p'_{\perp},\r''_{\perp}\right)}
{\mathcal{J}\left(\p''_{\perp},\r''_{\perp}\right)}
=\mathcal{J}\left(\p'_{\perp},\p''_{\perp}\right)\;,
\label{jactrans1}
\end{equation}
for the constant $\r'_{\perp}=\r^{\prime*}_{\perp}$,
one finally arrives at the phase space trace formula (\ref{pstrace}).

We need also the relation between generating functions
$\Phi_{\rm CT}$ and $\widehat{S}_{\rm CT}$ (Section IIF),
\be\l{phishat} 
\Phi_{\rm CT}= \widehat{S}_{\rm CT}-{\bf p}'\left({\bf r}'-{\bf r}''\right)\;.
\ee
where
\be\l{shats} 
\widehat{S}_{\rm CT}\left({\bf r}'',{\bf p}';\vareps\right)=
S_{\rm CT}\left({\bf r}',{\bf r}'';\vareps\right) +
{\bf p}'{\bf r}'\;.
\ee
%

\subsection{SCALING AND UNITS}
\l{appA3}

For convenience, let us consider classical dynamics in terms of the energy
$\overline{E}$, action ${\cal I}_i$, angular momentum $\Lambda$, radial
coordinate $\overline{r}$ and frequency $\overline{\omega}_r$ and curvature
$\overline{K}$ in
dimensionless units,
\begin{eqnarray}\label{units}
\overline{E}&=&E/V_0, \qquad \mathcal{I}_i=I_i/\sqrt{mR^2V_0}\;,\\
 \Lambda&=&
L/\sqrt{mR^2V_0}\;,\nonumber\\
\overline{r}&=&r/R\;, \qquad \overline{\omega}_r=
\omega_r \sqrt{mR^2/V_0}\;,\nonumber\\ 
\overline{K}&=&K \sqrt{mR^2V_0}\;. \nonumber
\end{eqnarray}
Due to a scaling property (\ref{scaling}),
for the classical dynamics in
the potential (\ref{potenra}) \cite{aritapap,arita2012} 
the energetic dependence
of the action $\mathcal{I}_r$ (\ref{actionvar}),
the angular momentum $\Lambda$, the frequency $\overline{\omega}_r$ 
(\ref{freq}) and
the curvature $\overline{K}$ (\ref{curv})
can be recovered in terms of the
``scaled energy'' $\epsi$,
\be\l{scalE}
\epsi = \kappa \overline{\vareps}^{1/\alpha+1/2},
\qquad \kappa=\sqrt{\frac{mR^2V_0}{\hbar^2}}\;.
\ee
In particular, one can express the classical quantities in
(\ref{units})
through their values at $\epsi=\kappa$ ($E=V_0$, or 
$\epsi=1$) and 
put $\kappa=1$ because there is no dependence on $\kappa$ in all
final results for the trace formula, 
\bea\l{scaliolk}
&\mathcal{I}_i=\mathcal{I}_i(1)\epsi,\qquad
\Lambda=\Lambda(1)\epsi\;,
\\
& \overline{\omega}_r^{-1}=
\overline{\omega}_r^{-1}(1)\epsi^{(2-\alpha)/(2+\alpha)},\quad
\overline{K}=\overline{K}(1)/\epsi\;, \nonumber
\eea
where $\mathcal{I}_i(1)$,
$\Lambda(1)$, $\overline{\omega}_r^{-1}(1)$ and $\overline{K}(1)$ are
the corresponding quantities of (\ref{scaliolk}) taken
at ``the scaled energy'' $\epsi=1$. Therefore, due to
the scaling invariance (\ref{scaling}) and (\ref{scaliolk}) we need
to calculate these classical dynamical quantities only at one
value of the energy $\epsi=1$ or $E=V_0$. In the following,
we shall omit the argument
$\epsi=1$ everywhere for simplicity.

The energy surface $\mathcal{I}_r$ for the potential (\ref{potenra})
can be expressed explicitly in terms of the frequencies $\overline{\omega}_r$
and $\overline{\omega}_\varphi$ (see left identity in (\ref{percond}))
\cite{MAFptp2006},
\begin{equation}
\mathcal{I}_r = \frac{2 \alpha}{\alpha+2}\,\overline{\omega}_r^{-1} - 
\Lambda \,f(\Lambda)\;. 
\label{actra}
\end{equation}
This relation is useful, in particular, for
the derivation of the curvature $K_{\rm D}$ (\ref{curvraD}) for diameters
as the limit of $\overline{K}(\Lambda)$ (\ref{curv}) at $\Lambda \to 0$.
By differentiating the identity (\ref{actra}) term by term 
over $\Lambda$ and using the definition for the ratio
of frequencies $f(\Lambda)$ (\ref{percond}), for the 
curvature (\ref{curv}) one finally obtains (\ref{curvraD}), 
see \cite{MAFptp2006} for details.
%

\subsection{JACOBIAN CALCULATIONS\\ 
FOR CIRCULAR ORBITS}
\l{appA4}

For the calculation of the Jacobian 
$\mathcal{J}_{M{\rm C}}^{(p)}$ (\ref{jacpC}),
one first transforms it to the invariant form. 
Using the general properties of the Jacobian transformations, for the
Jacobian $\mathcal{J}_{M{\rm C}}^{(p)}$ (\ref{jacpC}), one
obtains
\be\l{jacpCtrans}
\mathcal{J}_{M{\rm C}}^{(p)}=
-\left(\frac{\partial^2 \Phi_{\rm CT}}{\partial L^2}\right)^{}_{M{\rm C}}\;
\frac{\left(\partial I_r/\partial p_r^\prime\right)_{M{\rm C}}^2}{
\left(\partial I_r/\partial L \right)_{M{\rm C}}^2}\;,
\ee
where
\be\l{d2fiL2}
\left(\frac{\partial^2 \Phi_{\rm CT}}{\partial L^2}\right)^{}_{M{\rm C}}=
2 \pi M K_{\rm C},
\ee
$K_{\rm C}$ is the curvature (\ref{curv}) for the circular orbits, 
see (\ref{curvraC}) \cite{MAFptp2006}. 
According to the Legendre transformations (\ref{actionp}), 
as applied for the radial 
action $I_r$,
\be\l{irleg}
I_r=\int_{r^{\prime}}^{r^{\prime\prime}} p_r \d r=
-\int_{p_r^{\prime}}^{p_r^{\prime\prime}} r \d p_r + p_r^{\prime\prime}r^{\prime\prime}-
p_r^{\prime}r^{\prime}\;,
\ee   
for the derivative $\left(\partial I_r/\partial p_r^{\prime}\right)^{}_{C}$, 
one finds
\be\l{dirdprC}
\left(\frac{\partial I_r}{\partial p_r^{\prime}}\right)_{\rm C}=
\left(r^{\prime\prime}
 \frac{\partial p_r^{\prime\prime}}{\partial p_r^{\prime}}\right)^{}_{\rm C}
=r^{}_{\rm C}\;.
\ee
Using also the expression (\ref{percond}) for 
$\left(\partial I_r/\partial L \right)^{}_{M{\rm C}}$, one obtains
\be\l{dirdLC}
\left(\frac{\partial I_r}{\partial L}\right)^{}_{M{\rm C}}=
\left(\frac{\omega^{}_{\theta}}{\omega_{r}}\right)_{M{\rm C}}=
\frac{\omega^{}_{\rm C}}{\Omega_{\rm C}}=\frac{1}{\sqrt{\alpha+2}}\;,
\ee
see (\ref{omtcra}) for the frequency of motion of a
particle along the C PO $\omega^{}_{\rm C}$ and 
(\ref{Omrcra}) for the radial frequency $\Omega_{\rm C}$, also for their
ratio. Collecting 
(\ref{d2fiL2}), (\ref{dirdprC}) and (\ref{dirdLC}) in  
(\ref{jacpCtrans}), for the Jacobian
$\mathcal{J}_{M{\rm C}}^{(p)}$ (\ref{jacpC}), 
one results in (\ref{jacpCres}).

The arguments 
(\ref{ZplimCfin}) of the error functions in the amplitude (\ref{amp2ispC})
In terms of the dimensionless curvature and angular moment 
for the C orbits, 
see (\ref{units}) and (\ref{scaliolk}),
are given by
\bea\l{ZplimCfin1}
&&\hspace{-0.9cm}{\mathcal Z}_{p\;{M{\rm C}}}^{+}\!=\!
\Lambda_{\rm C}(1)\;\sqrt{-i \pi\,
(\alpha+2) M \overline{K}_{\rm C}(1)\; \epsi}\,,\\
&&{\mathcal Z}_{r\;{M{\rm C}}}^{-}\!=\!
-\sqrt{\frac{F_{M{\rm C}}\; \epsi}{
4 i \pi (\alpha+2)\, M \overline{K}_{\rm C}(1)}}\,. \nonumber
\eea
%

\subsection{THE JACOBIAN FOR \\
 DIAMETER TRACE FORMULAS}
\l{appA5}

For the calculation of the Jacobian $J_\perp^{(p)}$ (\ref{jacpD}),
we first transform it to the invariant Jacobian 
$\partial \Theta^{\prime\prime}/\partial L $ (see \cite{annphysik97}) 
through the relations,
\be\l{jacrelation1}
\delta y^{\prime\prime}=r^{\prime\prime} \delta \Theta^{\prime\prime}, \qquad
\delta L=r^\prime \delta p_y^{\prime}\;,
\ee
by using the standard Jacobian transformations as
\be\l{jacrelation2}
\hspace{-0.4cm}J_\perp^{(p)}=
\left(\frac{\delta y''}{\delta p_y'}\right)^{}_{\rm PO}\!
=\!\left(\frac{\delta y^{\prime\prime}}{\delta \Theta^{\prime\prime}}
\frac{\delta \Theta^{\prime\prime}}{\delta L}
\frac{\delta L}{\delta p_y^{\prime}}\right)^{}_{M{\rm D}}.
\ee
The Jacobian 
 $(\delta \Theta^{\prime\prime}/\delta L)^{}_{M{\rm D}}$
at the closed diameter orbit $M$D with $M$ repetition number
is invariant independent of the spacial coordinates, in particular of the 
radial coordinate $r^\prime=r^{\prime\prime}=r$. In \cite{annphysik97} a 
more general case of the axial symmetrical potential was considered
for the calculation of the Jacobian 
$(\delta \varphi''/\delta L_z)^{}_{\rm PO}$ 
at the PO with $\varphi''$ being the
azimuthal angle and $L_z$ the projection of the angular momentum of the 
particle onto a symmetry $z$ axis. Using the
axial symmetry, it was shown that this Jacobian is invariant 
independent of
the spacial coordinates. The specific expression for 
$(\delta \varphi''/\delta L_z)^{}_{\rm PO}$ was obtained for the POs in the 
spheroidal cavity. In order to derive the specific expression for
the Jacobian $(\delta \varphi''/\delta L_z)^{}_{\rm PO}$ in the middle
of (\ref{jacrelation2}) for $J_\perp^{(p)}$, let us use the solution
of the classical equations of motion (\ref{tetar1}) for $\theta(r)$  and 
the second identity in the middle of (\ref{percond}),
\be\l{jacinv}
\delta \Theta^{\prime\prime} =
- 2\pi \delta \frac{\d I_r(\vareps,L)}{\d L}=
- 2\pi M K_{\rm D} \delta L\;,
\ee
where $K_{\rm D}$ is the curvature (\ref{curvraD}) for the diameter PO.
Factor two takes into account that we have to calculate the Jacobian 
for the transformation of the variation of the perpendicular momentum 
$\delta p_y'$ to the variation 
of the final perpendicular momentum through the period $T$ of the particle 
motion along the diameter which differs from the period of the radial motion
$T_r$ by the factor 2, $T=2 T_r =2 \pi/\omega_r$:
\be\l{period}
T= \frac{2 \pi n_\theta}{\omega_\theta}=\frac{2 \pi n_\varphi}{\omega_\varphi}=
\frac{2 \pi n_r}{\omega_r}\;.
\ee
For the diameters, one has $n_\theta=n_\varphi=1$ and $n_r=2$.
Therefore, the Jacobian 
$\left(\delta \Theta^{\prime\prime}/\delta L\right)^{}_{M{\rm D}}$ takes 
the invariant form:
\be\l{invjac}
\left(\frac{\delta \Theta^{\prime\prime}}{\delta L}\right)^{}_{M{rm D}}
=-2\pi M K_{\rm D}\;.
\ee
From (\ref{jacrelation1}), (\ref{jacrelation2}) and (\ref{invjac}),
one finally arrives at (\ref{jacpD}) for the Jacobian
$J_\perp^{(p)}$ (\ref{jacrelation2}).

\subsection{AMPLITUDE OF THE DIAMETER \\ 
CONTRIBUTIONS}
\l{appA6}

The amplitude of the diameter contribution into the PO sum in the trace formula
(\ref{dengenD}) is proportional to the integral given by
\be\l{integrald}
I(M\zeta)= \int_0^1 \d u \mathcal{A}(u)\;e^{i M\zeta \Phi(u)}\;,
\ee
where
\be\l{phiu}
\Phi(u)= u^2(1-u^\alpha)\;,
\ee
\be\l{ampu}
\mathcal{A}(u)= \left(1-u^\alpha\right)^{-1/2}\;.
\ee

Let us evaluate this integral by using the ISPM, i.e. expanding 
the phase and 
pre-exponent amplitude factor near the stationary points $u_i^\ast$ in powers of
$u-u_n^\ast$ ($n$ numbers the stationary points) as solution of the following
equation:
\be\l{phasu}
\Phi^\prime(u) \equiv 2 u -(\alpha+2) u^{\alpha+1}=0\;.
\ee
For any $\alpha \geq 2$ under our consideration and a finite
$\zeta$, one has
the two stationary points:
\be\l{statpointsun}
u_1=\left(\frac{2}{\alpha+2}\right)^{1/\alpha},\qquad u_2=0\;.
\ee
Expanding the exponent phase $\Phi(u)$ and the pre-exponent 
amplitude $\mathcal{A}(u)$ of the integrand (\ref{integrald}) near the 
stationary points $u_n$ [$n=1,2$, (\ref{statpointsun})] in power 
Taylor series, one finds
\be\l{phiexp}
\Phi(u)=\Phi(u_n) + \frac12 \Phi^{\prime\prime}(u_n)\;(u-u_n)^2 + \cdots\;,
\ee
\be\l{ampexp}
\mathcal{A}(u)=\mathcal{A}(u_n)+\cdots\;.
\ee
Note that in the limit $\alpha \rightarrow \infty$ one has the two close 
stationary points, that is for the catastrophe situation like caustic
and turning points 
of Fedoriuk (\cite{fedoryuk_pr,maslov,fedoryuk_book1,MAFptp2006}) and 
the bifurcation points because they are at the boundary of the integration
region $u=1$. Therefore, following mainly \cite{MAFptp2006}, we 
should expand the phase and
amplitude up to higher order terms keeping the final integration limits.

According to the simplest ISPM, one 
can evaluate the
integral (\ref{integrald}) for large values of $|\zeta|$  
through the error functions
(\ref{errorf}). Then, we arrive at (\ref{erfampu}).
In these derivations, we used
\be\l{curvu}
\hspace{-0.2cm}\Phi_1''=\Phi''(u_1)\!=\!-2\alpha\;,\quad 
\Phi_2''=\Phi''(u_2)\!=\!2\;.
\ee
We transformed the integration variable $u$ to $z_1$ for the evaluation of
the contribution of the stationary point $u_1$ and to $z_2$ for  $u_2$
by
\bea\l{transformuz}
z_1&=&\sqrt{-i M\zeta \Phi_1''/2}(u-u_1)\;,\\
 z_2&=&\sqrt{-i M\zeta \Phi_2''/2}\; u\;. \nonumber
\eea
The integral $I(M\zeta)$ (\ref{integrald}) 
can be expressed in terms of the
error functions (\ref{errorf})  by (\ref{erfampu}) 
with the arguments (\ref{argerf}).

We show also several other helpful expressions derived as explained in the text
through the dimensionless quantities:
\be\l{rmaxa}
r^{}_{\rm max}=R \left(E/V_0\right)^{1/\alpha}=R \epsi^{2/(\alpha+2)},
\ee
\bea\l{zetasc}
\zeta=-2\pi \overline{K}_{\rm D}(1) \epsi\;, \\ 
(\kappa=\sqrt{mR^2V_0/\hbar^2}=1)\;, \nonumber
\eea
\bea\l{freq1}
\frac{1}{\omega_r}&=& \frac{1}{\overline{\omega}_r(1)}\;
\sqrt{\frac{mR^2}{V_0}}\epsi^{(2-a)/(2+a)}\;,\\
\frac{1}{\overline{\omega}_r(1)}&=&\frac{\Gamma\left(1+1/\alpha\right)}{
\sqrt{2 \pi}\Gamma\left(1/2+1/\alpha\right) }\;. \nonumber
\eea
The trace formula (\ref{dengenD}) with the amplitude (\ref{ampgenD})
can be written also in terms of the scaled quantities by using Appendix 
A6,
\be\l{ampdendsc}
{\cal A}_{M{\rm D}}^{(2)}=
-\frac{
m R^2 \epsi^{4/(\alpha+2)}\sqrt{2}}{2 \pi^2 i \hbar 
\overline{K}_{\rm D}(1) M}\,
\left[I(M\zeta) -I(0)\right],
\ee
\be\l{Iu0}
I(0)=\int_0^1 \frac{\d u}{\sqrt{1-u^\alpha}} =
\frac{\sqrt{\pi} \Gamma\left(1+ 1/\alpha\right)}{
\Gamma\left(1/2 + 1/\alpha\right)}\;.
\ee 
%

\subsection{STABILITY MATRIX FOR THE \\ 
A ORBIT}
\label{appA7}

For small energies $e$, the trace $\Tr\mathcal{M}_A$
can be expressed through Mathieu functions by using a general method
of solving Hill's equation for the Poincar\'e coordinate  $x(t)$ perpendicular 
to the A orbit directed along the $y$ axis. 
The perturbation $x(t)$ (in scaled variables (\ref{scaling})) near the 
A orbit is determined by Hill's equation
(\ref{eqmot}) for the HH Hamiltonian (\ref{scaling}),
\be\l{hilleq}
\ddot{x}(t) +\left[1+2 y_{A}(t)\right] x(t)=0,
\ee
where $ y_{A}(t)$ is the periodic solution for the A orbit 
\cite{bmtjpa2001,kaidelbrack,fmbpre2008},
\bea\l{ya}
y_A&=&y_1 + (y_2-y_1) {\rm sn}^2(z,k),\\ 
z&=&a_kt+F(\varphi,k)\;,\nonumber
\eea
${\rm sn}(z,k)$ is the Jacobi elliptic function \cite{byrdbook}
with argument $z$; its modulus $k$ and the constant $a_k$ are given by
\be\l{ka}
k=\sqrt{\frac{y_2-y_1}{y_3-y_1}}, \qquad a_k=\sqrt{\frac{y_3-y_1}{6}}; 
\ee
$y_1$ and $y_2$
are the lower and upper turning points,
\bea\l{yn}
y_1&=&\frac12 - \cos\left(\frac{\pi}{3}-\frac{\phi}{3}\right),\\ 
y_2&=&\frac12 - \cos\left(\frac{\pi}{3}+\frac{\phi}{3}\right),\nonumber\\
y_3&=&\frac12 + \cos\left(\frac{\phi}{3}\right), \qquad \cos\phi 
=1-2 e\;. \nonumber
\eea
 $F(\varphi,k)$ is the incomplete elliptic integral of first
kind as a function of 
\bea
&&\hspace{-1.5cm}\varphi=\arcsin \{[(y_0-y_1)/(y_2-y_1)]^{1/2}\}\;,\\ 
&&y_0=y_A(t=0) \nonumber
\eea
 is the initial value. Using the Fourier expansion of ${\rm sn}^2(z,k)$, 
one has \cite{milne}
\bea\l{foursn2}
&{\rm sn}^2(z,k)=
\frac{K(k)-E(k)}{k^2 K(k)} -\frac{2 \pi^2}{k^2 K^2(k)}\times \\
&\times\sum_{n=1}^{\infty} 
\frac{n s^n}{1-s^{2 n}}\;\cos\left(\frac{\pi n z}{K(k)}\right), \nonumber
\eea
where $s=\exp\left[-\pi K(\sqrt{1-k^2})/K(k)\right]$ is 
Jacobi's Nome \cite{byrdbook},
$K(k)$ and $E(k)$ are the complete elliptic integrals of first and second
kind, respectively. 

For small energies $e$ where Jacobi's Nome $s$ is small, 
$s \rightarrow k^2/16 \approx \sqrt{e}/(12 \sqrt{3})$ 
for $e \rightarrow 0$ [$k \rightarrow 0$, see (\ref{ka})], the convergence of
the Fourier series (\ref{foursn2}) is fast even for $e \siml 0.8$
($s\siml 0.04$).
For such energies, we may truncate the Fourier series approximately, keeping
only the first ($n=1$) harmonic term. After substitution of (\ref{ya})
with the expansion (\ref{foursn2}), a simple transformation of the time  
variable and the parameters in (\ref{hilleq}) leads to the standard 
Mathieu equation: 
\be
\frac{{\rm d}^2}{{\rm d} \tau^{2}}x(\tau) +
\left[a -2 b \cos\left(2 \tau\right)\right]x(\tau) =0\,,
\ee
with 
\bea\l{parammathieu}
&\tau= \pi z/[2 K(k)]\,,\\
&a=\left(\frac{2 K}{\pi a_k}\right)^2
\left\{1+2\left[y_1 + (y_2-y_1)\frac{K-E}{k^2 K}\right]\right\},
\nonumber\\
&b= 8 s(y_2-y_1)/[k^2 a_k (1-s^2)]\,. \nonumber
\eea
The solution of this second-order ordinary differential equation can be 
sought as a linear superposition of the fundamental set of the even  
$M_{C}(a,b,\tau)$ and odd $M_{S}(a,b,\tau)$  
Mathieu functions with arbitrary constants $C_1$ and 
$C_2$:
\be\l{mathsol}
x(\tau)= C_1 {\rm M}_{C}(a,b,\tau) + 
C_2  {\rm M}_{S}(a,b,\tau)\,.
\ee
Applying to (\ref{mathsol}) the boundary conditions for calculations 
of the stability matrix elements $\mathcal{M}_{xx}$ and  
$\mathcal{M}_{\dot{x}\dot{x}}$  as in \cite{fmbpre2008}, one 
obtains the constants $C_1$ and $C_2$ and the following diagonal 
matrix elements:
\bea\l{matrixelem}
&\hspace{-1.0cm}\mathcal{M}_{xx}=
\frac{x(T)}{x(0)}\Big|_{\dot{x}(0) \rightarrow 0}=
\frac{{\rm M}_{S,0}'{\rm M}_{C,T} - {\rm M}_{C,0}'{\rm M}_{S,T}}{
{\rm M}_{C,0}{\rm M}_{S,0}' - {\rm M}_{S,0}{\rm M}_{C,0}'}\,,\\
&{}\hspace{-4.0ex}\mathcal{M}_{\dot{x}\dot{x}}=
\frac{\dot{x}(T)}{\dot{x}(0)}\Big|_{x(0) \rightarrow 0}=
\frac{{\rm M}_{S,T}'{\rm M}_{C,0} - {\rm M}_{C,T}'{\rm M}_{S,0}}{
{\rm M}_{C,0}{\rm M}_{S,0}' - {\rm M}_{S,0}{\rm M}_{C,0}'}\,, \nonumber
\eea
where primes means the partial derivatives of the Mathieu
functions $M_{C}(a,b,\tau)$ and $M_{S}(a,b,\tau)$ 
with respect to $\tau$. The lower indices $0$ and $T$ show the values
at the initial $t=0$ and final $t=T$ times, and $T=T_A=2 K(k)/a_k$ is the 
period of motion of the particle along the A orbit.
For the trace $\Tr\mathcal{M}_A$, one finally finds 
\be\l{trma}
\Tr\mathcal{M}_A=\mathcal{M}_{xx}+\mathcal{M}_{\dot{x}\dot{x}}\,,
\ee
with the diagonal matrix elements given in (\ref{matrixelem}).

For comparison, we recall the solution for the trace $\Tr\mathcal{M}_A$ 
near the saddle $e \rightarrow 1$ obtained in \cite{bkwf2006,fmbpre2008} 
in terms of the Legendre functions by using in (\ref{ya}) the approximation
of  the Jacobi elliptic function, ${\rm sn}(z,k) \approx \tanh(z)$, i.e.,
by the zero-order term of its expansion near the saddle in powers of $1-k^2$:
\bea\l{snexpsad}
&{\rm sn}(z,k) \approx \tanh z\times\\
&\times\left[1 +
\frac14 (1-k^2)\left(1 - \frac{z}{\sinh z \cosh z}\right)\right].\;\nonumber
\eea
As shown in \cite{bkwf2006,fmbpre2008}, the trace $\Tr\mathcal{M}_A$ 
is in this approximation in good agreement with the numerical results 
\cite{bmtjpa2001} near the saddle $e \rightarrow 1$.

Generally speaking, for a more general solution,  
it is difficult to take into account exactly the next term of the expansion 
(\ref{snexpsad}) to get a simple analytical result similar to that presented 
explicitly in \cite{fmbpre2008}. 
However, we may use the approximate constant $r$ for the square brackets
in (\ref{snexpsad}), which effectively takes into account the correction to
$\tanh z$,
\be\l{rconst} 
r \approx 1+ r_{\rm corr}(1-k^2),\qquad r_{\rm corr}=1/4\;.
\ee
Within this approximation, one has again the result in 
terms of the Legendre functions $P_{\nu}^{\mu}$ and $Q_{\nu}^{\mu}$
with complex indices $\nu$ and $\mu$ depending on the energy $e$
\be\l{numu}
\mu\!=\!i \sqrt{A+B}\,, \,\,\, \nu\!=\!(-1+i\sqrt{4 A -1})/2\,,
\ee
where $B$ is the same as in \cite{fmbpre2008} but $A$ contains the 
additional constant factor $r$:
\be\l{ab}
A=12\,r\,k^2\,, \qquad B=(1+2 y_1)/a_k^2\;,
\ee
corresponding at $r$=1 (or $r_{\rm corr}=0$ in our notations)
to the results in \cite{fmbpre2008}.

The comparison of numerical calculations 
\cite{bracktanaka,bmtjpa2001} with our analytical results for the trace of the 
stability matrix $\Tr\mathcal{M}_A$ in the case of the A orbit was presented
in \cite{kkmabPS2015}.
The solution for  $\Tr\mathcal{M}_A$ in terms of
the Mathieu functions is in good agreement with the exact numerical 
results even at energies $e \siml 0.8$. We show 
in \cite{kkmabPS2015} also another approximation 
in terms of Legendre functions with the indices (\ref{numu}), 
improved at finite and small energies $e$ through the constant $A$ (\ref{ab}) 
with $r$ given in (\ref{rconst}) as compared to the result ($r=1$) obtained
earlier near the saddle [i.e., using only the leading term in the expansion 
(\ref{snexpsad}) for $e-1 \ll 1$] \cite{fmbpre2008}. Through a modification of 
only one constant $r$ (\ref{rconst}), one has a remarkable agreement between 
this improved Legendre approximation and the numerical results everywhere
from the saddle point $e$  to the harmonic oscillator limit 
$\Tr\mathcal{M}_A \rightarrow 2$ for $e \rightarrow 0$.

This approximation can be slightly improved changing the constant $r_{corr}$
in (\ref{rconst}) from $r_{\rm corr}=1/4$ ($z \rightarrow \infty$) to about 2/9
of finite values of $z$. For small energies $e$ ($k \rightarrow 0$), 
one can, again, formally use (\ref{snexpsad}): the correction to $\tanh z$ can 
be neglected for small times ($z \ll 1$) because it gives the dominating 
contribution to $\Tr M_A$ [equation (\ref{hilleq}) with (\ref{ya}) becomes
approximately the same at small $z$], and $\Tr M_A \rightarrow 2$ 
in all analytical 
approximations, in agreement with the numerical results. In the limit 
$e \rightarrow 0$ the Legendre function approximation converges, indeed, to the 
analytical Mathieu function solution. 
Note also that this agreement with the numerical results
is not sensitive to a variation of the constant 
$r_{corr}$ around the analytical value (\ref{rconst}). 
The particle moving near the 
A orbit spends much more time near the saddle where the function of $z \propto 
t$ in the circle brackets (\ref{snexpsad}) is almost constant with respect to 
the remaining part of the trajectory. 
However, at small energies $e \rightarrow 0$, one finds a smaller 
time region ($z \ll 1$) where the correction in (\ref{snexpsad}) becomes 
negligible for $\Tr M_A$, such that all 
approximations 
have the same correct 
harmonic-oscillator limit $2$. Thus, a rather complicated function 
of time in the correction to the leading (hyperbolic tangent) term of the 
expansion of the Jacobi function (\ref{snexpsad}) can be reduced to a form 
involving the same Legendre functions as in \cite{fmbpre2008}, 
but with modified 
indices by the constant $r$ (\ref{rconst}) through (\ref{ab}).


\clearpage
\section*{CAPTIONS TO FIGURES:}
{
\parindent=-5mm
\leftskip=5mm

{\bf Fig.\ 1. 
} Poincar\'e surfaces of sections (PSS) of the 
scaled H\'enon-Heiles Hamiltonian $h$ (\ref{scaling});
left column: {\it (a), (b)} and {\it (c)} plots show the PSS
at $u=0$ for the energies 
$e=0.5$, $0.75$ and $1.0$, respectively; right column: 
{\it (d), (e)} and {\it (f)} graphics are given for $v=0$
at the same energies.

\vspace{0.2cm}

{\bf Fig.\ 2. 
} The scaled H\'enon-Heiles potential 
of the Hamiltonian (\ref{scaling}). 
{\it Left:} Equipotential contour lines are given in scaled energies $e$
in the plane $(u,v)$. The dashed lines are the symmetry axes. The three
shortest orbits A, B, and C (evaluated at the energy $e=1$) are shown by
the heavy solid lines. 
{\it Right:} Cut along $u=0$ shows a barrier. (After \cite{hhprl,sclbook}.) 

\vspace{0.2cm}

{\bf Fig.\ 3. 
} Quantum-mechanical (QM, solid), 
semiclassical (ISPM, dashed), and
Gutzwiller (GUTZ, dots) shell-corrections level density versus energy
$E$ (in units of $\hbar\omega$). Only the primitive POs A, B and C
are included in the semiclassical calculations,
the Gaussian averaging width is $\gamma=0.25 \hbar \omega$.

\vspace{0.2cm}

{\bf Fig.\ 4. 
} Quantum and semiclassical shell-correction energy 
$\delta U$ (\ref{escscl})
(in units of the Fermi energy $\vareps_F$) versus particle number 
parameter $N^{1/2}$, with
$N = 2\int_0^{\vareps_F} \d \vareps  \,g(\vareps)$. 
The same primitive POs are included as explained in the text.

\vspace{0.2cm}

{\bf Fig.\ 5. 
} {\it Left:} 
Schematic double-humped fission barrier of a typical actinide nucleus. 
Note the lowering of the outer barrier due to left-right asymmetric shapes. 
{\it Right:} 
Maximum probability amplitudes (schematic) of the two leading s.p.\ states 
responsible for the asymmetry effect (after \cite{gumni}).

\vspace{0.2cm}

{\bf Fig.\ 6. 
} Axially symmetric nuclear shapes in the $(c,h,\alpha)$ 
parametrization of
\cite{fuhi}. Dashed lines for $\alpha\neq 0$; the sequence with $h=\alpha=0$
corresponds to the shapes obtained in the LDM \cite{cosw}.

\vspace{0.2cm}

{\bf Fig.\ 7. 
} Plot of quantum-mechanical shell-correction energy 
$\delta U$
versus cube-root of particle number, $N^{1/3}$, and elongation $c$ 
(along $h=\alpha=0$) in the cavity model.
The contours lines are for constant values of 
$\delta U$ ({\it white:} 
positive values, {\it gray to black:} negative values.
The heavy lines indicate the loci of constant actions of the leading
POs. {\it Dashed-dotted lines:} meridional triangular orbits (3,1,1)s; 
{\it narrow lines:} diameter orbits in equatorial planes (2,1)EQ (solid) 
and in parallel perpendicular planes (2,1)AQ (dashed); 
{\it broad lines:} triangular orbits in equatorial planes (3,1)EQ (solid) 
and in parallel perpendicular planes (3,1)AQ (dashed). The 
horizontal dotted line at $N\simeq 180$ corresponds to the situation 
with the isomer minimum at the correct deformation \cite{fuhi} 
$c\simeq 1.42$ of the real nucleus $^{240}$Pu.

\vspace{0.2cm}

{\bf Fig.\ 8. 
} Fourier spectra of the fission cavity model with
$h=\alpha=0$ for five values of $c$:
amplitude of Fourier transform of the quantum
spectrum versus length $L$ (in units of $R_0$) of the 
classical POs. {\it Short arrows:} POs lying in planes 
orthogonal to symmetry axis; {\it long arrows:} POs lying 
in meridional planes containing the symmetry axis (labels as in
Fig.\ 7; see text for more details).

\vspace{0.2cm}

{\bf Fig.\ 9. 
} Perspective view of the semiclassical outer fission
barrier versus
elongation $c$ and left-right asymmetry $\alpha$ for $h=0$. To the
left, the shapes corresponding to the points A, B and C in the
deformation energy surface are displayed; the vertical solid 
(dashed) lines indicate the planes containing the stable (unstable)
POs.

\vspace{0.2cm}

{\bf Fig.\ 10. 
} Contour plots of the shell-correction energy 
$\delta U$ versus
$c$ and $\alpha$. {\it Upper panels:} for $h=0$, {\it lower panels:}
for $h=-0.075$. {\it Left panels:} results of quantum-mechanical
SCM calculations with realistic nuclear shell model potentials 
\cite{fuhi} (shown is the shell correction of the neutrons); 
{\it right panels:} semiclassical POT results with the fission 
cavity model described above. 

\vspace{0.2cm}

{\bf Fig.\ 11.
} The scaled periods $\tau^{}_{\rm PO}$ (horizontal axis) of
some short
periodic orbits PO plotted as functions of the power parameter
$\alpha$ (vertical axis).  Thin solid curves are circle orbits $M$C,
dashed green curves are diameter orbits $M$(2,1), and thick solid  
curves are polygon-like orbits $M(n_r,n_\varphi)$ $(n_r>2n_\varphi)$ which
bifurcate from the circle orbits $M$ C at the bifurcation points
indicated by open circles.

\vspace{0.2cm}

{\bf Fig.\ 12. 
} Moduli of the Fourier transform $|F(\tau)|$ of 
the quantum scaled-energy level density (\ref{fourierpower}) 
plotted for several values of $\alpha$.

\vspace{0.2cm}

{\bf Fig.\ 13. 
} Comparison of the  quantum-mechanical (QM, solid) 
and semiclassical [ISPM (dashed) and SSPM (dots)]
shell-correction scaled-energy
level density $\delta \mathcal{G}_\gamma(\epsi)$, see (\ref{scdentot}) and
(\ref{deltadenstotprlp}), divided by $\epsi$ [see (\ref{scldenstyra1})],
as function of the scaled-energy $\epsi$ for $\alpha=6.0$ 
and averaging 
parameter $\gamma=0.6$ [upper panel {\it (a)}]; contributions of the 
$\mathcal{K}=3$ (PISP, thick solid), circular (CISP, dots),
and diameter (DISP, approximately DSSP, dashed) 
POs [lower panel {\it (b)}].

\vspace{0.2cm}

{\bf Fig.\ 14. 
} Same as in Fig.\ 13 at $\alpha=6.0$ 
but for the averaging 
parameter $\gamma=0.2$.

\vspace{0.2cm}

{\bf Fig.\ 15. 
} Same as in Fig.\ 13 at $\alpha=6.0$ 
but for the averaging 
parameter $\gamma=0.03$.

\vspace{0.2cm}

{\bf Fig.\ 16. 
} Same as in Fig.\ 13 at $\alpha=7.0$ 
but for the averaging 
parameter $\gamma=0.6$ (without CISP POs).

\vspace{0.2cm}

{\bf Fig.\ 17. 
} Same as in Fig.\ 13 at $\alpha=7.0$ 
but for the averaging 
parameter $\gamma=0.2$ (without CISP POs).

\vspace{0.2cm}

{\bf Fig.\ 18. 
} Same as in Fig.\ 13 at $\alpha=7.0$ but for the averaging 
parameter $\gamma=0.1$ (without CISP POs).

\vspace{0.2cm}

{\bf Fig.\ 19. 
} Shell-correction energy $\delta  \mathcal{U}$ 
(scaled and divided by $\epsi^{}_F$) as function of the particle number 
parameter $N^{1/3}$ at $\alpha=6.0$ [upper panel {\it (a)}]; 
contributions of the 
$\mathcal{K}=3$ (PISP, thick solid), circular (CISP, dots),
and diameter (DISP, approximately DSSP, dashed) 
POs [lower panel {\it (b)}].

\vspace{0.2cm}

{\bf Fig.\ 20. 
} Same as in Fig.\ 19 but at $\alpha=7.0$ 
[upper panel {\it (a)}]; [lower panel {\it (b)}]: same as Fig.\ 19
but without CISP POs.

\vspace{0.2cm}

{\bf Fig.\ 21. 
} The equi-potential surfaces of the spheroidal
($\delta=0.5$, solid curve) and quadrupole ($\beta_2=0.4$, broken
curve) deformations in the meridian plane.

\vspace{0.2cm}

{\bf Fig.\ 22. 
} Single-particle level diagrams of the  
power-law potential model
with $\alpha=5.0$.  Scaled energy eigenvalues $\varepsilon_i$
in the PLP
potential (\ref{PLP}) are
plotted as functions of the deformation parameters: (a) 
for the spheroidal deformation $\delta$ (\ref{fspheroid}) 
and (b) for the quadrupole deformation  $\beta_2$ (\ref{fquadrupole}). 
Solid and
broken lines represent positive and negative parity levels, respectively.

\vspace{0.2cm}

{\bf Fig.\ 23. 
} Poincare\'{e} surfaces of section 
$(x,p_x)$ of the meridian-plane
trajectories in the 
power-law potential (\ref{PLP}) with $\alpha=5.0$ and
deformations shown in Fig.\ 21.

\vspace{0.2cm}

{\bf Fig.\ 24. 
} Some short meridian-plane orbits in the 
power-law potential (\ref{PLP}) 
with $\alpha=5.0$ and deformations shown in Fig.\ 21.

\vspace{0.2cm}

{\bf Fig.\ 25. 
} 
PO scaled periods
as functions of the deformation parameter $\delta$ for the
power parameter $\alpha=3.0$;  $M$X (long-dashed lines)
are those of the equatorial diameter POs, $M$Z (short-dashed ones)
are symmetry-axis diameter PO.
Dots indicate the
bifurcation points of the meridian bridge orbits, and their
scaled periods are shown by solid and dotted lines
for $M$C and for the other meridian bridges $M$B, respectively.
$M$EC (dash-dotted lines) are equatorial circular orbit,
whose bifurcations are omitted to avoid complication.

\vspace{0.2cm}

{\bf Fig.\ 26. 
} Illustration of the growth of the bridge orbit C with
increasing 
power parameter $\alpha$.  Scaled periods of the periodic
orbits C (meridian oval), X (equatorial diameter)
and Z (symmetry-axis diameter) are plotted as functions of
deformation parameter
$\delta$ for  
$\alpha=2.5$, 3.0 and 5.0.
Bifurcation points are indicated by the dots.

\vspace{0.2cm}

{\bf Fig.\ 27. 
} Color map of the Fourier amplitude $|F(\tau;\delta)|$ as function 
of $\delta$ and
$\tau$.  Lines represent the scaled periods $\tau^{}_{\rm PO}$ of the
classical periodic orbits as functions of the deformation parameter
$\delta$.  Dots indicate the bifurcation points. 

\vspace{0.2cm}

{\bf Fig.\ 28. 
} Contour map of the shell-correction energies 
in the $(\delta,N^{1/3})$ plane.  
The negative and positive shell-correction energies are shown by
the red (solid) 
and blue (broken) 
contour lines, respectively.  The thick
lines are the constant-action lines of the bridge orbit C.

\vspace{0.2cm}

{\bf Fig.\ 29. 
} Shapes of the equi-potential surfaces for three values of the
tetrahedral deformation parameter $\beta_{\rm td}=0.1$, 0.5 and 0.9.
The tetrahedron corresponding to
$\beta_{\rm td}=1$
is also drawn
with dotted lines in all panels.

\vspace{0.2cm}

{\bf Fig.\ 30. 
} Single-particle level diagram of
power-law potential model for
the tetrahedral deformation (a) $\alpha=4.0$ and (b) $6.0$.  Scaled energy
levels $\epsi_i$ are plotted as functions of the tetrahedral
deformation parameter $\beta_{\rm td}$.  Dotted, dashed and solid
lines represent the levels belonging to the A, E and F irreps of the
$T_d$ group.

\vspace{0.2cm}

{\bf Fig.\ 31. 
} Some short periodic orbits in the tetrahedral-like
deformed power-law potential model with $\alpha=6.0$.  The top and
middle 6 panels are for $\beta_{\rm td}=0.3$, and the bottom 3 panels
are for the written values of $\beta_{\rm td}$.  Thick solid lines
represent the orbits and the thick dashed lines represent their
projections onto $(x,y)$, $(y,z)$ and $(z,x)$ planes.  Thin dotted
lines represent the tetrahedron which has the same symmetry of the
equi-potential surface.

\vspace{0.2cm}

{\bf Fig.\ 32. 
} The scaled periods $\tau^{}_{\rm PO}$ of the classical
periodic orbits for the power parameter $\alpha=6.0$ as functions of
the tetrahedral-deformation parameter $\beta_{\rm td}$.  Dots indicate
the bifurcation points.  The bottom panel is the enlargement of the
top panel in the region indicated by the dotted rectangle.

\vspace{0.2cm}

{\bf Fig.\ 33. 
}  
Color map of the Fourier amplitude 
$|F(\tau;\beta_{\rm td})|$ of the
quantum scaled-energy level density in the $(\beta_{\rm td},\tau)$
plane.  Solid lines represent the scaled periods $\tau^{}_{\rm PO}$ of
the classical periodic orbits as functions of $\beta_{\rm td}$.  Dots
indicate their bifurcation points.

\vspace{0.2cm}

{\bf Fig.\ 34. 
} Shell-correction energies $\delta U$ plotted in
units of $V_0$ as functions of the cubic root of the particle number
$N^{1/3}$.  Solid lines represent quantum mechanical results.  Dashed
lines in the two bottom panels represent the semiclassical results
based on the Gutzwiller trace formula.
}

\widetext

\begin{figure} 
\begin{center}
\includegraphics[width=0.8\textwidth,clip]{fig01.eps}
\end{center}
\caption{
 }
\label{fig1}
\end{figure}
\begin{figure}
\begin{center}
\includegraphics[width=0.8\textwidth,bb=30 40 760 380,clip]{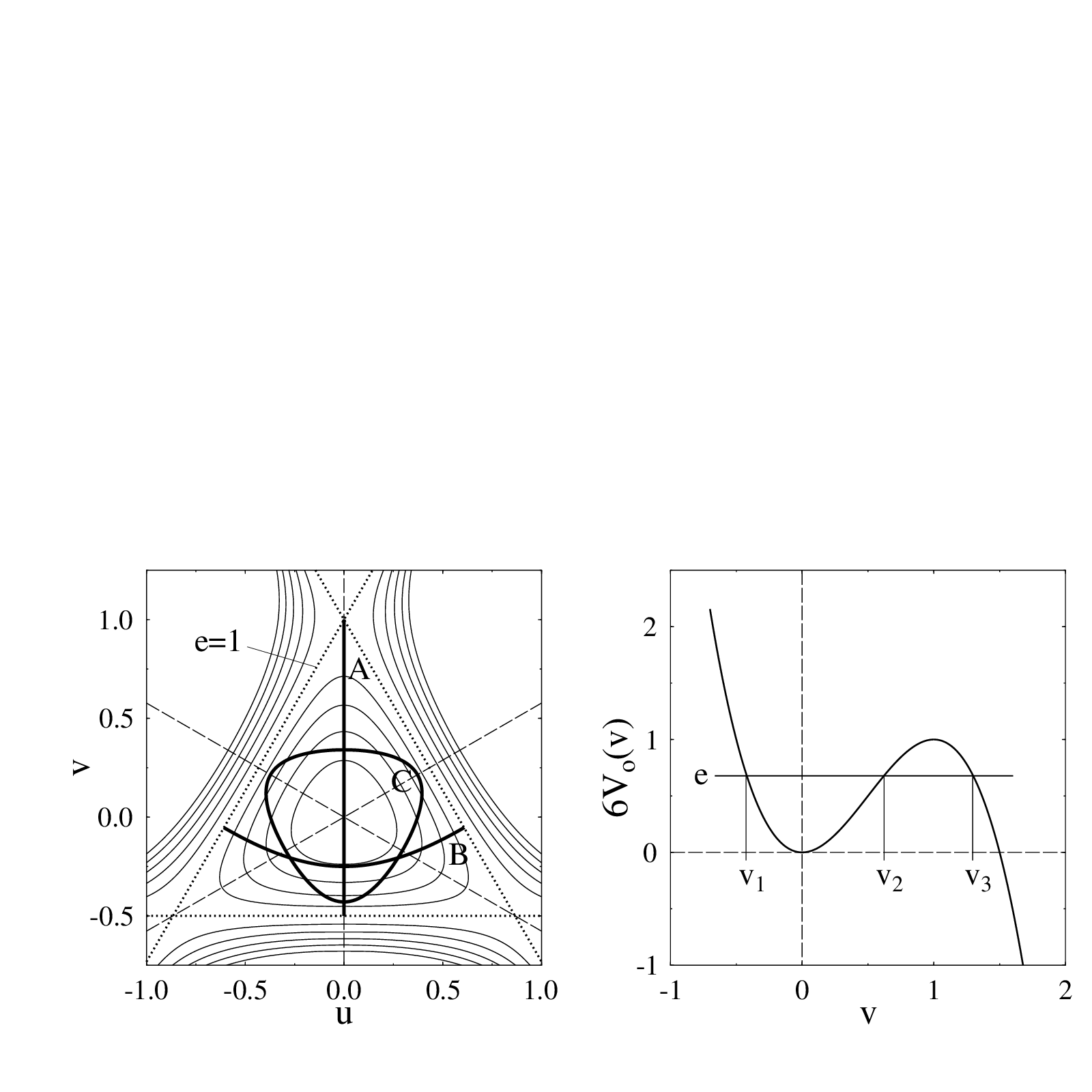}
\end{center}
\caption{ 
}
\label{fig2}
\end{figure}

%
\begin{figure}
\begin{center}
\includegraphics[width=0.8\textwidth,clip]{fig03.eps}
\end{center}
\caption{
}
\label{fig3}
\end{figure}
\begin{figure}
\begin{center}
\includegraphics[width=0.8\textwidth,clip]{fig04.eps}
\end{center}
\caption{
}
\label{fig4}
\end{figure}

%
\begin{figure}
\hspace{0.5cm}
\begin{center}
\includegraphics[height=0.75\textwidth,angle=270,clip=true]{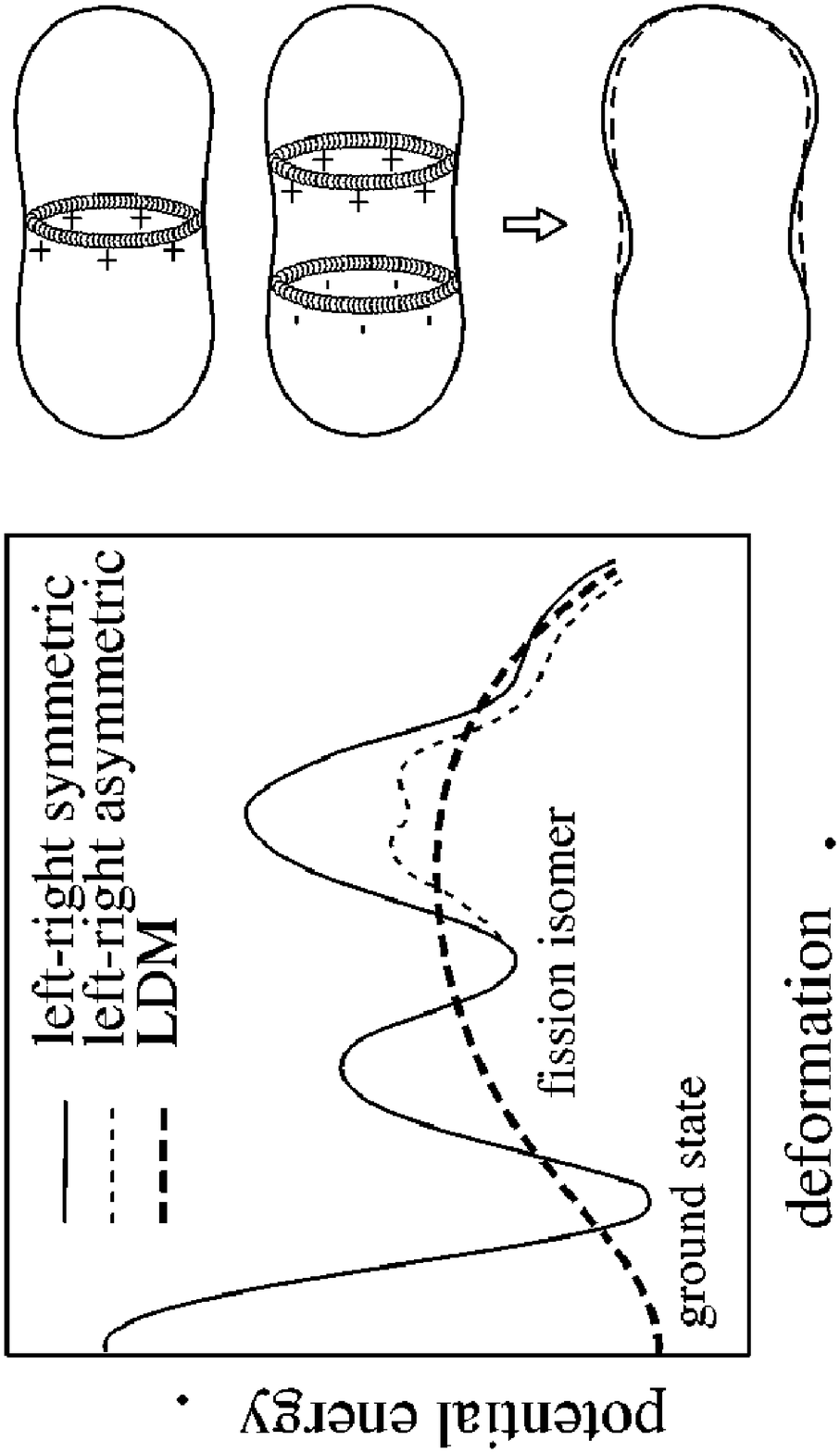}
\end{center}
\caption{
}
\label{fig5} 
\end{figure}

\begin{figure}
\hspace{0.5cm}
\begin{center}
\includegraphics[width=0.8\textwidth,clip=true]{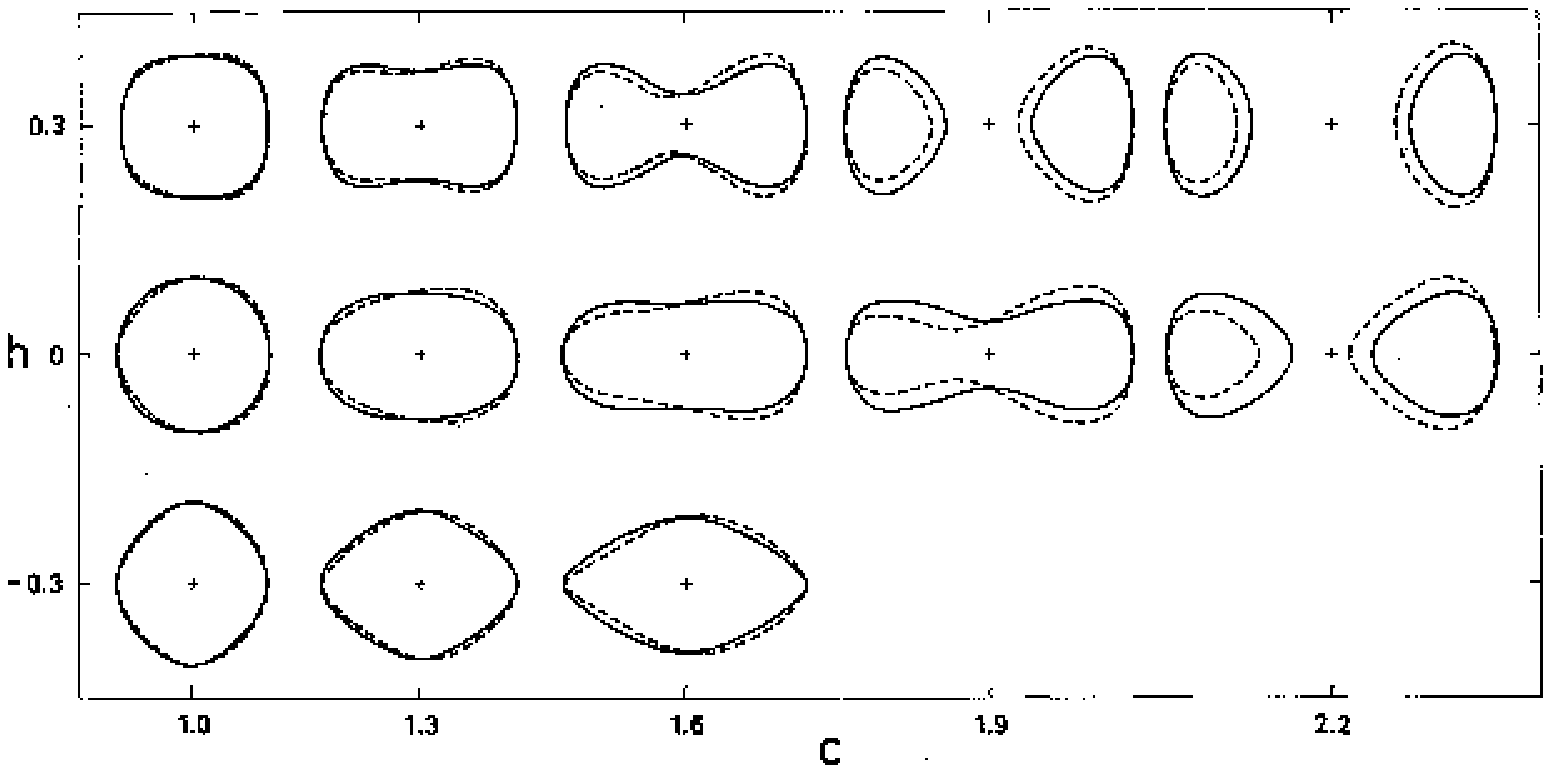}
\caption{
}
\label{fig6} 
\end{center}
\end{figure}
\begin{figure}
\begin{center}
\includegraphics[width=0.8\textwidth,clip=true]{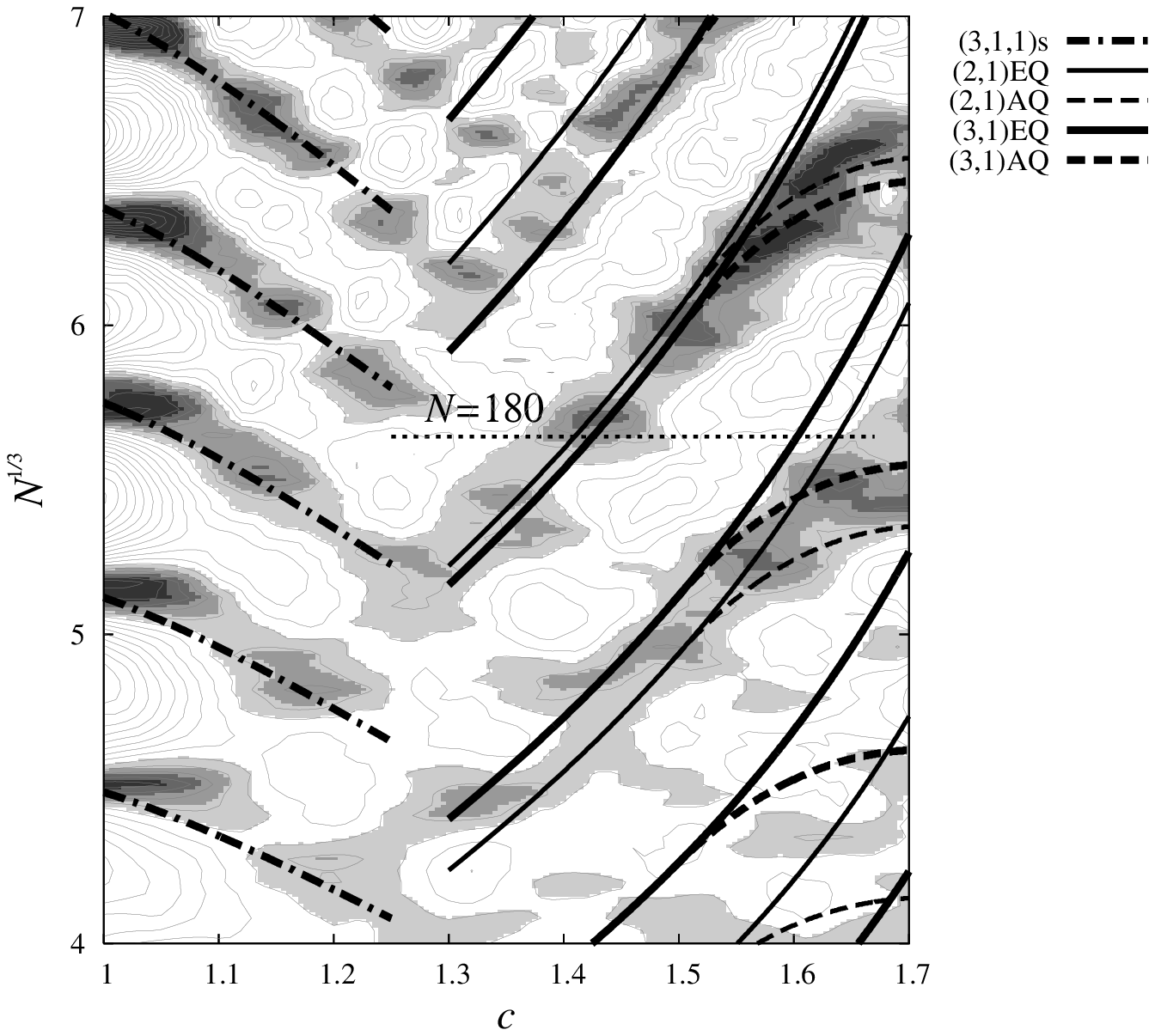}
\end{center}
\caption{
}
 \label{fig7}
\end{figure}
\begin{figure}
\begin{center}
\includegraphics[width=0.8\textwidth,clip=true]{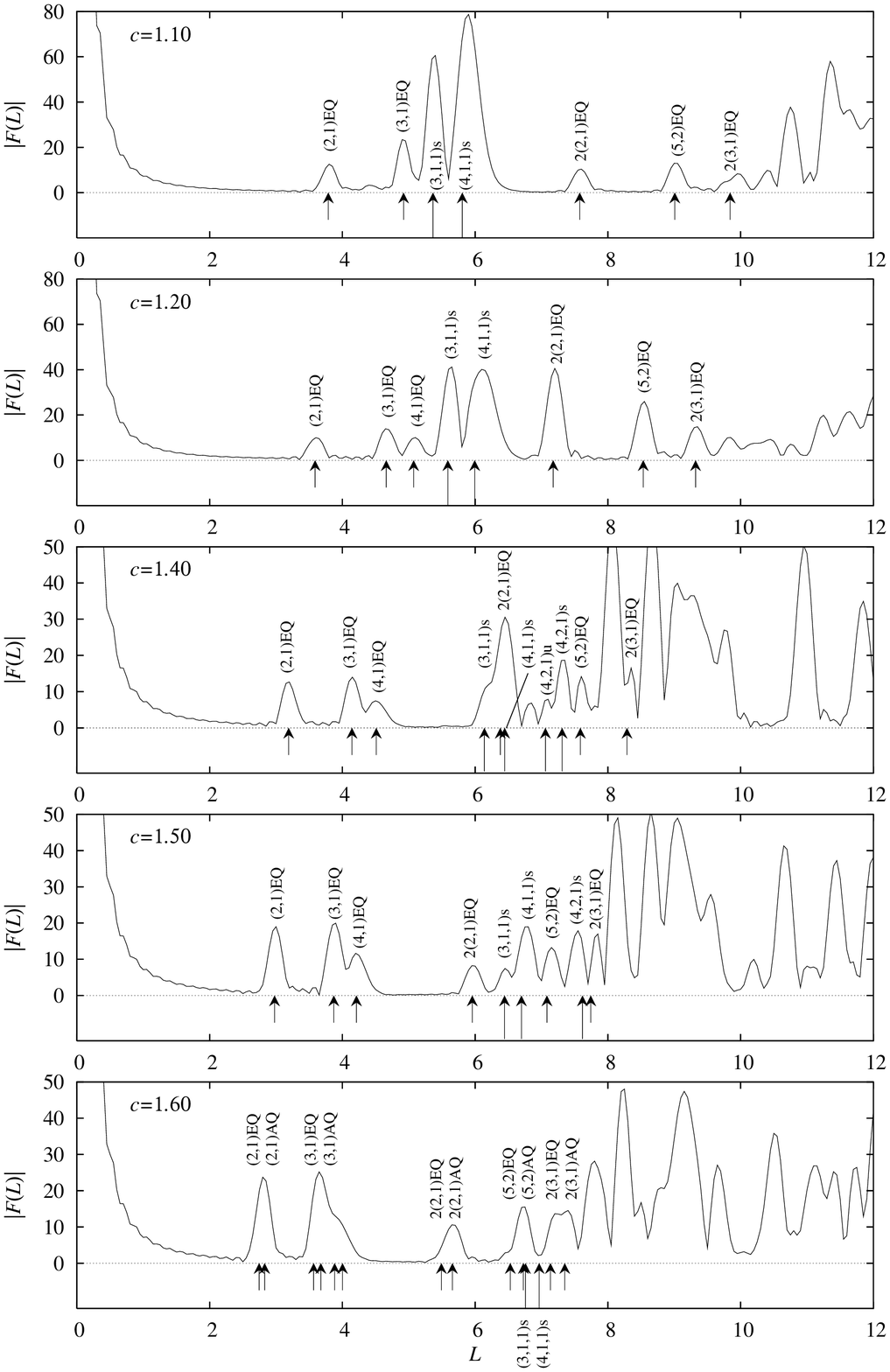}
\end{center}
\caption{
}
\label{fig8} 
\end{figure}
%

\begin{figure}
\begin{center}
\includegraphics[width=0.8\textwidth,clip=true]{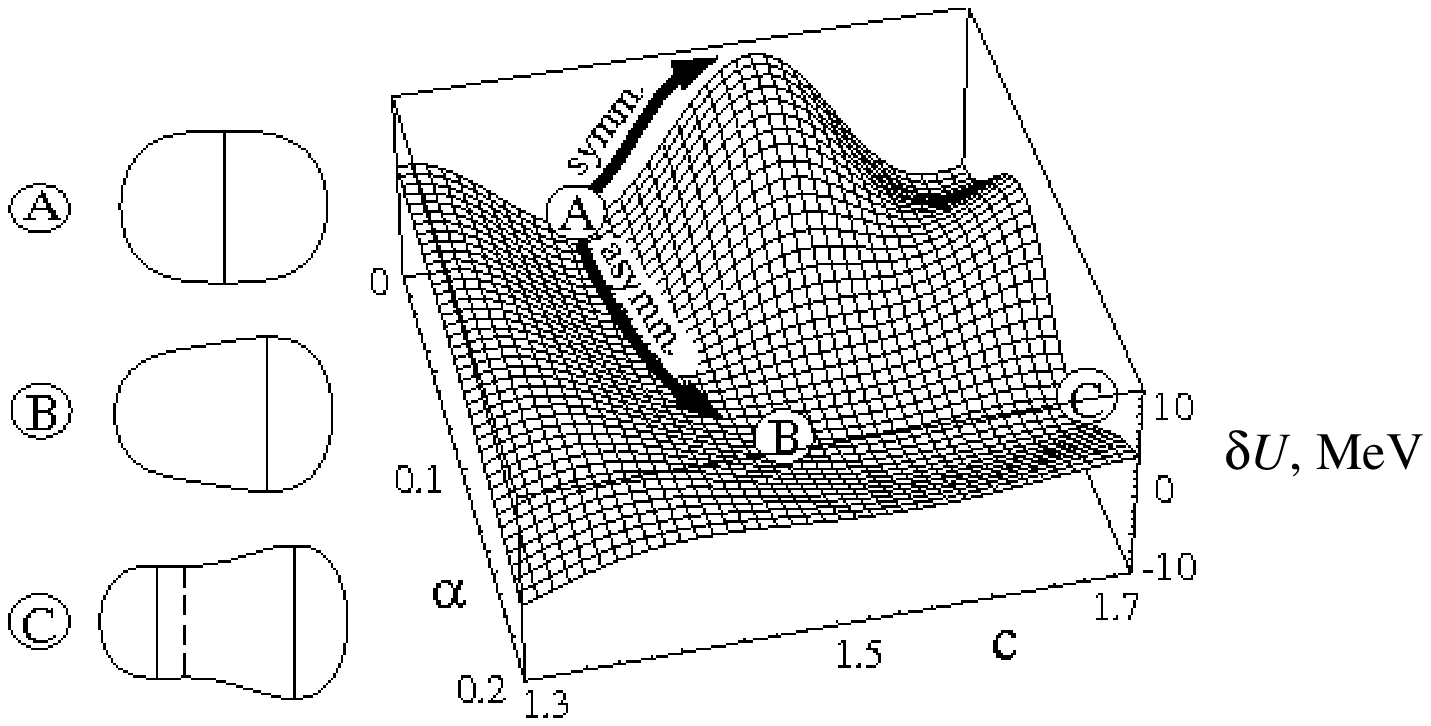}
\end{center}
\caption{
}
\label{fig9} 
\end{figure}
%

\begin{figure}
\begin{center}
\includegraphics[width=0.75\textwidth,clip=true]{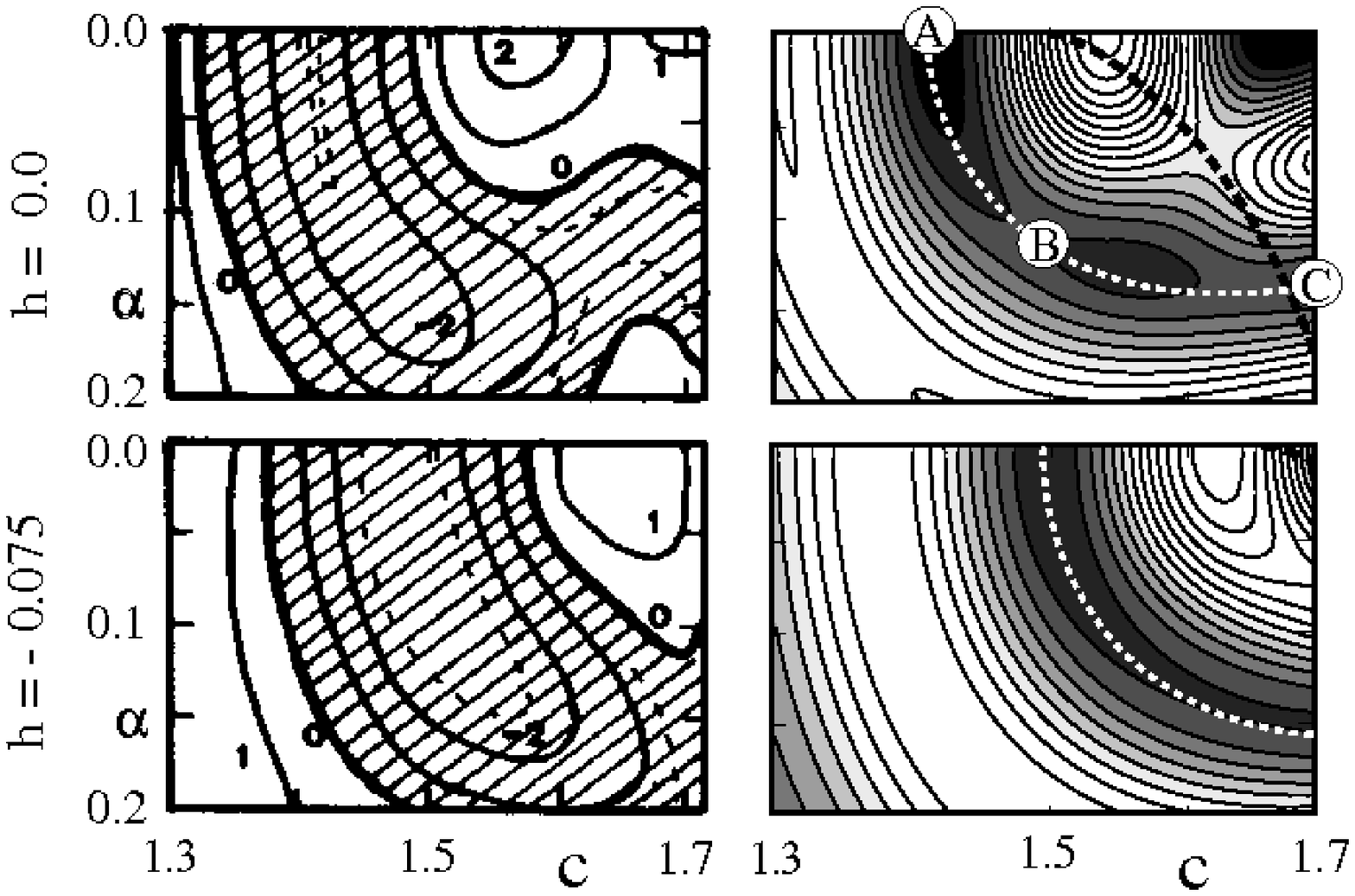}
\end{center}
\caption{
} 
\label{fig10} 
\end{figure}
%

\clearpage
\begin{figure}
\begin{center}
\includegraphics[width=0.8\textwidth,clip=true]{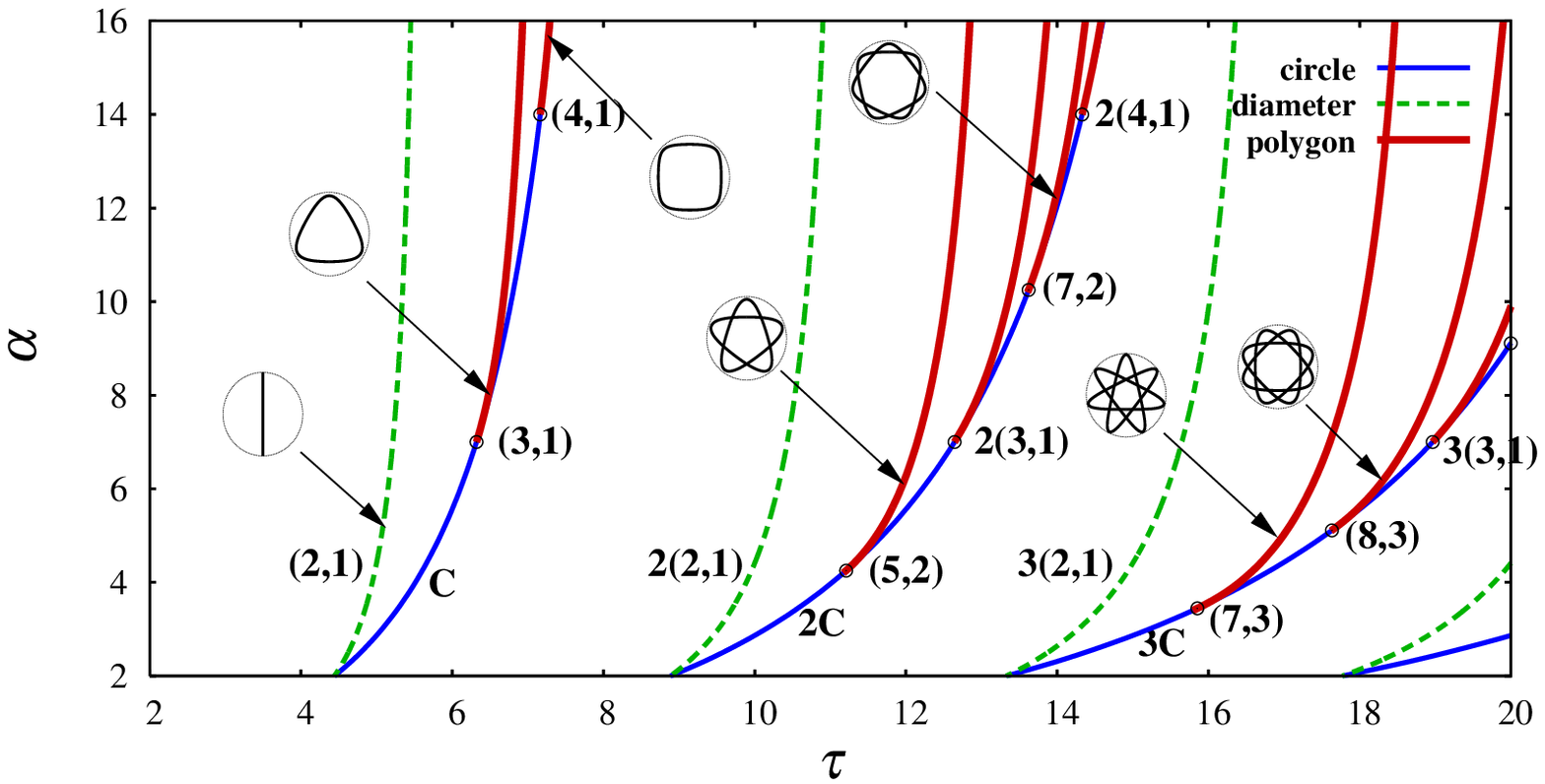}
\end{center}
\caption{
}
\label{fig11}
\end{figure}
%

%
\begin{figure}
\begin{center}
\includegraphics[width=0.8\textwidth,clip=true]{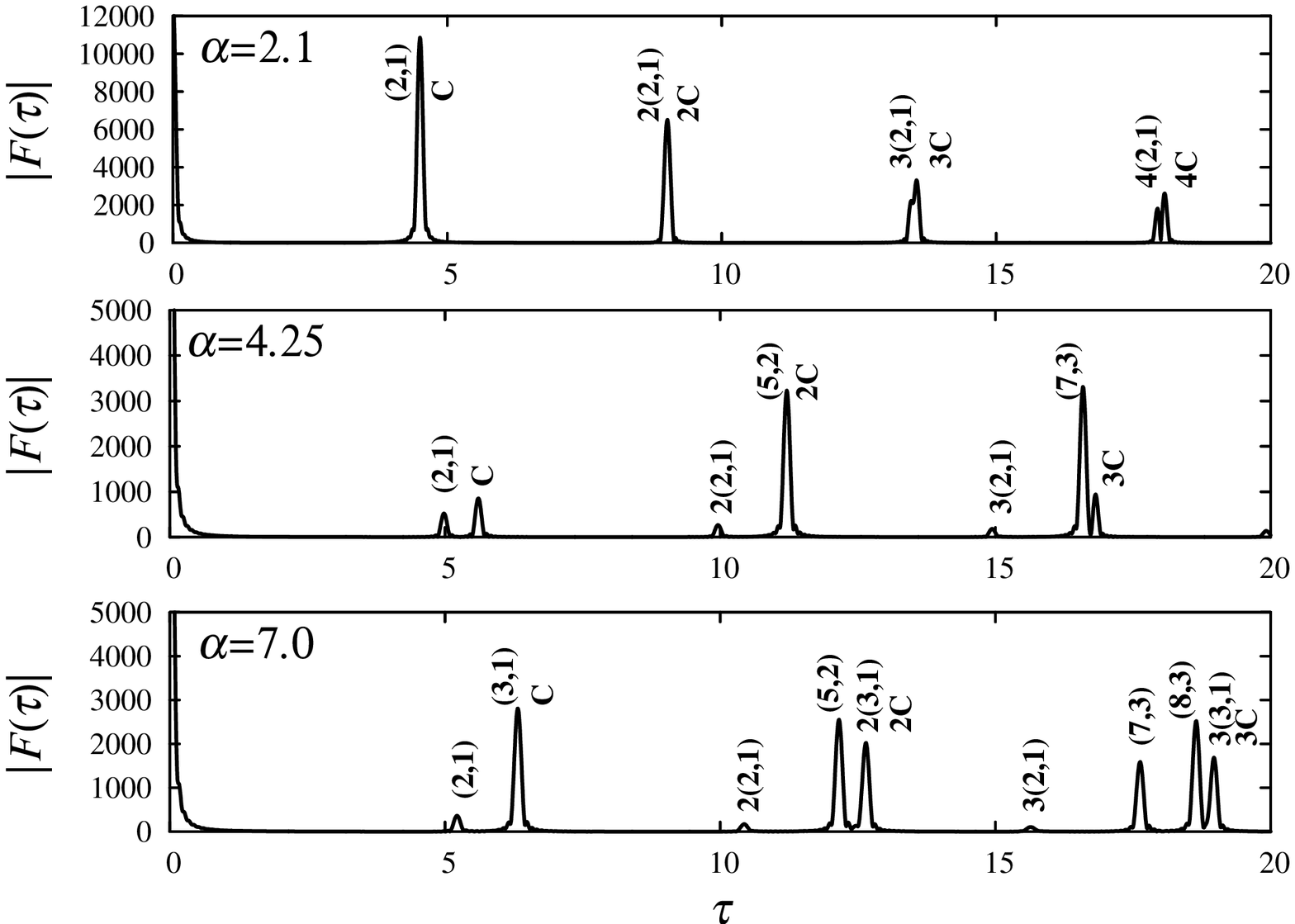}
\end{center}
\caption{
}
\label{fig12}
\end{figure}
\begin{figure}
\begin{center}
\includegraphics[width=0.8\textwidth,clip=true]{fig13.eps}
\end{center}
\caption{
}
\label{fig13}
\end{figure}
\begin{figure}
\begin{center}
\includegraphics[width=0.8\textwidth,clip=true]{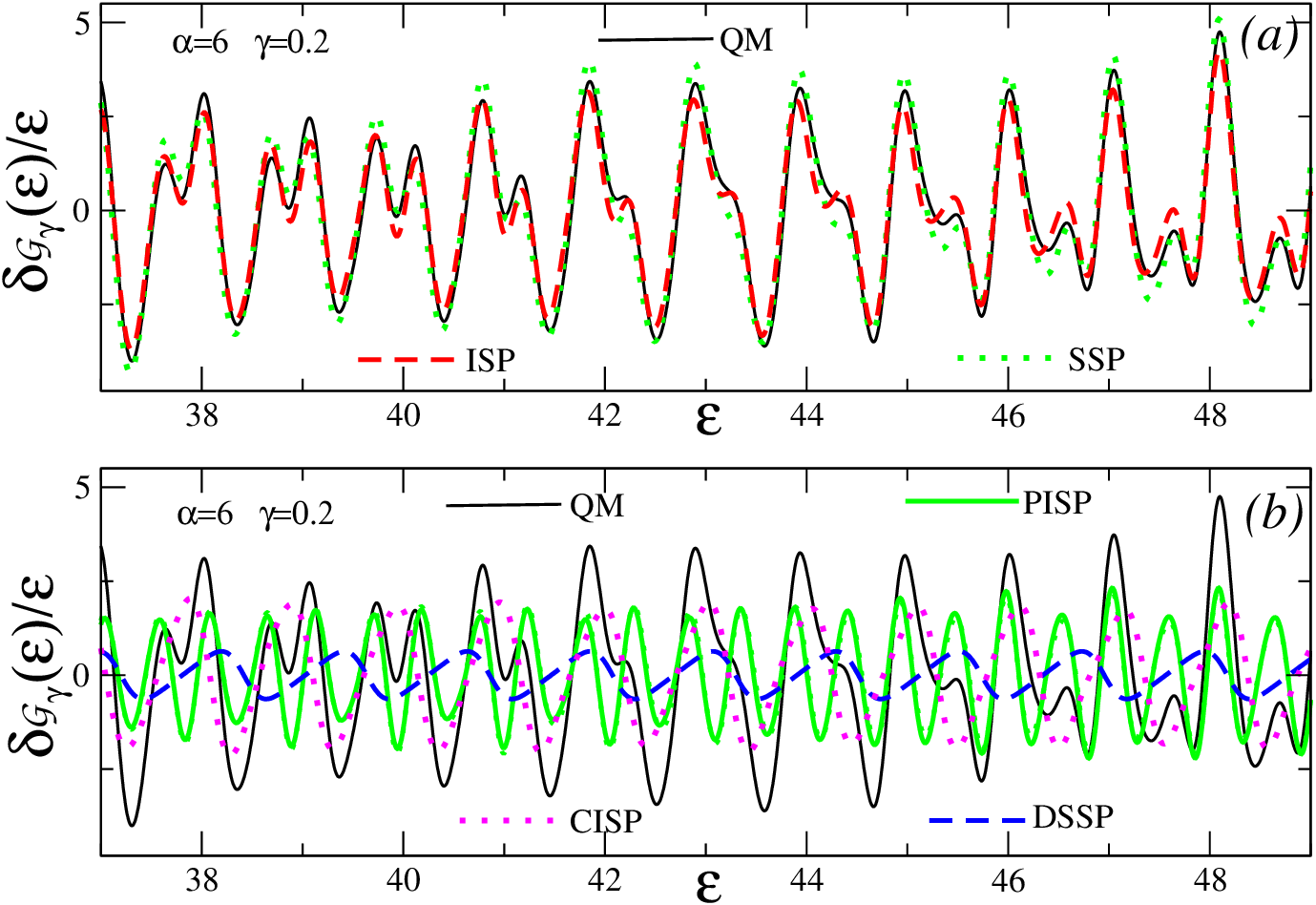}
\end{center}
\caption{ 
}
\label{fig14}
\end{figure}
\begin{figure}
\begin{center}
\includegraphics[width=0.8\textwidth,clip=true]{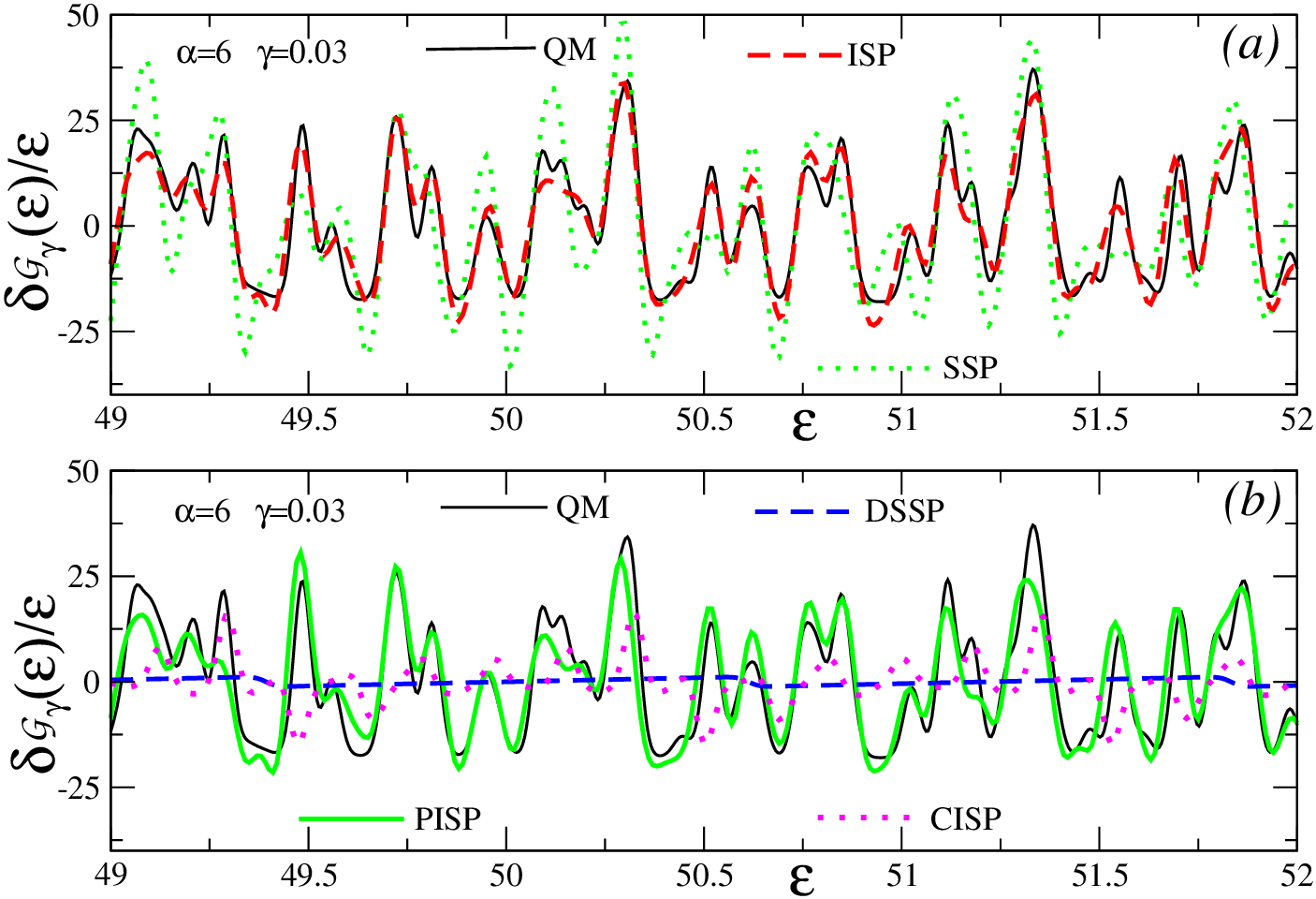}
\end{center}
\caption{ 
}
\label{fig15}
\end{figure}
\begin{figure}
\begin{center}
\includegraphics[width=0.8\textwidth,clip=true]{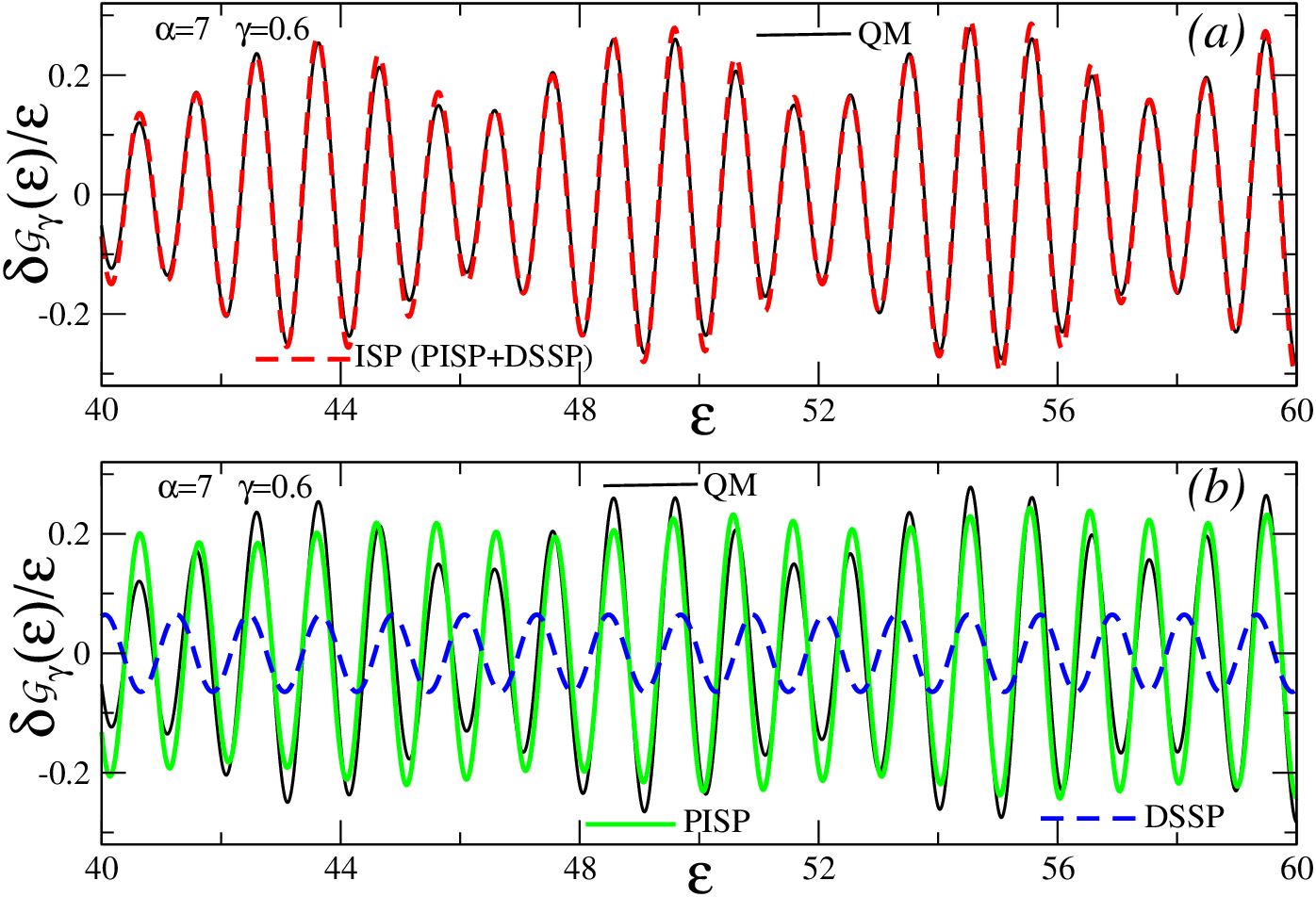}
\end{center}
\caption{ 
}
\label{fig16}
\end{figure}
\begin{figure}
\begin{center}
\includegraphics[width=0.8\textwidth,clip=true]{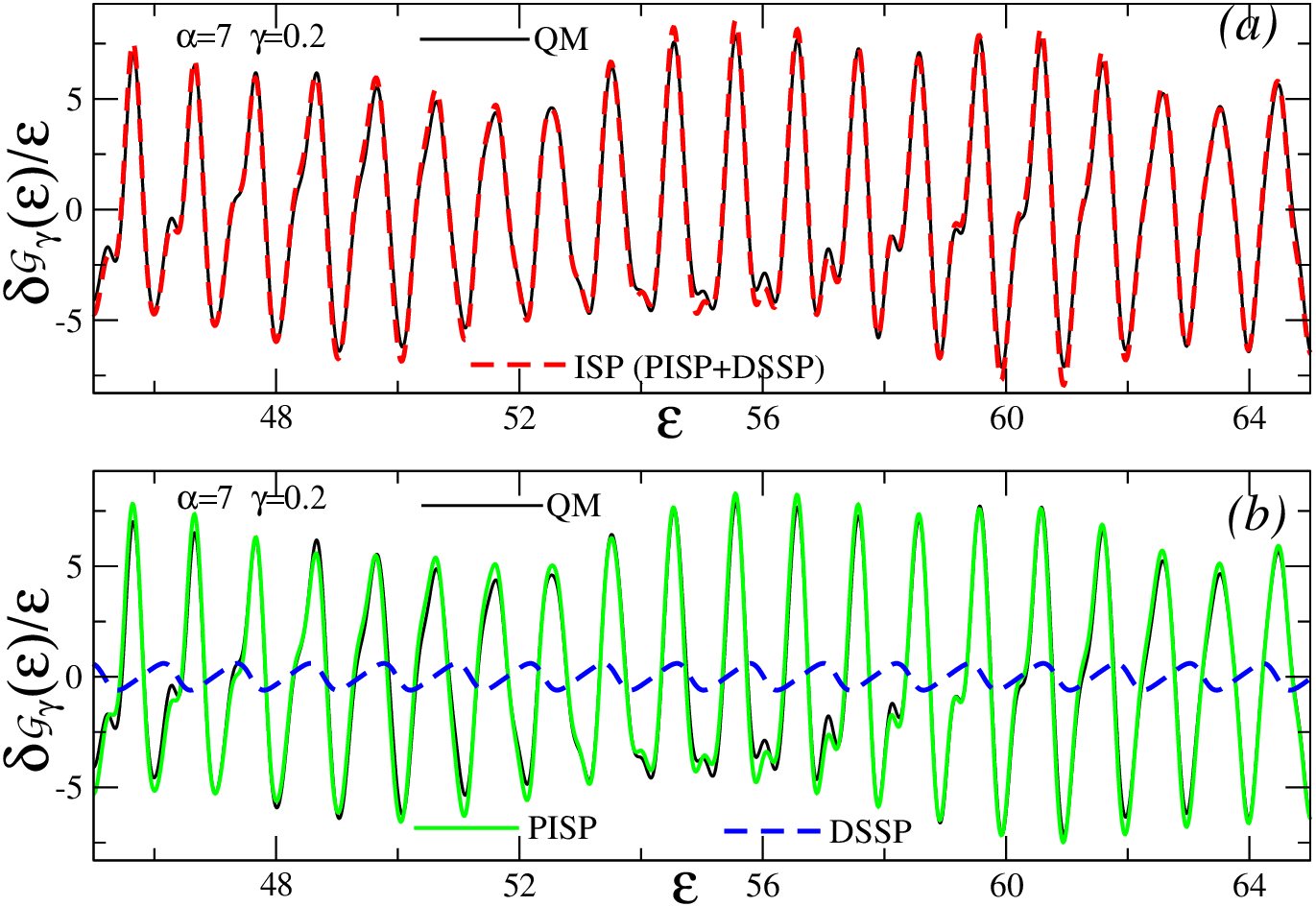}
\end{center}
\caption{ 
}
\label{fig17}
\end{figure}

\begin{figure}
\begin{center}
\includegraphics[width=0.8\textwidth,clip=true]{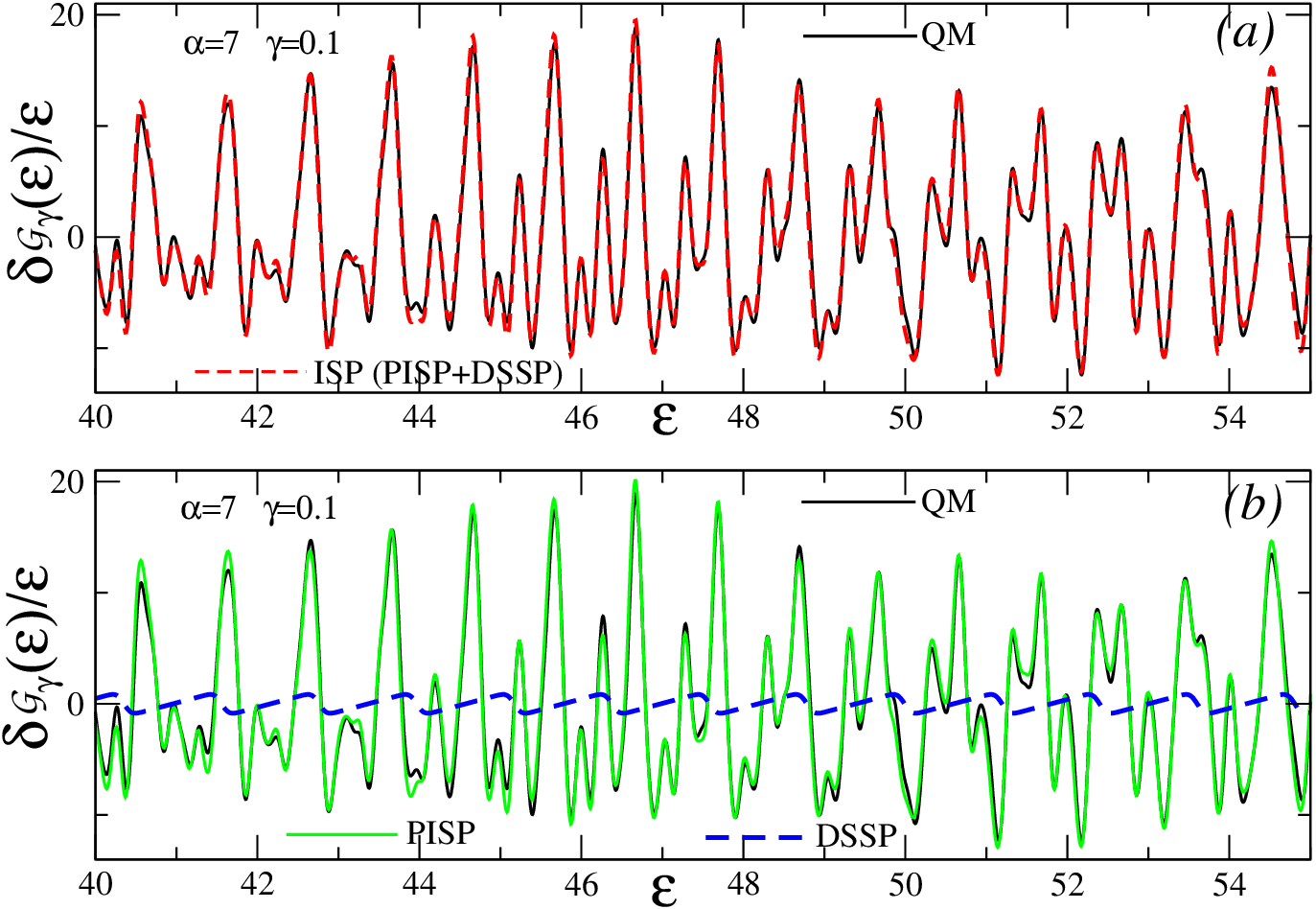}
\end{center}
\caption{ 
}
\label{fig18}
\end{figure}
\begin{figure}
\begin{center}
\includegraphics[width=0.8\textwidth,clip=true]{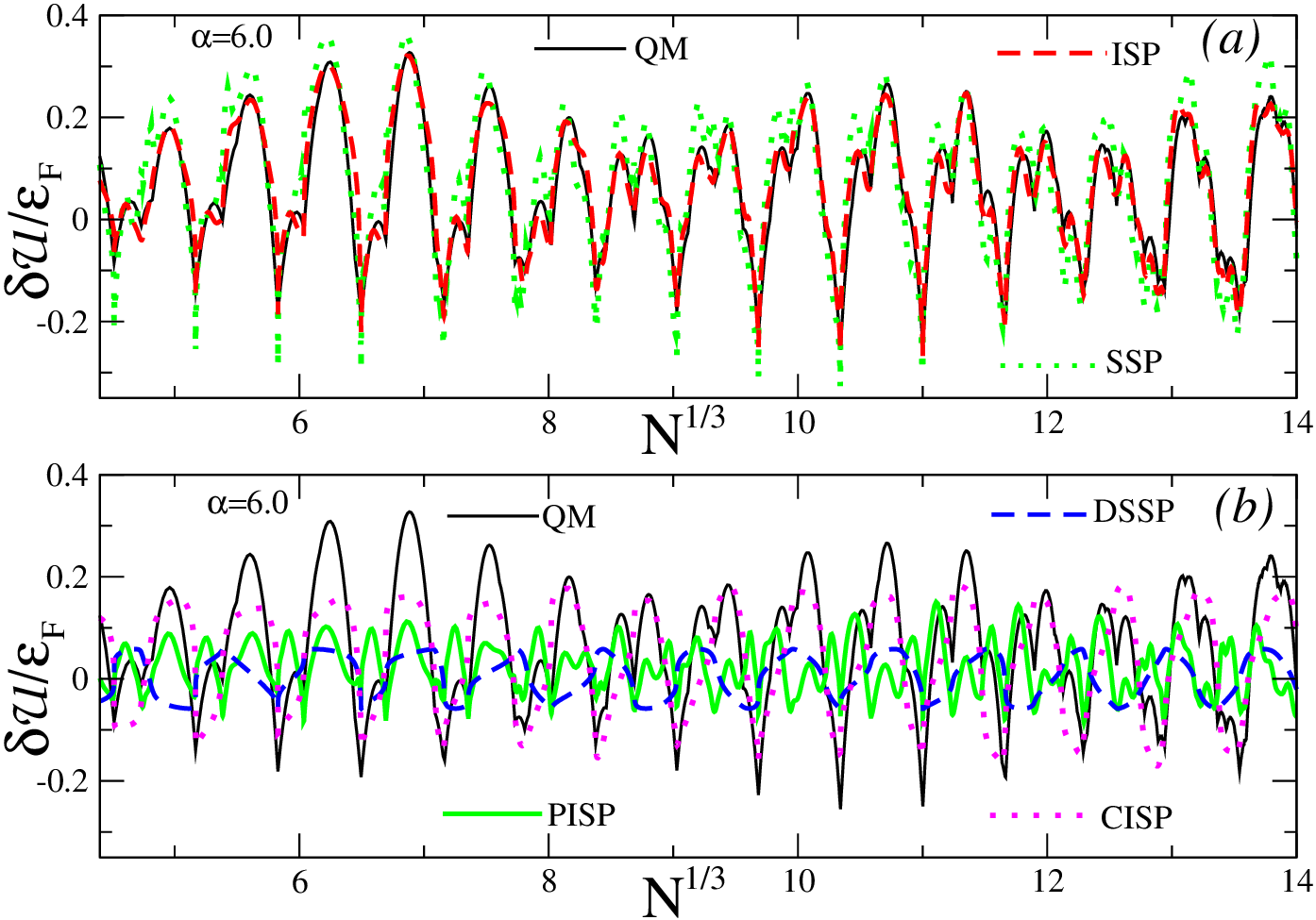}
\end{center}
\caption{ 
}
\label{fig19}
\end{figure}

\begin{figure}
\begin{center}
\includegraphics[width=0.8\textwidth,clip=true]{fig20.eps}
\end{center}
\caption{ 
}
\label{fig20}
\end{figure}

\clearpage
\begin{figure}
\begin{center}
\includegraphics[width=0.6\textwidth,clip=true]{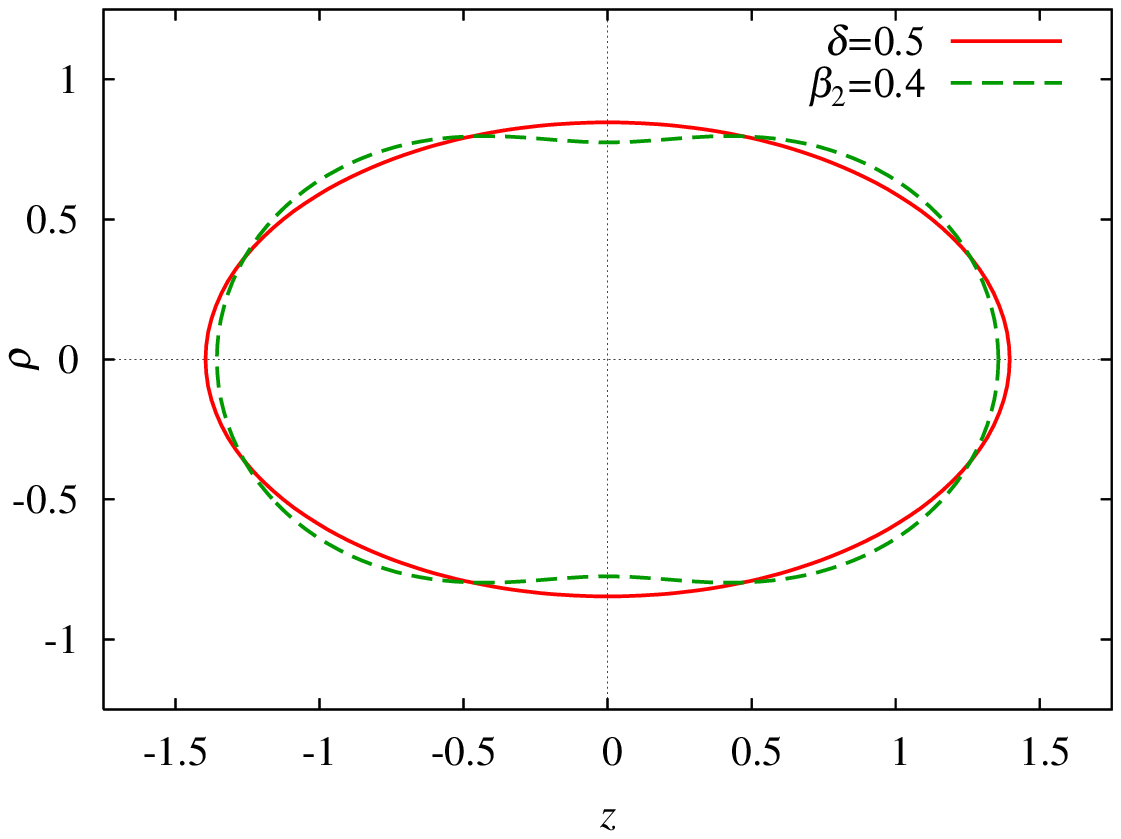}
\end{center}
\caption{
}
\label{fig21}
\end{figure}
\begin{figure}
\begin{center}
\begin{minipage}{.48\textwidth}
\leftline{(a)}\vspace{-\baselineskip}
\hfill\includegraphics[width=.96\linewidth]{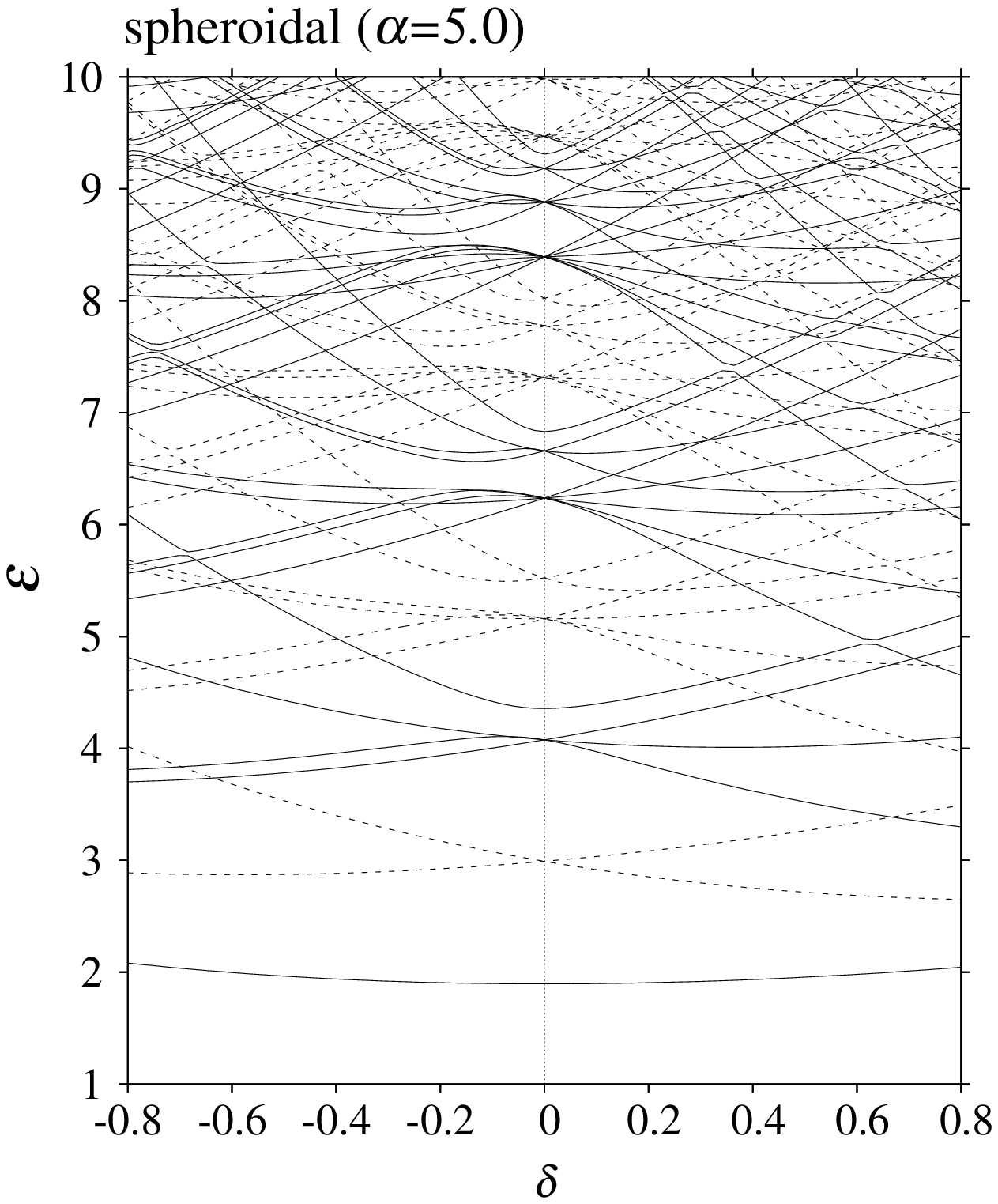}
\end{minipage}
\hfill
\begin{minipage}{.48\textwidth}
\leftline{(b)}\vspace{-\baselineskip}
\hfill\includegraphics[width=.96\linewidth]{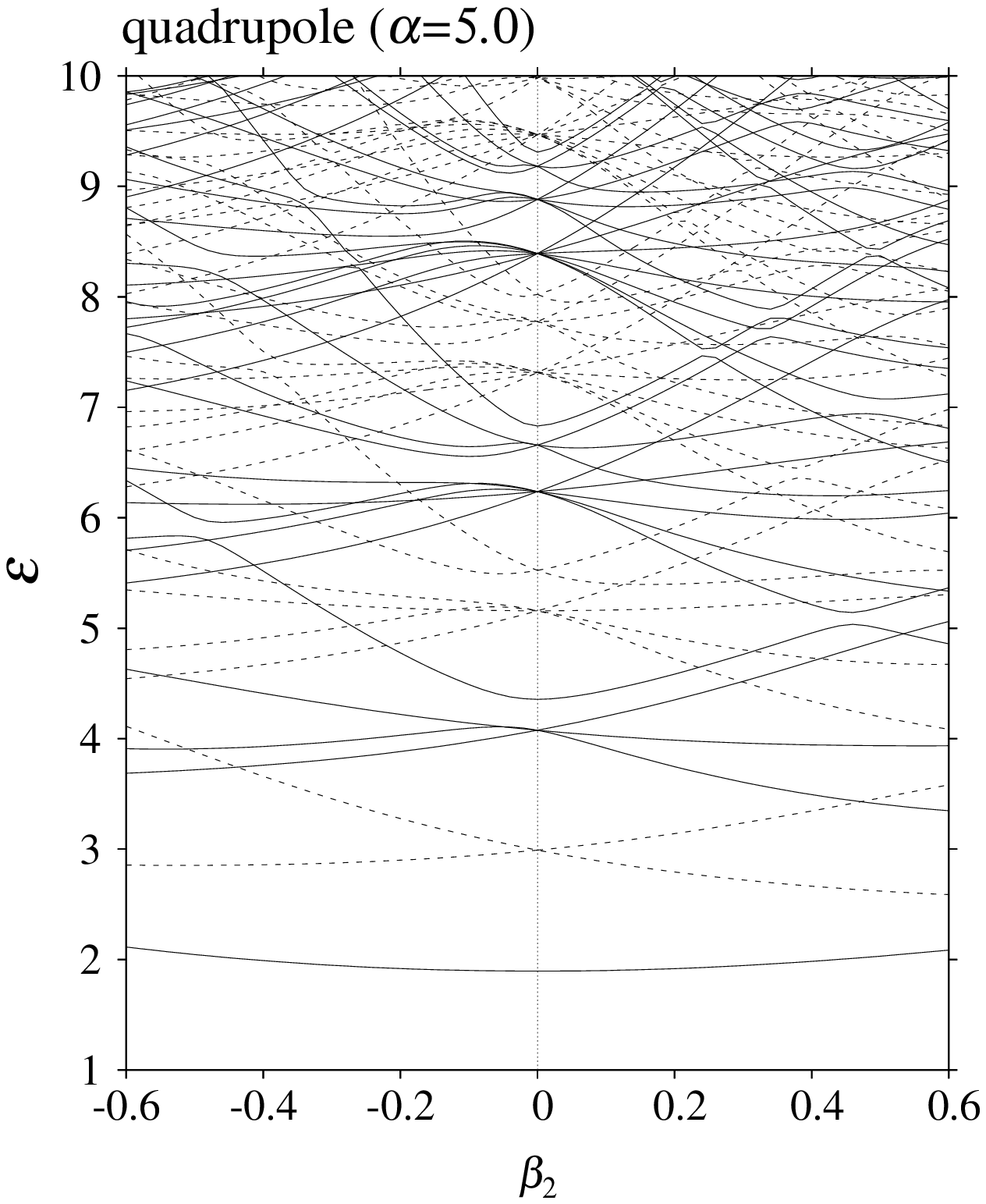}
\end{minipage}
\end{center}
\caption{
}
\label{fig22}
\end{figure}
\begin{figure}
\begin{center}
\begin{minipage}{.4\textwidth}
\leftline{(a)}\vspace{-\baselineskip}
\includegraphics[width=\linewidth]{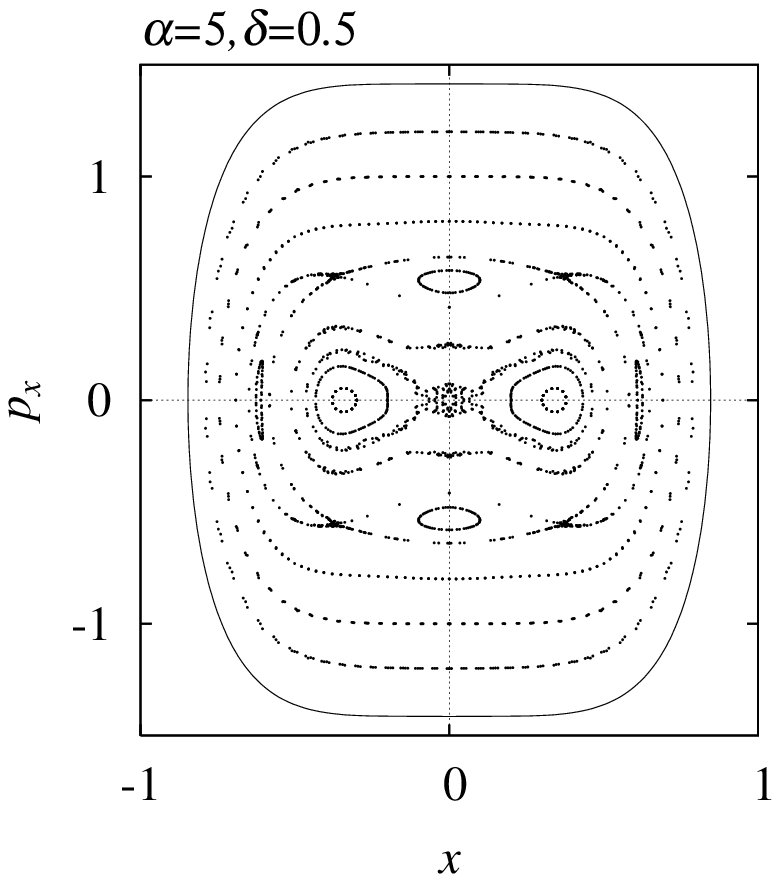}
\end{minipage}
\hspace{.05\textwidth}
\begin{minipage}{.4\textwidth}
\leftline{(b)}\vspace{-\baselineskip}
\includegraphics[width=\linewidth]{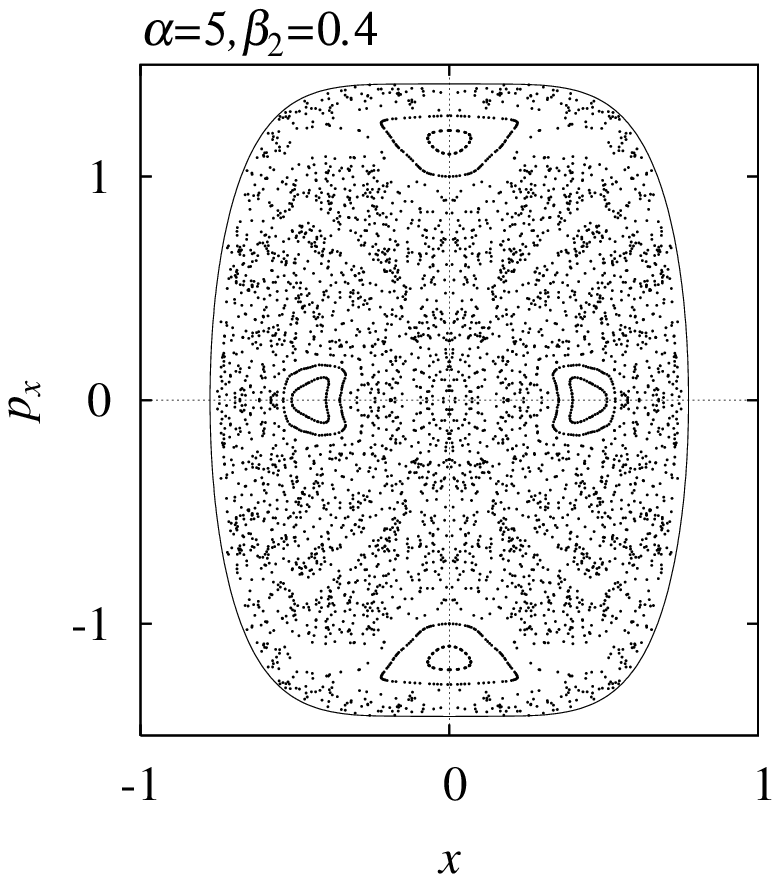}
\end{minipage}
\end{center}
\caption{
}
\label{fig23}
\end{figure}
\begin{figure}
\begin{center}
\begin{minipage}{.25\textwidth}
\includegraphics[bb=70 80 290 220,width=\linewidth]{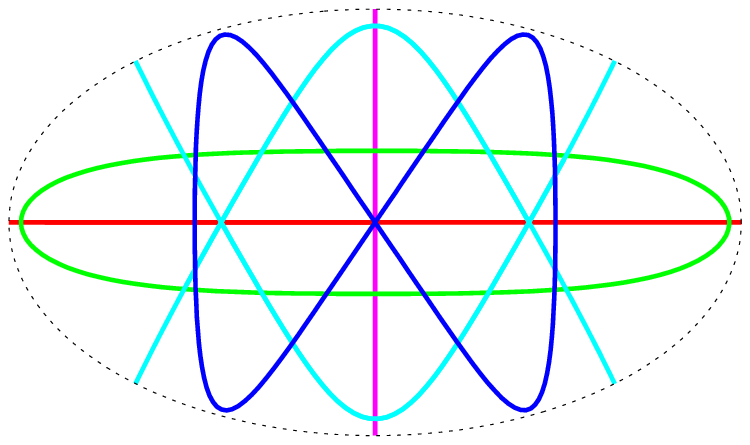}
\end{minipage}
\hspace{.05\textwidth}
\begin{minipage}{.25\textwidth}
\includegraphics[bb=70 80 290 220,width=\linewidth]{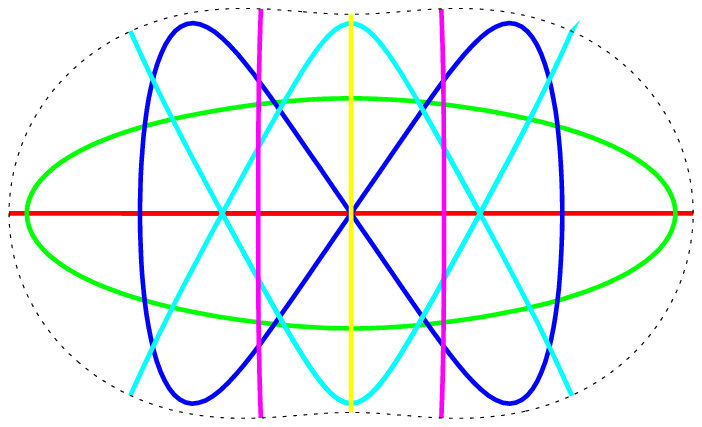}
\end{minipage}
\end{center}
\caption{
}
\label{fig24}
\end{figure}

\begin{figure}
\begin{center}
\includegraphics[width=.6\textwidth]{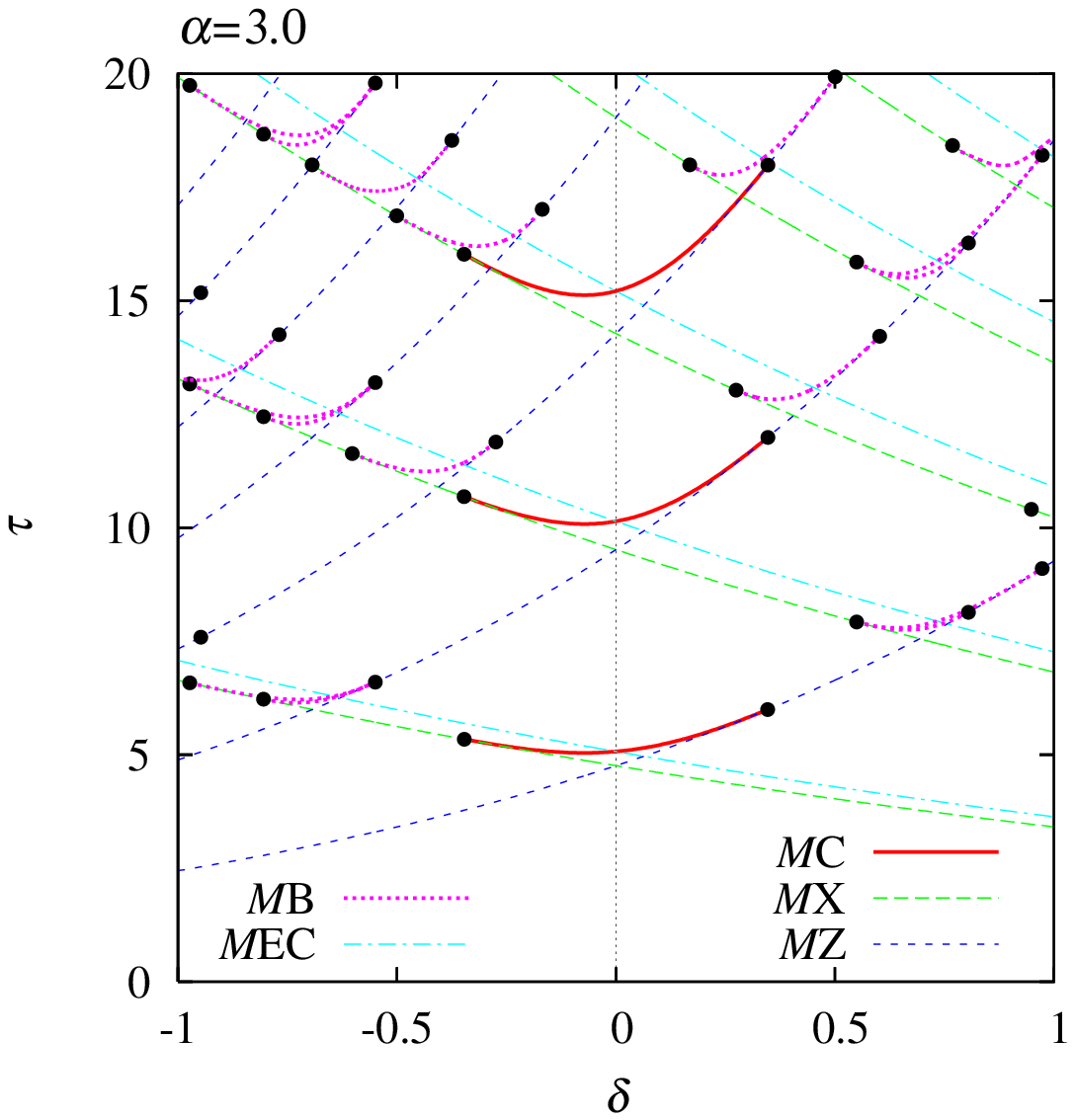}
\end{center}
\caption{
}
\label{fig25}
\end{figure}
\begin{figure}
\begin{center}
\includegraphics[width=.6\textwidth]{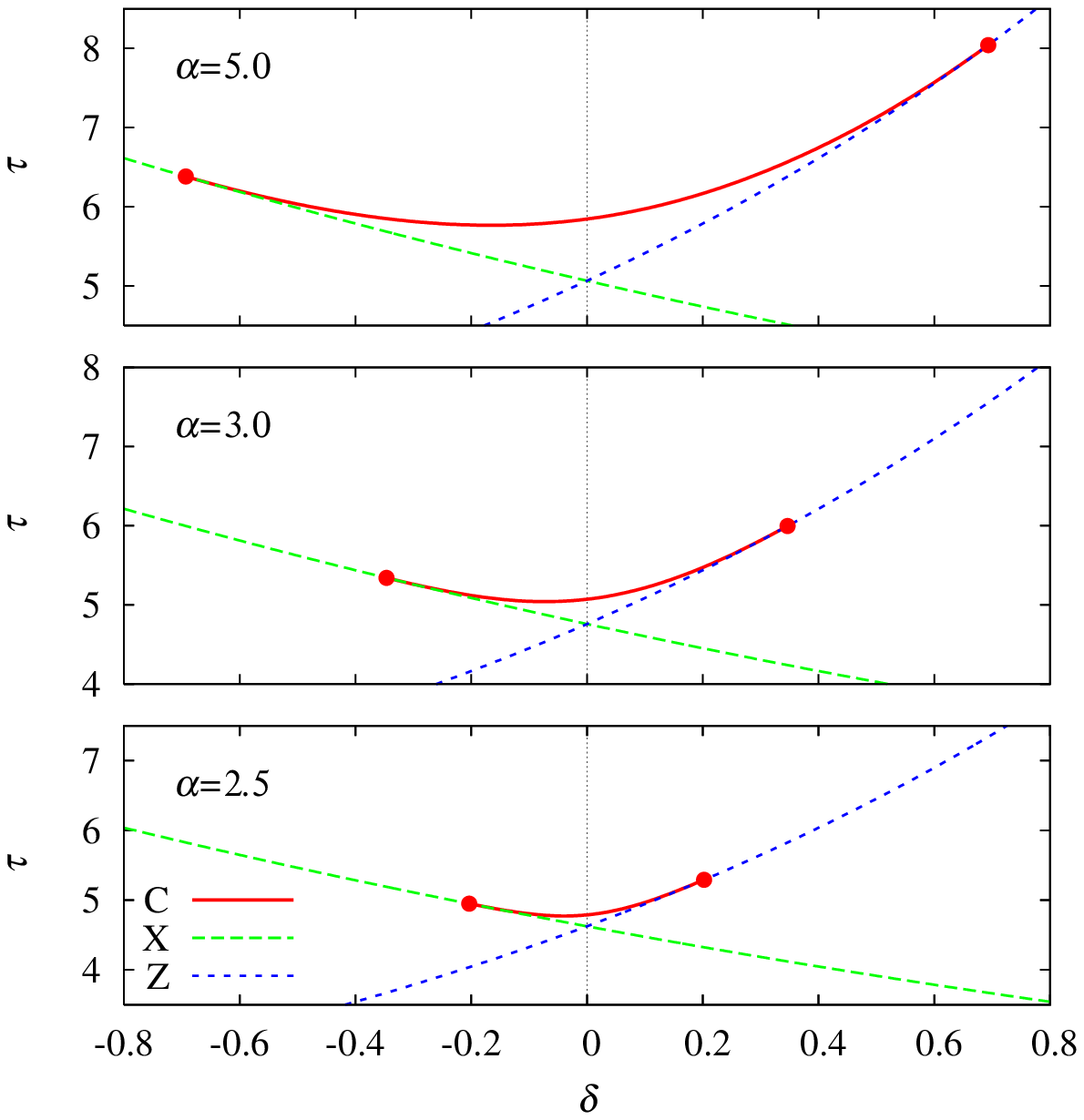}
\end{center}
\caption{
}
\label{fig26}
\end{figure}
\begin{figure}
\begin{center}
\includegraphics[width=.6\textwidth]{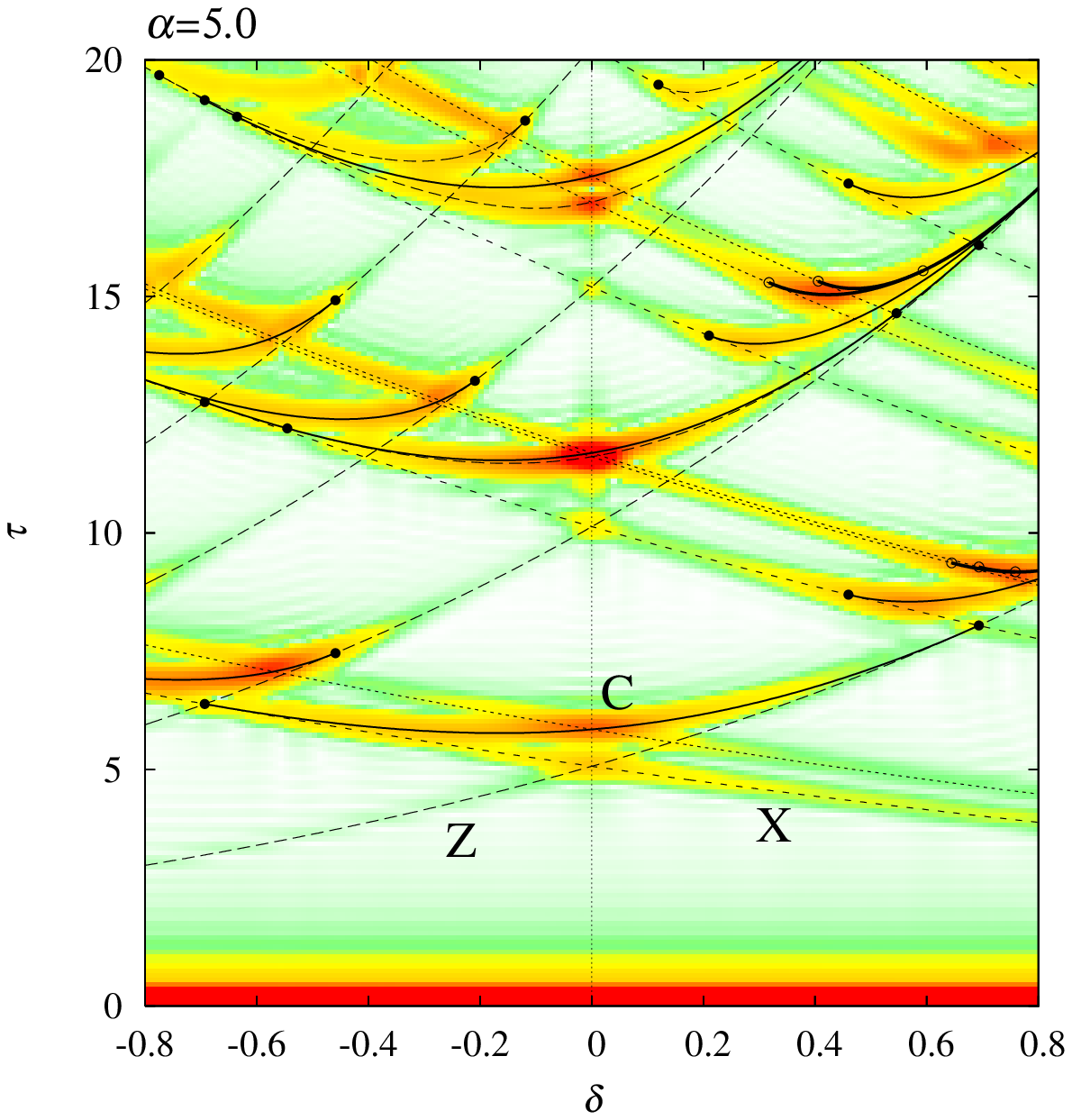}
\end{center}
\caption{
}
\label{fig27}
\end{figure}
\begin{figure}[tbh]
\begin{center}
\includegraphics[width=.6\textwidth]{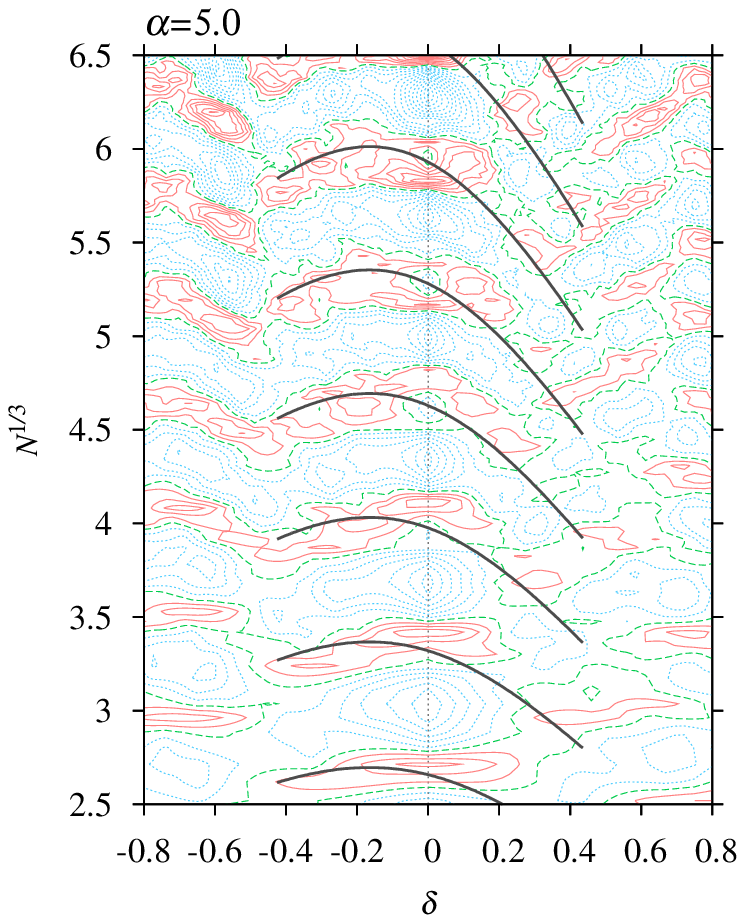}
\end{center}
\caption{
}
\label{fig28}
\end{figure}

%
\begin{figure}
\begin{center}
\includegraphics[width=.8\textwidth]{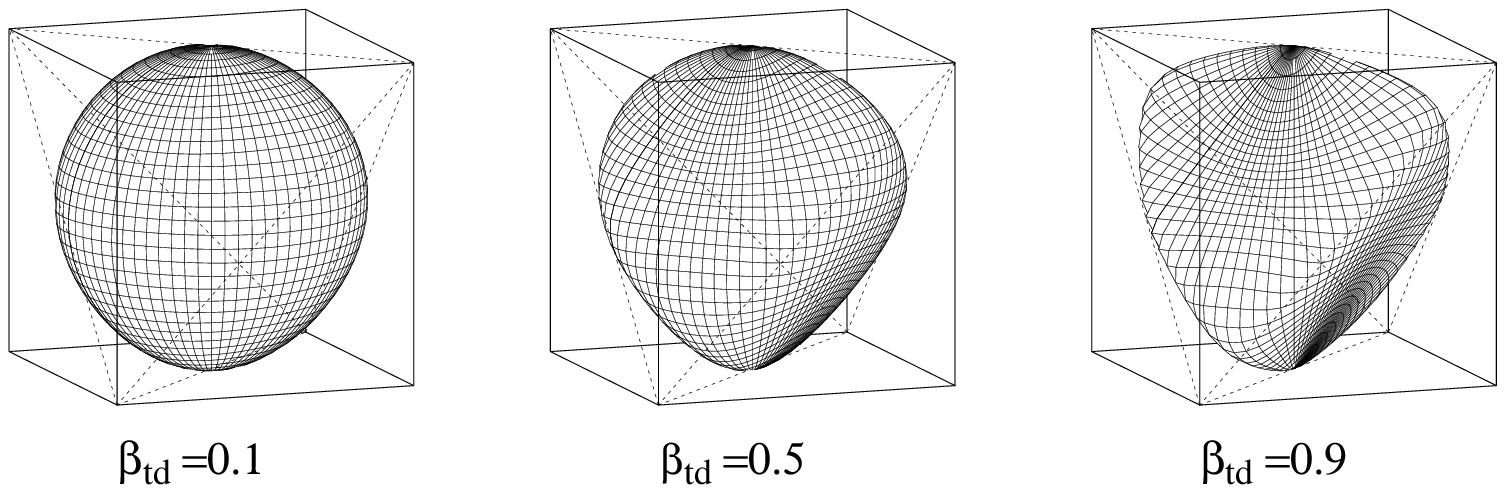}
\end{center}
\caption{
}
\label{fig29}
\end{figure}
\begin{figure}
\begin{center}
\begin{minipage}{.49\textwidth}
\leftline{(a)}\vspace{-\baselineskip}
\hfill\includegraphics[width=.96\linewidth]{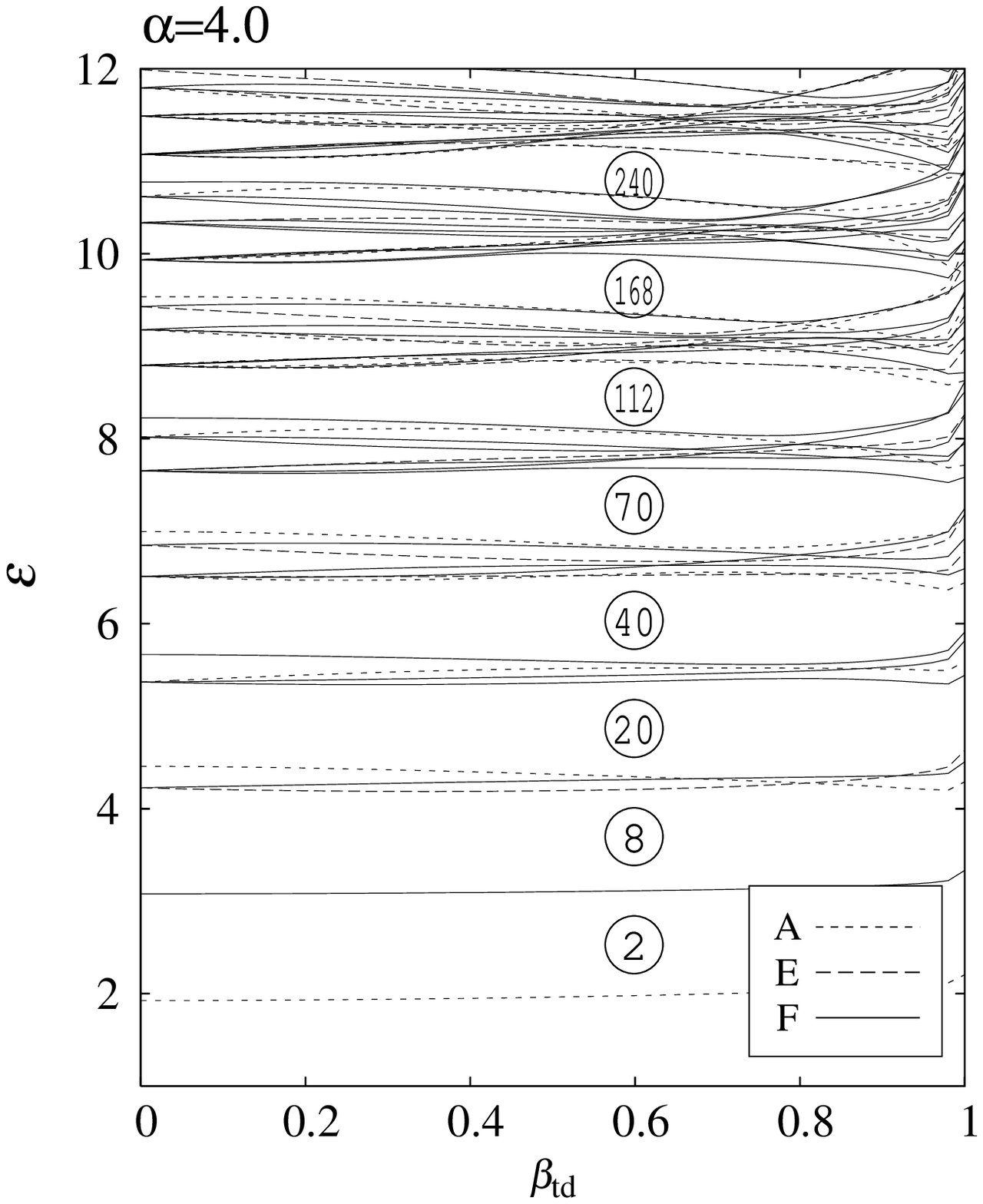}
\end{minipage}
\hfill
\begin{minipage}{.49\textwidth}
\leftline{(b)}\vspace{-\baselineskip}
\hfill\includegraphics[width=.96\linewidth]{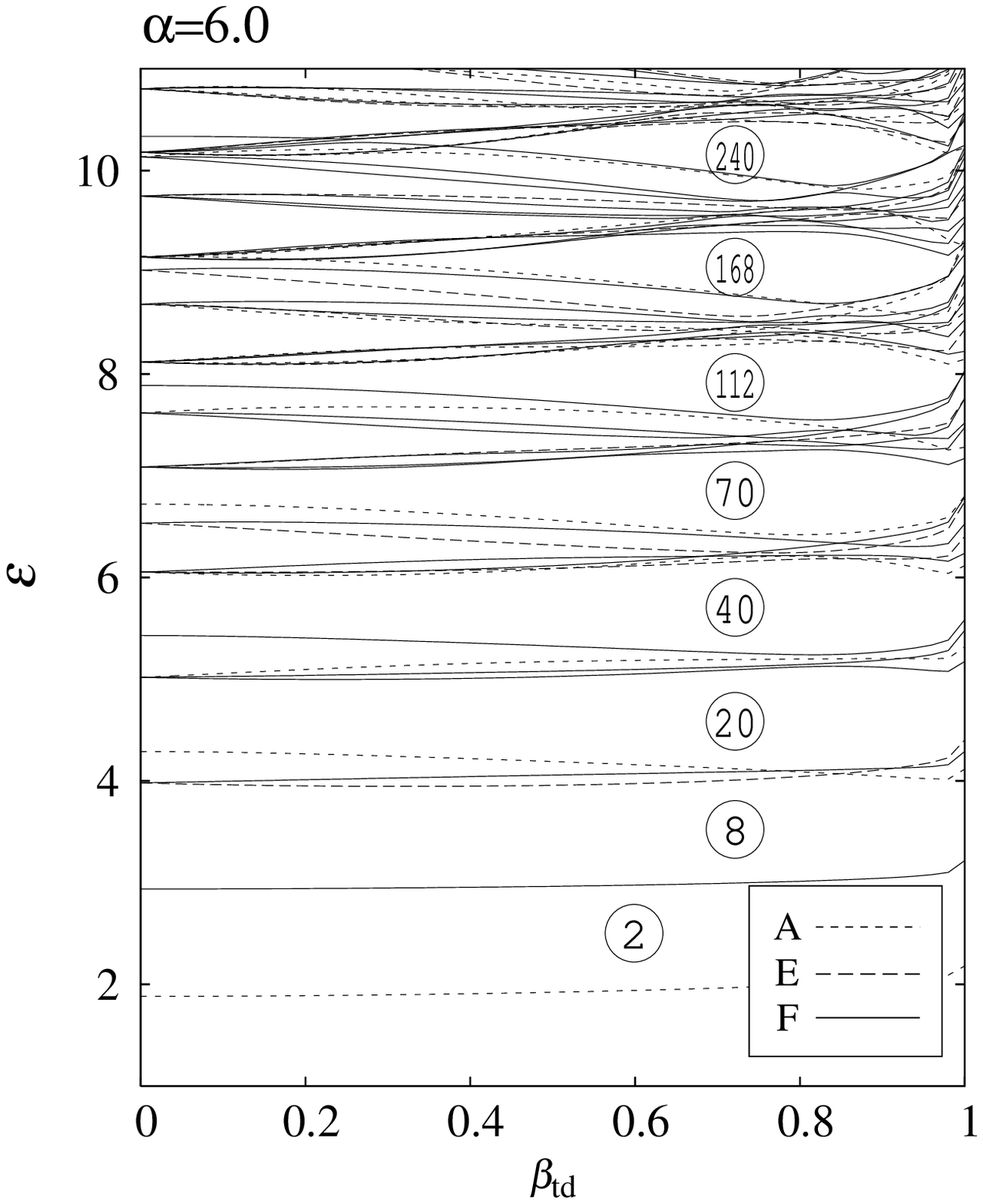}
\end{minipage}
\end{center}
\caption{
}
\label{fig30}
\end{figure}
\begin{figure}
\begin{center}
\includegraphics[width=.6\textwidth]{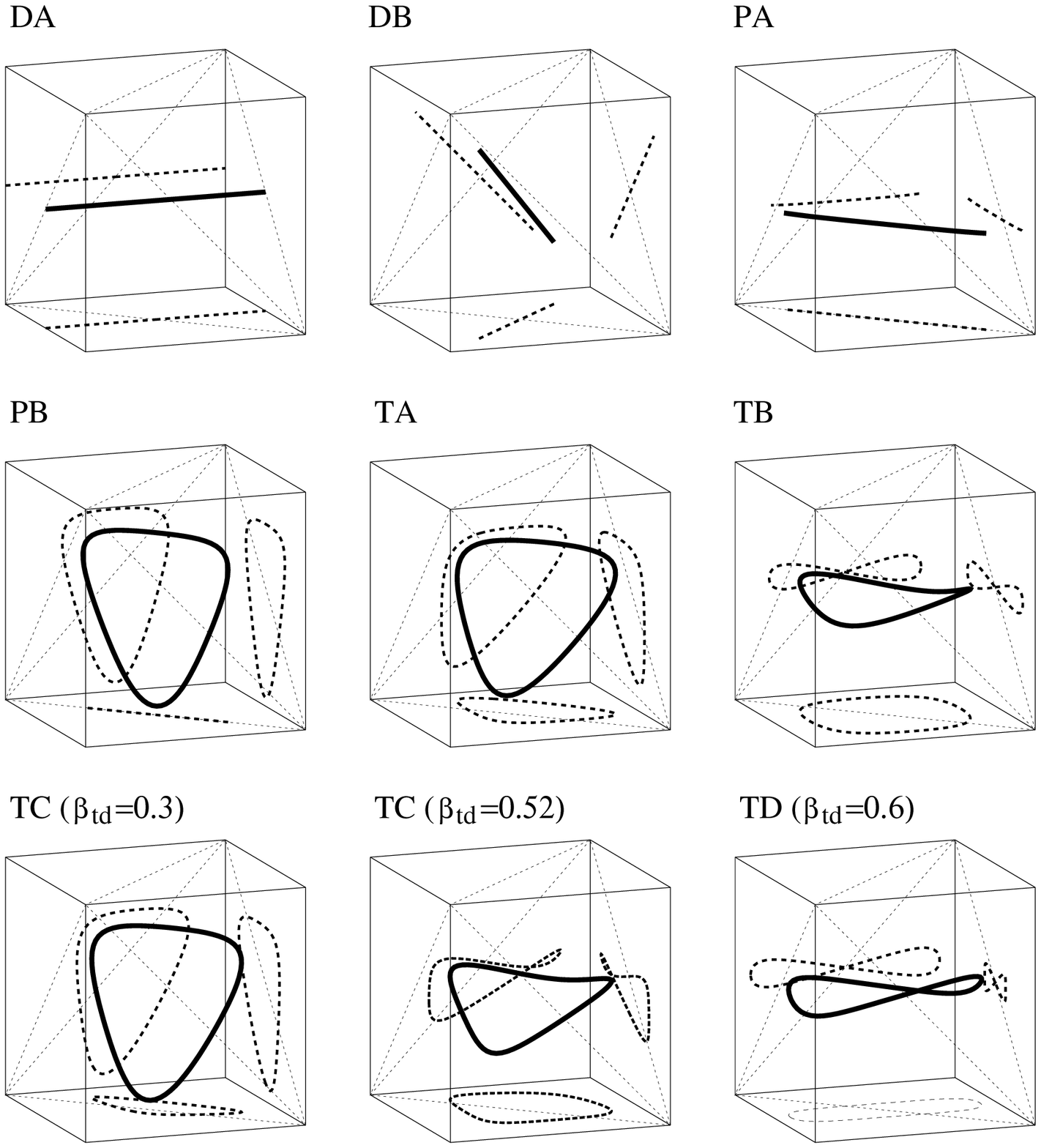}
\end{center}
\caption{
}
\label{fig31}
\end{figure}
\begin{figure}
\begin{center}
\includegraphics[width=.5\textwidth]{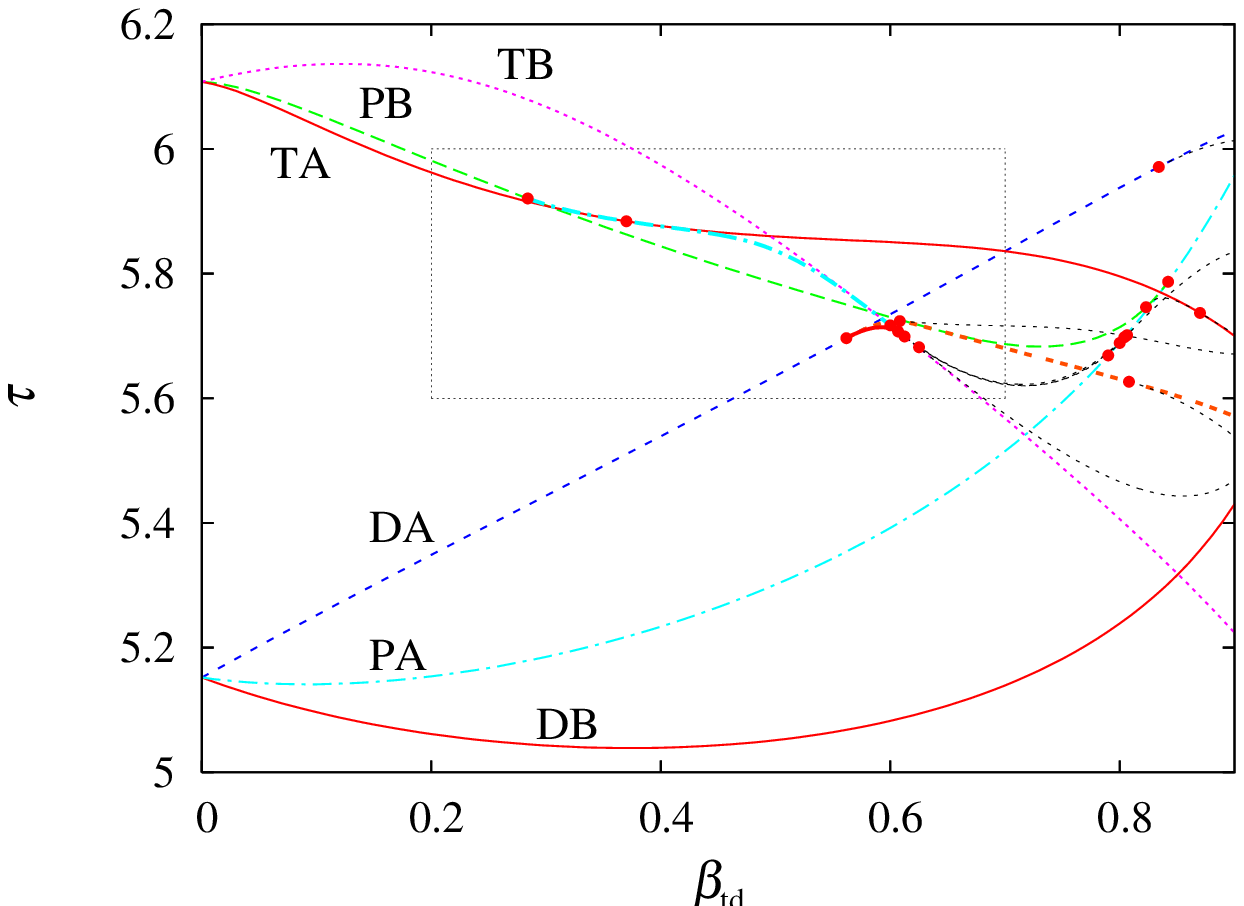}\\
\includegraphics[width=.5\textwidth]{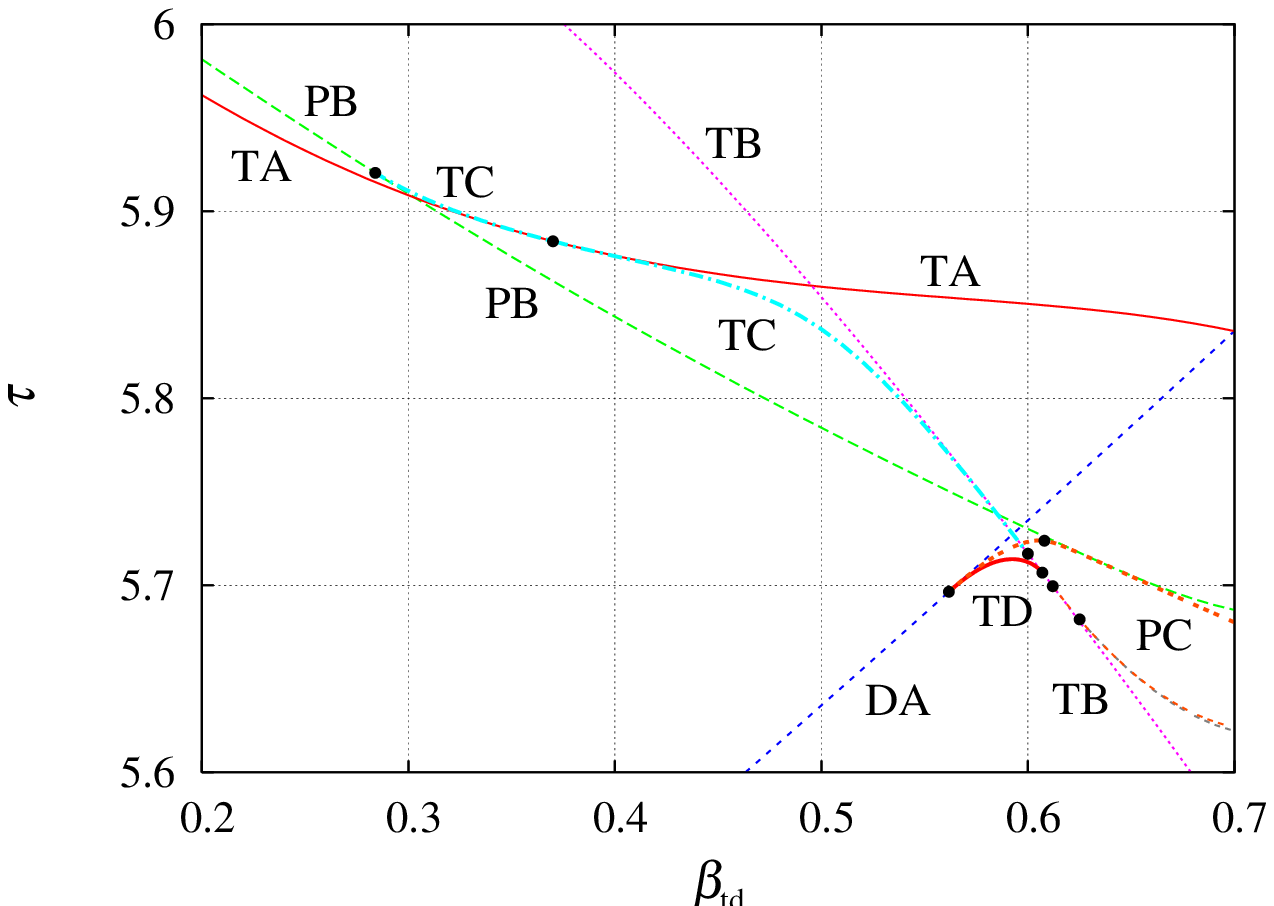}
\end{center}
\caption{
}
\label{fig32}
\end{figure}
\begin{figure}
\begin{center}
\includegraphics[width=.6\textwidth]{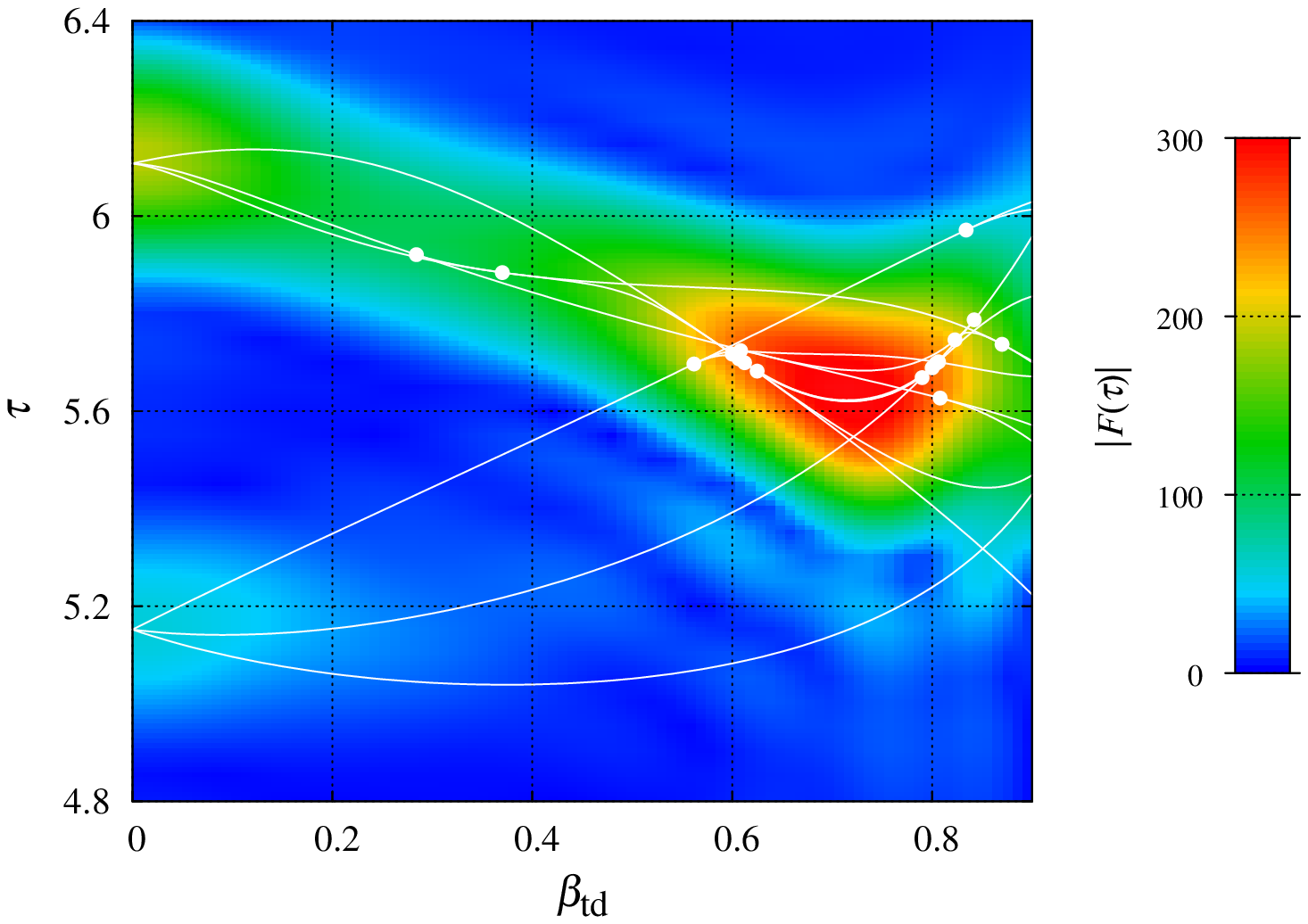}
\end{center}
\caption{
}
\label{fig33}
\end{figure}
\begin{figure}
\begin{center}
\includegraphics[width=.6\textwidth]{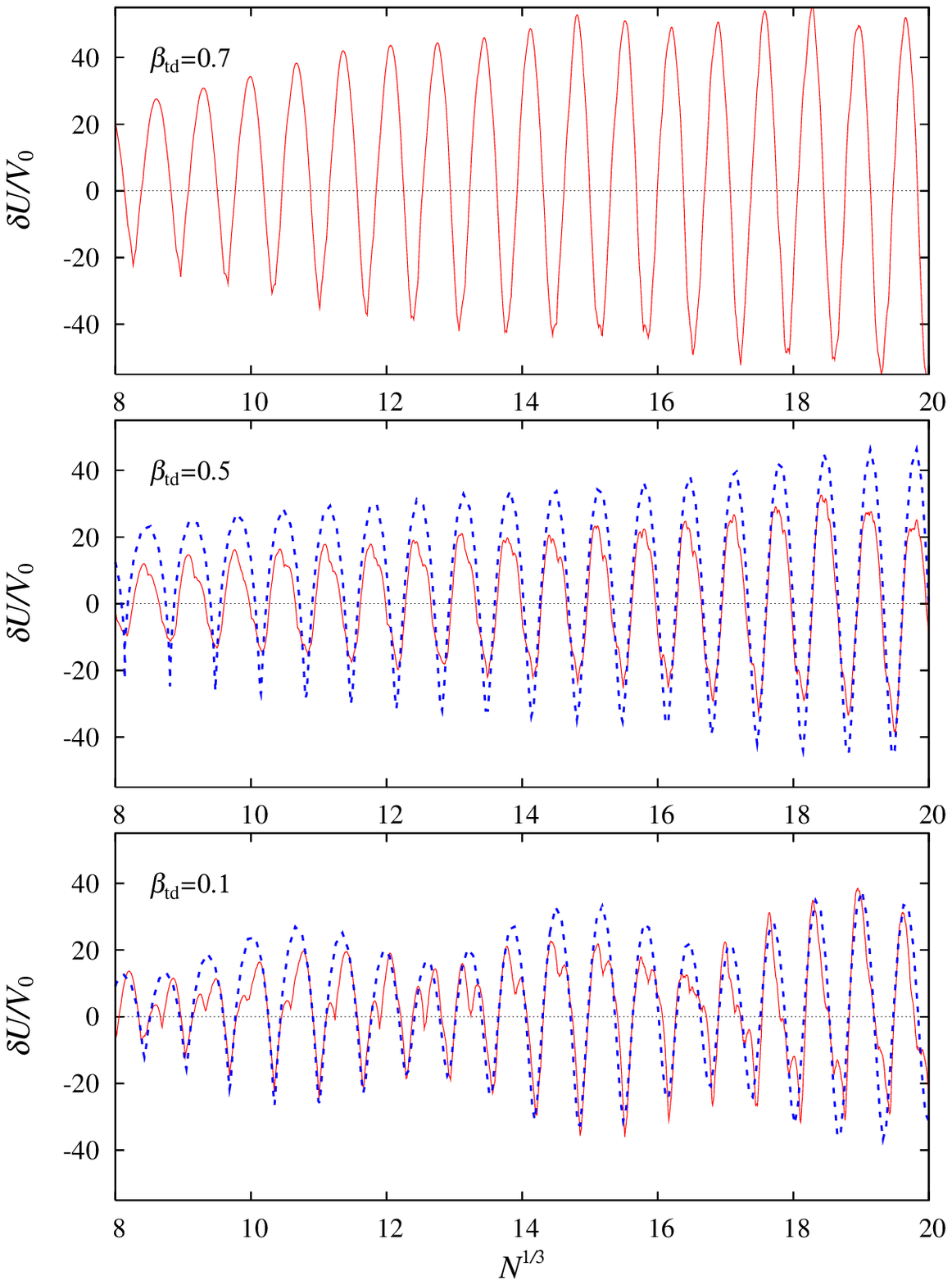}
\end{center}
\caption{
}
\label{fig34}
\end{figure}

\end{document}